\documentclass[amsart,11pt,english]{article}
\usepackage{amssymb,amsmath,amsfonts, amsmath,mathtools,hyperref,
amsthm,euscript,mathrsfs
,MnSymbol
,verbatim
,enumerate,multirow
,bbding,slashed
,multicol,color,array, 
esint,babel, tikz,tikz-cd,tikz-3dplot,tkz-graph,pgfplots,ytableau,graphicx,float,rotating,hyperref,geometry,mathdots,savesym,cite, wasysym,amscd,graphicx,pifont,float,setspace,multirow,wrapfig,picture,subfigure, amsthm,
hepth,
 enumitem, enumerate, booktabs
}
\usepackage{adjustbox}
\usepackage[utf8]{inputenc}
\usepackage{thm-restate}
\usepackage{thmtools}
\usepackage{prettyref}
\usepackage[T1]{fontenc}
\usepackage{datetime}
\numberwithin{equation}{section}
\usetikzlibrary{arrows,positioning,decorations.pathmorphing,   decorations.markings, matrix, patterns,shapes}
\hypersetup{colorlinks=true}
\hypersetup{linkcolor=black}
\hypersetup{citecolor=black}
\hypersetup{urlcolor=black}
\theoremstyle{plain}
\newtheorem*{thm*}{Theorem}
\theoremstyle{plain}
\newtheorem{thm}{Theorem}[section]

\newtheorem{lem}[thm]{Lemma}
\newtheorem{prop}[thm]{Proposition}

\theoremstyle{definition}
\newtheorem{defn}[thm]{Definition}
\newtheorem*{defn*}{Definition}

\newtheorem{rem}[thm]{Remark}

\usepackage[normalem]{ulem}
\tikzset{
  big arrow/.style={
    decoration={markings,mark=at position 1 with {\arrow[scale=1.5,#1]{>}}},
    postaction={decorate},
    shorten >=0.4pt},
  big arrow/.default=black}

\numberwithin{equation}{section}

\begin{document}
\thispagestyle{empty}
\begin{titlepage}
\begin{center}
\vspace{4cm}
{\Huge\bfseries  
  Flopping and Slicing:  SO($4$) and Spin($4$)-models\\
  }
\vspace{1cm}
{%
\LARGE  Mboyo Esole$^{\diamondsuit}$ and Monica Jinwoo Kang$^\clubsuit$ 
\\}
\vspace{1cm}

{\large $^{\diamondsuit}$ Department of Mathematics, Northeastern University, Boston, MA, 02115, USA}\par
{\large $^\clubsuit$ Department of Physics, Harvard University, Cambridge, MA 02138, USA}\par
 \scalebox{.95}{\tt  j.esole@northeastern.edu,   jkang@fas.harvard.edu 
  }\par
\vspace{3cm}
{ \bf{Abstract:}}\\
\end{center}
We study the geometric engineering of gauge theories with gauge group Spin($4$) and SO($4$) using  crepant resolutions of Weierstrass models. 
The corresponding elliptic fibrations realize a collision of singularities corresponding to two fibers with dual graphs $\widetilde{\text{A}}_1$. 
There are eight different ways to engineer such collisions using decorated Kodaira fibers. 
The  Mordell--Weil group of the elliptic fibration is required to be trivial for Spin($4$) and   $\mathbb{Z}/2\mathbb{Z}$ for SO($4$). 

Each of these models have two possible crepant resolutions connected by a flop. 
We also compute a generating function for the Euler characteristic of such elliptic fibrations over a base of arbitrary dimensions. 
In the case of a threefold, we also compute the triple intersection numbers of the fibral divisors. 
In the case of Calabi-Yau threefolds, we also compute their Hodge numbers, and check the cancellations of anomalies in a six-dimensional supergravity theory. 

\vfill 

\end{titlepage}

\thispagestyle{empty}

\tableofcontents
\thispagestyle{empty}

\setcounter{page}{1}

\section{Introduction}\label{sec:intro}

Crepant resolutions of Weierstrass models and their network of flops is a topic of interest for both mathematicians and physicists \cite{Anderson:2017zfm,Matsuki.Weyl,Nakayama,IMS}. 
 Such crepant resolutions form a rich setup to study elliptic fibrations since any elliptic fibration over a smooth base is birational to a Weierstrass model, and in many important cases, the birational transformation to the Weierstrass model preserves the canonical class of the elliptic fibration (see also \cite{Esole:2012tf}).   A crepant resolution of a Weierstrass model is a minimal model in the sense of Mori's theory, and minimal models are connected by flops \cite{Matsuki.Weyl,Nakayama}.

In F-theory and M-theory, elliptic fibrations are used as compactifying spaces  to geometrically engineer gauge theories  in space-times of dimension less than or equal to eight \cite{Morrison:1996na,Morrison:1996pp,Bershadsky:1996nh}.  
The F-theory's  algorithm attaches to a given elliptic fibration a  Lie algebra $\mathfrak{g}$, a Lie group $G$, and a representation $\mathbf{R}$ of $G$. 
 The Lie algebra $\mathfrak{g}$ is determined by the singular fibers over generic points of the discriminant locus. As the fundamental group of $G$ is isomorphic to the Mordell--Weil group of the elliptic fibration, the global structure of the Lie group $G$ depends on not only the Lie algebra but also the Mordell--Weil group of the fibration \cite{Aspinwall:1998xj,Mayrhofer:2014opa,Vafa:1996xn}. The representation $\mathbf{R}$ is derived from the singular fibers over codimension-two points of the base. In particular, the weights of $\mathbf{R}$ are geometrically computed by intersection of rational curves composing the singular fibers and fibral divisors that project to irreducible components of the discriminant (see section \ref{Sec:Rep}) .  An elliptic fibration for which the F-theory algorithm returns a   Lie group $G$ is called a $G$-model. The geometry of $G$-models are well understood mostly when $G$ is a simple group. 
 The semi-simple case is more subtle  because of the appearance of  colliding singularities.

\subsection{Colliding singularities}
The theory of elliptic surfaces is well understood since the seminal work of Kodaira and N\'eron \cite{Kodaira, Neron}. 
For an elliptic surface, the discriminant locus (i.e. the locus of singular fibers) is composed of isolated points and the singular fibers are classified by Kodaira symbols (see section \ref{Sec:Kodaira}). 

When the elliptic fibration is of dimension three or higher,  the discriminant locus can have  intersecting components. 
Kodaira fibers now classify the type of the geometric fiber over the generic point of a normal irreducible component of the discriminant locus. 
Singularities of the discriminant locus are called {\em collisions of singularities} \cite{Miranda}. 
A typical example of collision of singularities are intersection points of irreducible components of the discriminant locus. 
 Let $S_1$ and $S_2$ be two smooth irreducible components of the discriminant locus with generic fibers of Kodaira type $T_1$ and $T_2$.  
The generic fiber at the intersection of $S_1$ and $S_2$ is denoted by $T_1+T_2$. 
The type of the generic fiber at the collision does not have to be one of Kodaira's types \cite{Miranda,EY}. But it is usually a contraction of a Kodaira type or obtained by  letting some of the nodes of a Kodaira fiber to coincide. 
We will assume throughout this paper that we work over the complex numbers.

In the early 1980s, Miranda introduced a systematic  regularization procedure for elliptic threefolds defined by Weierstrass models \cite{Miranda}. He considers  collisions of type $T_1+T_2$ in which  $T_1$ and $T_2$ are  Kodaira fibers  having the same $j$-invariant  and such that the supporting divisors  are smooth divisors intersecting transversally. 
Mi randa's regularization produces  elliptic fibrations that are flat fibrations and their  $j$-invariant is a morphism. 
 Miranda's work on threefolds was generalized to elliptic $n$-folds in Szydlo's  Ph.D. thesis \cite{Szydlo.Thesis}. 
Elliptic fibrations resulting from Miranda's regularization are called {\em Miranda models}  \cite{MR1242006}. 
Miranda models were used by Dolgachev and Gross to study the Tate-Shafarevich group of elliptic threefolds \cite{MR1242006}. 
Using tools from the Minimal Model Program, Hodge theory, and toric geometry, Nakayama studied in \cite{Nakayama} the local fibration structure of elliptic threefolds defining a collision $\text{I}_n+\text{I}_m$ with normal transverse divisors in a nonsingular surface.

Miranda regularization is not usually a crepant resolution.  But when it is, it provides interesting examples of 
elliptic Calabi-Yau threefolds yielding non-Abelian semi-simple gauge theories. Applications of Miranda model to F-theory was first studied by Bershadsky and Johansen \cite{Bershadsky:1996nu} 
 and have applications in the classification of six-dimensional Conformal Field Theories (CFTs).

\subsection{SO($4$) and Spin($4$)-models}
In F-theory and M-theory, collisions of singularities  are crucial for geometric engineering of  gauge theories with semi-simple Lie groups, and matters charged under some representation of the Lie group \cite{Morrison:1996na,Morrison:1996na,Vafa:1996xn}.  
In recent years, we have improved our understanding of crepant resolutions of elliptic fibrations corresponding to simple Lie groups \cite{EY,ESY1,ESY2,ES,F4,G2}. 
We have extended these methods to the study of elliptic fibrations corresponding to semi-simple groups \cite{EKY1}, including the case in which the gauge group is not simply connected. 
Following a point of view that started in \cite{AE1,AE2,EFY,EKY1}, we work relatively to an arbitrary base and do not impose the Calabi-Yau condition. 
This allows us to understand the geometry of these elliptic fibrations in a larger setting before to specialize to the particular cases relevant to string theory. 
In the case of Calabi-Yau threefolds, such analyses are closely related to basic questions on five dimensional and six dimensional supergravity theories since they are obtained by compactifications of 
M-theory and F-theory. 
In the case of the five-dimensional theory, we  determine the matter content, the structure of the Coulomb branch, and the Chern-Simons levels. 
In its six-dimensional theory uplift, we determine the tensor branch, the matter content, and the fine detail of cancellations of gravitational, gauge, and mixed anomalies. 

In this paper, we study  the geometry of SO($4$) and Spin($4$)-models. 
The SO($n$) and Spin($n$)-models for $n\neq 4, n\geq 3$ are associated with gauge theories with simple gauge groups and  do not require a collision of singularities. 
 SO($4$) and Spin($4$)-models are associated with the semi-simple Lie algebra of type D$_2$, which is 
the    unique reducible semi-simple Lie algebra  $\mathfrak{g}$ of rank two, and  
the direct sum of two Lie algebras of type A$_1$:
$$
\text{D}_2\cong \text{A}_1\oplus \text{A}_1.
$$ 
In this sense, the SO($4$) and Spin($4$)-models are the simplest $G$-models with $G$ a semi-simple group.\footnote{ As discussed in section \ref{sec:Spin}, there are also other compact groups with Lie algebra D$_2$, namely the half-spin groups HSpin$^\pm$($4$) which are isomorphic to SO($3$)$\times$ SU($2$) and SU($2$) $\times$ SO($3$). However, these groups are not compatible with the typical representation ($\mathbf{2},\mathbf{2}$) observed at the collisions of two fibers with dual graphs  $\widetilde{\text{A}}_1$.}

 Geometrically, SO($4$) and Spin($4$)-models  are elliptic fibrations whose discriminant locus contains two smooth irreducible components  ($S$ and $T$) such that the dual graphs of the  fibers over the  generic points of $S$ and $T$   is the affine Dynkin diagram $\widetilde{A}_1$, while the fiber over other generic points of the  discriminant locus are  irreducible singular fibers. 
 Since Spin($4$) has a trivial fundamental group and the  fundamental group of  SO($4$) is  $\mathbb{Z}/2\mathbb{Z}$,  the  Mordell--Weil group of a  Spin($4$)-model is trivial while the one of an SO($4$)-model is $\mathbb{Z}/2\mathbb{Z}$.

The SO($4$)-model as a collision $\text{I}_2^{\text{ns}}+\text{I}_2^{\text{ns}}$ in an elliptic fibration with Mordell--Weil group $\mathbb{Z}/2\mathbb{Z}$ has been discussed in \cite{Morrison:2014era}   in the case of a Calabi-Yau threefold elliptically fibered over a  $\mathbb{P}^2$ base. 
The SO($4$)-model was  studied in more detail using  toric methods in   \cite{Mayrhofer:2014opa}. 
 The first appearing of elliptic fibrations with a $\mathbb{Z}/2\mathbb{Z}$ Mordell--Weil group in F-theory is in \cite{Bershadsky:1998vn}  in relation to the CHL model 
 (see also \cite{Sen:1997kw}) 
  and shortly after in \cite{Aspinwall:1998xj}  in the study of  non-simply connected Lie groups in F-theory.

\begin{figure}[htb]
\begin{center}
\scalebox{1.1}{\begin{tikzpicture}
\node (S) at (-6,0) {\includegraphics[scale=1]{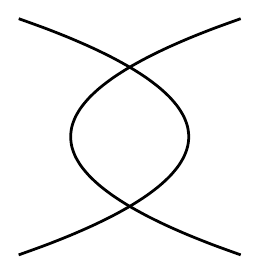}};
\node (T) at (6,0) {\includegraphics[scale=1]{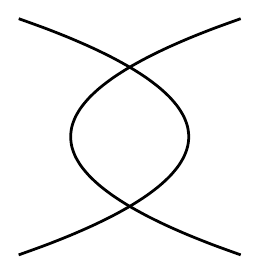}};

\node (ST) at (0,-5) {
\begin{tikzpicture}[every node/.style={circle,draw, minimum size= 10 mm}, scale=.5]
 \node (A1) at (0,3) {};
 \node (A2) at (-3,0) {};
 \node (A3) at (3,0) {};
 \node (A4) at (0,-3) {};
 \draw (A1)--(A2)--(A4)--(A3)--(A1);
 \draw[red,dashed] (-3.3,1.7) arc (90:270:1.5cm);
 \draw[red,dashed] (3.3,1.7) arc (90:-90:1.5cm);
 \draw[red,dashed]  (-3.3,1.7)--(3.3,1.7);
  \draw[red,dashed] (-3.3,-1.3)--(3.3,-1.3);
\end{tikzpicture}
};

\node (S1) at (-6,-5) {\includegraphics[scale=1.8]{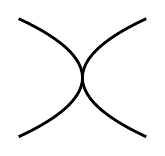}};
\node (S1) at (6,-5) {\includegraphics[scale=1.8]{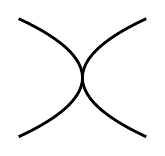}};
 \node[style={circle,draw, minimum size= 10 mm}, label distance=-3mm,  scale=1] (A1) at (0,-10) {};
 \node[style={circle,draw, minimum size= 10 mm}, label distance=-3mm,  scale=1,label=right:2] (A2) at (0,-12) {};
 \node[style={circle,draw, minimum size= 10 mm}, label distance=-3mm,  scale=1] (A3) at (0,-14) {};
 \draw (A1)--(A2)--(A3);
			\draw[->,>=stealth',thick=1mm]  (S)--($(ST)+(-2,2)$);
			\draw[->,>=stealth',thick=1mm]  (T)--($(ST)+(2,2)$);
			\draw[->,>=stealth',thick=1mm]  (0,-7.5)--(0,-9);
			\draw[->,>=stealth',thick=1mm]  (-6,-1.5)--(-6,-4);
						\draw[->,>=stealth',thick=1mm]  (6,-1.5)--(6,-4);
						\draw[<-,>=stealth',thick=1mm
						]  (1,-11)--(5,-7);
												\draw[->,>=stealth',thick=1mm]  (-5,-7)--(-1,-11);
												
												\node  at (-2,-2) {$\displaystyle{T=0}$};\node  at (-4.5,-9) {$\displaystyle{T=0}$};
																							\node  at (2,-2) {$\displaystyle{S=0}$};	\node  at (4.5,-9) {$\displaystyle{S=0}$};
																							\node  at (-6.7,-3) {$\displaystyle{a_2=0}$};	\node  at (6.7,-3) {$\displaystyle{a_2=0}$};
																							\node  at (.8,-8) {$\displaystyle{a_2=0}$};
	\end{tikzpicture}}
	\end{center}
	\caption{Fiber structure of an SO($4$)-model with collision I$_2^{\text{ns}}+\text{I}_2^{\text{ns}}$ and Mordell--Weil group $\mathbb{Z}/2\mathbb{Z}$. 
\label{Fig:SO4I2nsI2ns}}

\end{figure}

\begin{figure}[htb]  
  \begin{center}
\scalebox{.65}{\begin{tikzpicture}
\node (S) at (-6,0) {\includegraphics[scale=1]{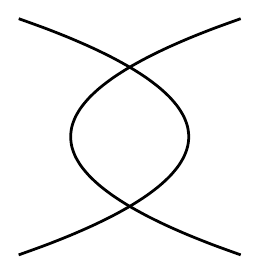}};
\node (T) at (6,0) {\includegraphics[scale=1]{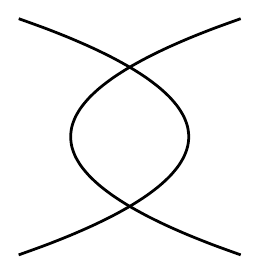}};

\node (ST) at (0,-5) {
\begin{tikzpicture}[every node/.style={circle,draw, minimum size= 10 mm}, scale=.4]
 \node (A1) at (0,3) {};
 \node (A2) at (-3,0) {};
 \node (A3) at (3,0) {};
 \node (A4) at (0,-3) {};
 \draw (A1)--(A2)--(A4)--(A3)--(A1);
\end{tikzpicture}
};

\node (S1) at (-6,-5) {\includegraphics[scale=1.8]{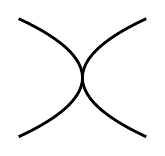}};
\node (S1) at (6,-5) {\includegraphics[scale=1.8]{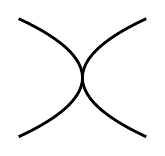}};
 \node[style={circle,draw, minimum size= 10 mm}, label distance=-3mm,  scale=1] (A1) at (0,-10) {};
 \node[style={circle,draw, minimum size= 10 mm}, label distance=-3mm,  scale=1,label=right:2] (A2) at (0,-12) {};
 \node[style={circle,draw, minimum size= 10 mm}, label distance=-3mm,  scale=1] (A3) at (0,-14) {};
 \draw (A1)--(A2)--(A3);
			\draw[->,>=stealth',thick=1mm]  (S)--($(ST)+(-2,2)$);
			\draw[->,>=stealth',thick=1mm]  (T)--($(ST)+(2,2)$);
			\draw[->,>=stealth',thick=1mm]  (0,-7.5)--(0,-9);
			\draw[->,>=stealth',thick=1mm]  (-6,-1.5)--(-6,-4);
						\draw[->,>=stealth',thick=1mm]  (6,-1.5)--(6,-4);
						\draw[<-,>=stealth',thick=1mm
						]  (1,-11)--(5,-7);
												\draw[->,>=stealth',thick=1mm]  (-5,-7)--(-1,-11);
												
												\node  at (-2,-2) {\scalebox{1.2}{$\displaystyle{T=0}$}};\node  at (-4.5,-9) {\scalebox{1.2}{$\displaystyle{T=0}$}};
																							\node  at (2,-2) {\scalebox{1.2}{$\displaystyle{S=0}$}};	\node  at (4.5,-9) {\scalebox{1.2}{$\displaystyle{S=0}$}};
																							\node  at (-7.2,-3) {\scalebox{1.2}{$\displaystyle{a_1^2+\widetilde{a}_2t=0}$}};	\node  at (6.7,-3) {\scalebox{1.2}{$\displaystyle{a_1=0}$}};
																							\node  at (.8,-8) {\scalebox{1.2}{$\displaystyle{a_2=0}$}};
	\end{tikzpicture}}
	\end{center}
	\caption{Fiber structure of an SO($4$)-model with collision I$_2^{\text{ns}}+\text{I}_2^{\text{s}}$ and Mordell--Weil group $\mathbb{Z}/2\mathbb{Z}$. 
\label{Fig:SO4I2nsI2s}}
  \begin{center}
\scalebox{.6}{\begin{tikzpicture}
\node (S) at (-6,0) {\includegraphics[scale=1]{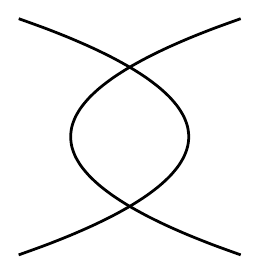}};
\node (T) at (6,0) {\includegraphics[scale=1]{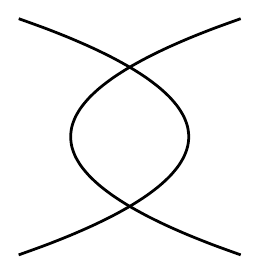}};

\node (ST) at (0,-5) {
\begin{tikzpicture}[every node/.style={circle,draw, minimum size= 10 mm}, scale=.4]
 \node (A1) at (0,3) {};
 \node (A2) at (-3,0) {};
 \node (A3) at (3,0) {};
 \node (A4) at (0,-3) {};
 \draw (A1)--(A2)--(A4)--(A3)--(A1);
\end{tikzpicture}
};

\node (S1) at (-6,-5) {\includegraphics[scale=1.8]{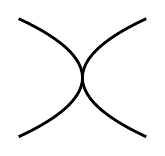}};
\node (S1) at (6,-5) {\includegraphics[scale=1.8]{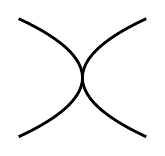}};
 \node[style={circle,draw, minimum size= 10 mm}, label distance=-3mm,  scale=1] (A1) at (0,-10) {};
 \node[style={circle,draw, minimum size= 10 mm}, label distance=-3mm,  scale=1,label=right:2] (A2) at (0,-12) {};
 \node[style={circle,draw, minimum size= 10 mm}, label distance=-3mm,  scale=1] (A3) at (0,-14) {};
 \draw (A1)--(A2)--(A3);
			\draw[->,>=stealth',thick=1mm]  (S)--($(ST)+(-2,2)$);
			\draw[->,>=stealth',thick=1mm]  (T)--($(ST)+(2,2)$);
			\draw[->,>=stealth',thick=1mm]  (0,-7.5)--(0,-9);
			\draw[->,>=stealth',thick=1mm]  (-6,-1.5)--(-6,-4);
						\draw[->,>=stealth',thick=1mm]  (6,-1.5)--(6,-4);
						\draw[<-,>=stealth',thick=1mm
						]  (1,-11)--(5,-7);
												\draw[->,>=stealth',thick=1mm]  (-5,-7)--(-1,-11);
												
												\node  at (-2,-2) {\scalebox{1.3}{$\displaystyle{T=0}$}};\node  at (-4.5,-9) {\scalebox{1.3}{$\displaystyle{T=0}$}};
																							\node  at (2,-2) {\scalebox{1.3}{$\displaystyle{S=0}$}};	
																							\node  at (4.5,-9) {\scalebox{1.3}{$\displaystyle{S=0}$}};
																							\node  at (-6.8,-3) {\scalebox{1.3}{$\displaystyle{a_1=0}$}};	
																							\node  at (6.7,-3) {\scalebox{1.3}{$\displaystyle{{a}_1=0}$}};
																							\node  at (.8,-8) {\scalebox{1.3}{$\displaystyle{a_1=0}$}};
	\end{tikzpicture}}
	\end{center}
	\caption{Fiber structure of an SO($4$)-model with collision I$_2^{\text{s}}+\text{I}_2^{\text{s}}$ and Mordell--Weil group $\mathbb{Z}/2\mathbb{Z}$. 
\label{Fig:SO4I2sI2s}}
	\end{figure}

\begin{figure}[htb]
\begin{center}
\scalebox{.7}{\begin{tikzpicture}

\node (S) at (-5,2) {\includegraphics[scale=1]{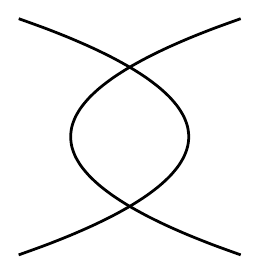}};
\node (T) at (6,2) {\includegraphics[scale=1]{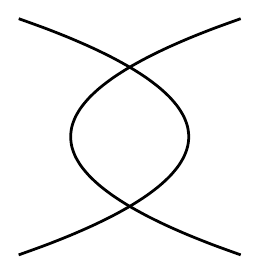}};

\node (ST) at (0,-3) {
\begin{tikzpicture}[every node/.style={circle,draw, minimum size= 10 mm}, scale=.5]
 \node (A1) at (0,3) {};
 \node (A2) at (-3,0) {};
 \node (A3) at (3,0) {};
 \node (A4) at (0,-3) {};
 \draw (A1)--(A2)--(A4)--(A3)--(A1);
\end{tikzpicture}
};

\node (S1) at (-10,-4) {\includegraphics[scale=1.8]{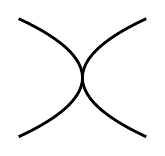}};
\node (T1) at (10,-3.5) {\includegraphics[scale=1.8]{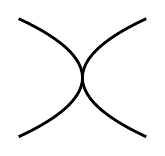}};

\node (S2) at (-5,-3.5) {
\scalebox{.8}{
\begin{tikzpicture}[every node/.style={circle,draw, minimum size= 10mm}, scale=.5]
 \node (A1) at (0,4) {};
 \node (A2) at (-2.8,0) {};
 \node (A3) at (2.8,0) {};
 \draw (A1)--(A2)--(A3)--(A1);
 \draw[red,dashed] (-2.9,1.8) arc (90:270:1.6cm);
 \draw[red,dashed] (2.9,-1.4) arc (-90:90:1.6cm);
 \draw[red,dashed] (-2.9,-1.4)--(2.9,-1.4);
 \draw[red,dashed] (-2.9,1.8)--(2.9,1.8);
\end{tikzpicture}
}};

\node (T2) at (6,-3) {
\scalebox{.8}{
\begin{tikzpicture}[every node/.style={circle,draw, minimum size= 10mm}, scale=.5]
 \node (A1) at (0,4) {};
 \node (A2) at (-2.8,0) {};
 \node (A3) at (2.8,0) {};
 \draw (A1)--(A2)--(A3)--(A1);
 \draw[red,dashed] (-2.9,1.8) arc (90:270:1.6cm);
 \draw[red,dashed] (2.9,-1.4) arc (-90:90:1.6cm);
 \draw[red,dashed] (-2.9,-1.4)--(2.9,-1.4);
 \draw[red,dashed] (-2.9,1.8)--(2.9,1.8);
\end{tikzpicture}
}};

\node (ST1)  at (3.5,-11) {
\scalebox{.8}{
\begin{tikzpicture}[every node/.style={circle,draw, minimum size= 10 mm}, label distance=-2mm,  scale=.5]
 \node (A1) at (0,4) {};
 \node (A2) at (-4.3,0.8) {};
 \node (A3) at (-2.5,-4) {};
 \node (A4) at (2.5,-4) {};
 \node (A5) at (4.3,0.8) {};
 \draw (A1)--(A2)--(A3)--(A4)--(A5)--(A1);
 \draw[red,dashed] (-2.8,-2.3) arc (90:270:1.5cm);
 \draw[red,dashed] (2.8,-2.3) arc (90:-90:1.5cm);
 \draw[red,dashed] (-2.8,-2.3)--(2.8,-2.3);
 \draw[red,dashed] (-2.8,-5.3)--(2.8,-5.3);
\end{tikzpicture}
}};

\node (ST1b) at (-3,-10.5)
{\scalebox{.8}{
\begin{tikzpicture}[every node/.style={circle,draw, minimum size= 10 mm}, label distance=-3mm,  scale=.5]
 \node (A1) at (0,0) {};
 \node[label=right:2] (A2) at (0,-4) {};
 \node (A3) at (0,-8) {};
 \draw (A1)--(A2)--(A3);
\end{tikzpicture}
}};

\node (ST2) at (0,-18) {\scalebox{.9}{
\begin{tikzpicture}[every node/.style={circle,draw, minimum size= 10 mm}, label distance=-3mm,  scale=.5]
 \node (A0) at (90:7) {};
 \node (A1) at (90:3) {$2$};
 \node (A2) at (225:3) {};
 \node (A3) at (-45:3) {};
 \draw (A1)--(0,0);
 \draw (A2)--(0,0);
 \draw (A3)--(0,0);
 \draw (A0)--(A1);
 \draw[red,dashed] (-2.9,-.9) arc (90:270:1.3cm);
 \draw[red,dashed] (2.9,-.9) arc (90:-90:1.3cm);
 \draw[red,dashed] (-2.9,-.9)--(2.9,-.9);
 \draw[red,dashed] (-2.9,-3.5)--(2.9,-3.5);
\end{tikzpicture}
}};

\node (ST3) at (0,-24) {\scalebox{.9}{
\begin{tikzpicture}[every node/.style={circle,draw, minimum size= 10 mm}, label distance=-3mm,  scale=.5]
 \node (A1) at (0,3.5) {};
 \node (A2) at (0,0) {$2$};
 \node (A3) at (0,-3.5) {$2$};
 \draw (A1)--(A2)--(A3);
\end{tikzpicture}
}};

\node (S3) at (-10,-11) {\scalebox{1}{
\begin{tikzpicture}[every node/.style={circle,draw, minimum size= 10 mm}, label distance=-3mm,  scale=.5]
 \node (A1) at (90:3) {};
 \node (A2) at (225:3) {};
 \node (A3) at (-45:3) {};
 \draw (A1)--(0,0);
 \draw (A2)--(0,0);
 \draw (A3)--(0,0);
  \draw[red,dashed] (-2.9,-.9) arc (90:270:1.3cm);
   \draw[red,dashed] (2.9,-.9) arc (90:-90:1.3cm);
\draw[red,dashed] (-2.9,-.9)--(2.9,-.9);
 \draw[red,dashed] (-2.9,-3.5)--(2.9,-3.5);
 \end{tikzpicture}
}};

\node (T3) at (10,-10) {\scalebox{1}{
\begin{tikzpicture}[every node/.style={circle,draw, minimum size= 10 mm}, label distance=-3mm,  scale=.5]
 \node (A1) at (90:3) {};
 \node (A2) at (225:3) {};
 \node (A3) at (-45:3) {};
 \draw (A1)--(0,0);
 \draw (A2)--(0,0);
 \draw (A3)--(0,0);
   \draw[red,dashed] (-2.9,-.9) arc (90:270:1.3cm);
   \draw[red,dashed] (2.9,-.9) arc (90:-90:1.3cm);
\draw[red,dashed] (-2.9,-.9)--(2.9,-.9);
 \draw[red,dashed] (-2.9,-3.5)--(2.9,-3.5);

\end{tikzpicture}
}};

\node (S4) at (-10,-18) {\scalebox{1}{
\begin{tikzpicture}[every node/.style={circle,draw, minimum size= 10 mm}, label distance=-3mm,  scale=.5]
 \node (A1) at (0,0) {};
 \node (A2) at (0,-5) {$2$};
 \draw (A1)--(A2);
\end{tikzpicture}
}};

\node (T4) at (10,-18) {\scalebox{1}{
\begin{tikzpicture}[every node/.style={circle,draw, minimum size= 10 mm}, label distance=-3mm,  scale=.5]
 \node (A1) at (0,0) {};
 \node (A2) at (0,-5) {$2$};
 \draw (A1)--(A2);
\end{tikzpicture}
}};

												\draw[->,>=stealth',thick=1mm]  (S1)--(S3) node [midway,right=.3cm, fill=white,text=black] {$  \widetilde{a}_4=0$};
												\draw[->,>=stealth',thick=1mm]  (S3)--(S4)node [midway,right=.3cm, fill=white,text=black] {$\widetilde{a}_6=0$};
												\draw[->,>=stealth',thick=1mm]  (S2)--(S3);
												\node[fill=white,text=black]   at ($(S2)+(-1.2,-2.2)$) {$a_1=0$};

	\draw[->,>=stealth',thick=1mm]  (T1)--(T3) node [midway,right=.3cm, fill=white,text=black] {$\widetilde{a}_4=0$};
	\draw[->,>=stealth',thick=1mm]  (T3)--(T4) node [midway,right=.3cm, fill=white,text=black] {$\widetilde{a}_6=0$};
	
	\draw[->,>=stealth',thick=1mm]  (T2)--(T3) node [midway, fill=white,text=black] {$a_1$};
		\draw[->,>=stealth',thick=1mm]  (ST)--(ST1);
		\node[fill=white,text=black] at  ($(ST1)+(-2,5)$)  {$\widetilde{a}_4^2+2a_1\widetilde{a}_6=0$};
	\draw[->,>=stealth',thick=1mm]  (ST)--(ST1b);
	 \node [fill=white,text=black]  at  ($(ST1)+(-4.5,5.2)$)   {$a_1=0$};
			\draw[->,>=stealth',thick=1mm]  (S2)--(ST1);
			\node[fill=white,text=black]  at  ($(ST1)+(-4,3.5)$)  {$T=0$};

		\draw[->,>=stealth',thick=1mm]  (ST1)--(ST2) node [midway=.5cm,right=.3cm, fill=white,text=black] {$  a_1=0$};
	\draw[->,>=stealth',thick=1mm]  (ST2)--(ST3) node [midway,right=.3cm, fill=white,text=black] {$ \widetilde{a}_6=0$};
				\draw[->,>=stealth',thick=1mm]  (T2)--(ST1)node [midway,right=.5cm, fill=white,text=black] {$S=0$};
				\draw[->,>=stealth',thick=1mm]  (S3)--(ST2) node [midway, fill=white,text=black] {$T=0$};
				\draw[->,>=stealth',thick=1mm]  (T3)--(ST2) node [midway,right=.3cm, fill=white,text=black] {$S=0$};
				\draw[->,>=stealth',thick=1mm]  (S4)--(ST3) node [midway, fill=white,text=black] {$T=0$};
				\draw[->,>=stealth',thick=1mm]  (T4)--(ST3) node [midway,right=.3cm, fill=white,text=black] {$S=0$};

				\draw[->,>=stealth',thick=1mm]  (S1)--(ST1b) node [midway,fill=white,text=black] {$T=0$};
				\draw[->,>=stealth',thick=1mm]  (T1)--(ST1b) node [midway, fill=white,text=black] {$S=0$};
				\draw[->,>=stealth',thick=1mm]  (ST1b)--(ST2) node [midway,right=.3cm, fill=white,text=black] {$\widetilde{a}_4=0$};

\draw[->,>=stealth',thick=1mm]  (S)--(S1) node [midway,left=.3cm, fill=white,text=black] {$ a_1=0$};
				\draw[->,>=stealth',thick=1mm]  (S)--(S2) node [midway, fill=white,text=black] {$ \widetilde{a}_4^2+2a_1\widetilde{a}_6=0$};
				\draw[->,>=stealth',thick=1mm]  (S)--(ST) node [midway, fill=white,text=black] {$T=0$};

\draw[->,>=stealth',thick=1mm]  (T)--(T1) node [midway, fill=white,text=black] {$a_1=0$};
				\draw[->,>=stealth',thick=1mm]  (T)--(T2) node [midway, fill=white,text=black] {$ \widetilde{a}_4^2+2a_1\widetilde{a}_6=0$};
				
				\draw[->,>=stealth',thick=1mm]  (T)--(ST) node [midway, fill=white,text=black] {$S=0$};

	\end{tikzpicture}}
	\end{center}
	\caption{Fiber structure of a Spin($4$)-model with I$_2^{\text{s}}+$I$_2^{\text{s}}$. 
\label{Fig:Spin4}}
	\end{figure}

\begin{figure}[htb]
\begin{center}
\scalebox{.7}{\begin{tikzpicture}

\node (T) at (5,2) {\includegraphics[scale=1]{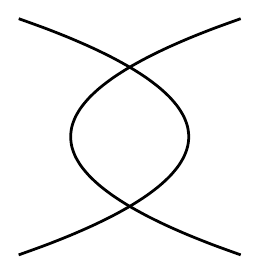}};

\node (S1) at (-6,2) {\includegraphics[scale=1.8]{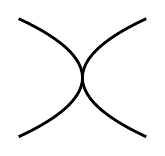}};
\node (T1) at (5,-3) {\includegraphics[scale=1.8]{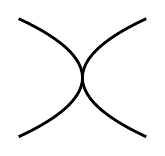}};

\node (T2) at (10,-3) {
\scalebox{.8}{
\begin{tikzpicture}[every node/.style={circle,draw, minimum size= 10mm}, scale=.5]
 \node (A1) at (0,4) {};
 \node (A2) at (-2.8,0) {};
 \node (A3) at (2.8,0) {};
 \draw (A1)--(A2)--(A3)--(A1);
 \draw[red,dashed] (-2.9,1.8) arc (90:270:1.6cm);
 \draw[red,dashed] (2.9,-1.4) arc (-90:90:1.6cm);
 \draw[red,dashed] (-2.9,-1.4)--(2.9,-1.4);
 \draw[red,dashed] (-2.9,1.8)--(2.9,1.8);
\end{tikzpicture}
}};

\node (ST1b) at (0,-4)
{\scalebox{.9}{
\begin{tikzpicture}[every node/.style={circle,draw, minimum size= 10 mm}, label distance=-3mm,  scale=.5]
 \node (A1) at (0,0) {};
 \node (A2) at (0,-4) {$2$};
 \node (A3) at (0,-8) {};
 \draw (A1)--(A2)--(A3);
\end{tikzpicture}
}};

\node (ST2) at (0,-11) {\scalebox{.9}{
\begin{tikzpicture}[every node/.style={circle,draw, minimum size= 10 mm}, label distance=-3mm,  scale=.5]
 \node (A0) at (90:7) {};
 \node (A1) at (90:3) {$2$};
 \node (A2) at (225:3) {};
 \node (A3) at (-45:3) {};
 \draw (A1)--(0,0);
 \draw (A2)--(0,0);
 \draw (A3)--(0,0);
 \draw (A0)--(A1);
 \draw[red,dashed] (-2.9,-.9) arc (90:270:1.3cm);
 \draw[red,dashed] (2.9,-.9) arc (90:-90:1.3cm);
 \draw[red,dashed] (-2.9,-.9)--(2.9,-.9);
 \draw[red,dashed] (-2.9,-3.5)--(2.9,-3.5);
\end{tikzpicture}
}};

\node (ST3) at (0,-18) {\scalebox{1}{
\begin{tikzpicture}[every node/.style={circle,draw, minimum size= 11 mm}, label distance=-3mm,  scale=.5]
 \node (A1) at (0,3.5) {};
 \node (A2) at (0,0) {$2$};
 \node (A3) at (0,-3.5) {$2$};
 \draw (A1)--(A2)--(A3);
\end{tikzpicture}
}};

\node (S3) at (-10,-4) {\scalebox{1}{
\begin{tikzpicture}[every node/.style={circle,draw, minimum size= 10 mm}, label distance=-3mm,  scale=.5]
 \node (A1) at (90:3) {};
 \node (A2) at (225:3) {};
 \node (A3) at (-45:3) {};
 \draw (A1)--(0,0);
 \draw (A2)--(0,0);
 \draw (A3)--(0,0);
  \draw[red,dashed] (-2.9,-.9) arc (90:270:1.3cm);
   \draw[red,dashed] (2.9,-.9) arc (90:-90:1.3cm);
\draw[red,dashed] (-2.9,-.9)--(2.9,-.9);
 \draw[red,dashed] (-2.9,-3.5)--(2.9,-3.5);
 \end{tikzpicture}
}};

\node (T3) at (10,-10) {\scalebox{1}{
\begin{tikzpicture}[every node/.style={circle,draw, minimum size= 10 mm}, label distance=-3mm,  scale=.5]
 \node (A1) at (90:3) {};
 \node (A2) at (225:3) {};
 \node (A3) at (-45:3) {};
 \draw (A1)--(0,0);
 \draw (A2)--(0,0);
 \draw (A3)--(0,0);
   \draw[red,dashed] (-2.9,-.9) arc (90:270:1.3cm);
   \draw[red,dashed] (2.9,-.9) arc (90:-90:1.3cm);
\draw[red,dashed] (-2.9,-.9)--(2.9,-.9);
 \draw[red,dashed] (-2.9,-3.5)--(2.9,-3.5);

\end{tikzpicture}
}};

\node (S4) at (-10,-10) {\scalebox{1}{
\begin{tikzpicture}[every node/.style={circle,draw, minimum size= 10 mm}, label distance=-3mm,  scale=.5]
 \node (A1) at (0,0) {};
 \node (A2) at (0,-5) {$2$};
 \draw (A1)--(A2);
\end{tikzpicture}
}};

\node (T4) at (10,-15) {\scalebox{1}{
\begin{tikzpicture}[every node/.style={circle,draw, minimum size= 10 mm}, label distance=-3mm,  scale=.5]
 \node (A1) at (0,0) {};
 \node (A2) at (0,-5) {$2$};
 \draw (A1)--(A2);
\end{tikzpicture}
}};

\draw[->,>=stealth',thick=1mm]  (T)--(T1) node [midway, fill=white,text=black] {$\widetilde{a}_1^2 s+4\widetilde{a}_2=0$};
				\draw[->,>=stealth',thick=1mm]  (T)--(T2) node [midway,right=.6cm, fill=white,text=black] {$ \widetilde{a}_4^2-\widetilde{a}_6 (\widetilde{a}_1^2 s+4\widetilde{a}_2)=0$};;
				\draw[->,>=stealth',thick=1mm]  (T)--(ST1b) node [midway,left=.3cm, fill=white,text=black] {$S=0$};;
												
												\draw[->,>=stealth',thick=1mm]  (S1)--(S3) node [midway,left=.3cm, fill=white,text=black] {$\widetilde{a}_4=0$};;
												\draw[->,>=stealth',thick=1mm]  (S3)--(S4) node [midway,left=.3cm, fill=white,text=black] {$ \widetilde{a}_6=0$};;

	\draw[->,>=stealth',thick=1mm]  (T1)--(T3) node [midway,right=.3cm, fill=white,text=black] {$\widetilde{a}_4=0$};;;
	\draw[->,>=stealth',thick=1mm]  (T3)--(T4)node [midway,right=.3cm, fill=white,text=black] {$ \widetilde{a}_6=0$};;;
	\draw[->,>=stealth',thick=1mm]  (T2)--(T3)node [midway,right=.3cm, fill=white,text=black] {$ \widetilde{a}_1^2 s+4\widetilde{a}_2=0$};;;
	
	\draw[->,>=stealth',thick=1mm]  (T2)--(ST2)node [midway,left=.6cm, fill=white,text=black] {$S=0$};;;
	\draw[->,>=stealth',thick=1mm]  (ST2)--(ST3)node [midway,left=.3cm, fill=white,text=black] {$\widetilde{a}_6=0$};;;
				\draw[->,>=stealth',thick=1mm]  (S3)--(ST2) node [midway,left=.6cm, fill=white,text=black] {$T=0$};;;
				\draw[->,>=stealth',thick=1mm]  (T3)--(ST2)node [midway, fill=white,text=black] {$S=0$};;;
				\draw[->,>=stealth',thick=1mm]  (S4)--(ST3)node [midway,left=.3cm, fill=white,text=black] {$T=0$};;;
				\draw[->,>=stealth',thick=1mm]  (T4)--(ST3)node [midway,right=1cm, fill=white,text=black] {$S=0$};;;

				\draw[->,>=stealth',thick=1mm]  (S1)--(ST1b)node [midway,left=.3cm, fill=white,text=black] {$T=0$};;;
				\draw[->,>=stealth',thick=1mm]  (T1)--(ST1b)node [midway,fill=white,text=black] {$S=0$};;;
				\draw[->,>=stealth',thick=1mm]  (ST1b)--(ST2)node [midway,left=.3cm, fill=white,text=black] {$ \widetilde{a}_4=0$};;

	\end{tikzpicture}}
	\end{center}
	\caption{Fiber structure of a Spin($4$)-model with III+I$_2^{\text{ns}}$. 
\label{Fig:Spin4IIIns}}
	\end{figure}

\begin{figure}[htb]
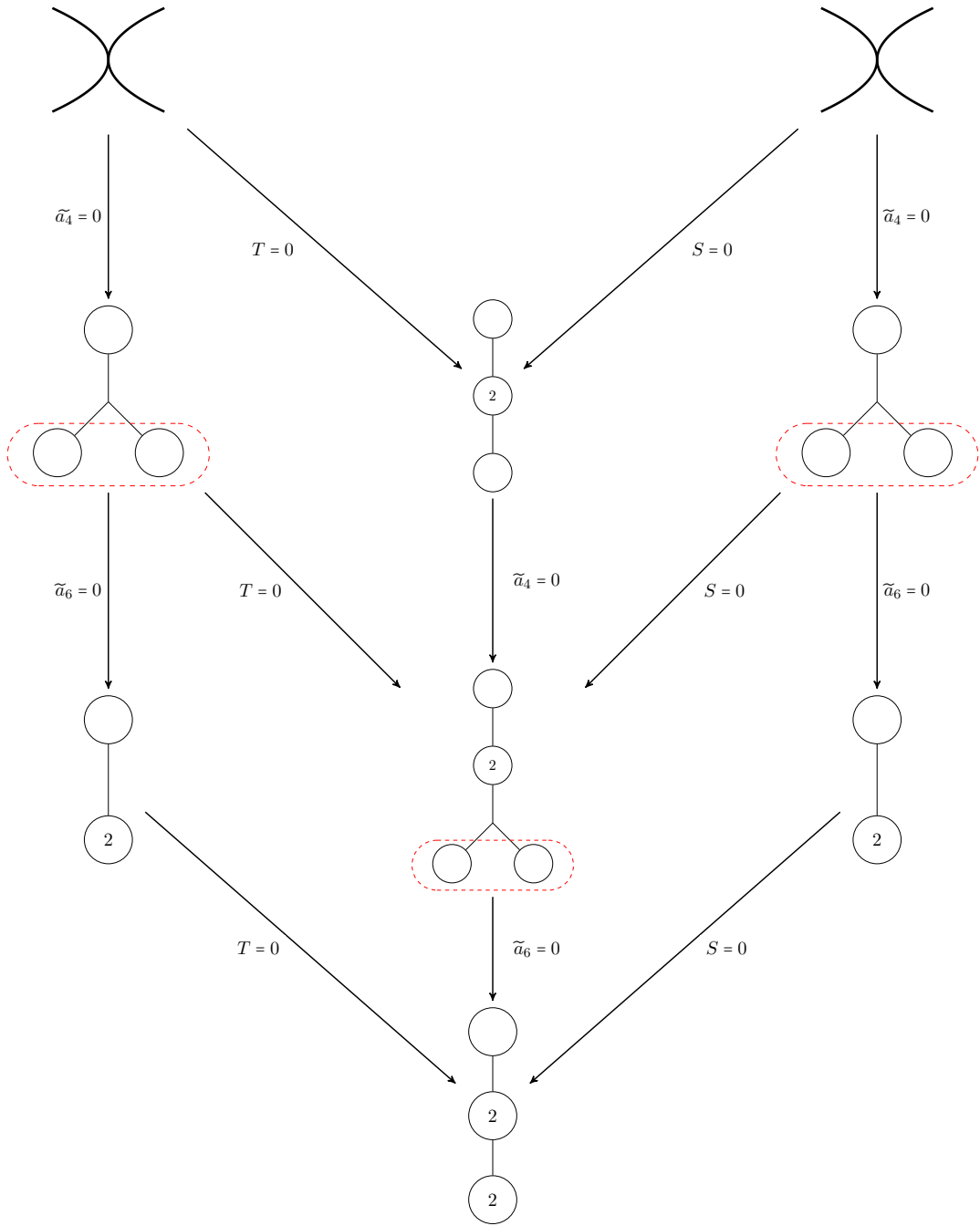

\begin{center}
\scalebox{.7}{\begin{tikzpicture}

\node (S1) at (-8,-3) {\includegraphics[scale=1.8]{III}};
\node (T1) at (8,-3) {\includegraphics[scale=1.8]{III}};

\node (ST1b) at (0,-10)
{\scalebox{.8}{
\begin{tikzpicture}[every node/.style={circle,draw, minimum size= 10 mm}, label distance=-3mm,  scale=.5]
 \node (A1) at (0,0) {};
 \node (A2) at (0,-4) {$2$};
 \node (A3) at (0,-8) {};
 \draw (A1)--(A2)--(A3);
\end{tikzpicture}
}};

\node (ST2) at (0,-18) {\scalebox{.8}{
\begin{tikzpicture}[every node/.style={circle,draw, minimum size= 10 mm}, label distance=-3mm,  scale=.5]
 \node (A0) at (90:7) {};
 \node (A1) at (90:3) {$2$};
 \node (A2) at (225:3) {};
 \node (A3) at (-45:3) {};
 \draw (A1)--(0,0);
 \draw (A2)--(0,0);
 \draw (A3)--(0,0);
 \draw (A0)--(A1);
 \draw[red,dashed] (-2.9,-.9) arc (90:270:1.3cm);
 \draw[red,dashed] (2.9,-.9) arc (90:-90:1.3cm);
 \draw[red,dashed] (-2.9,-.9)--(2.9,-.9);
 \draw[red,dashed] (-2.9,-3.5)--(2.9,-3.5);
\end{tikzpicture}
}};

\node (ST3) at (0,-25) {\scalebox{1}{
\begin{tikzpicture}[every node/.style={circle,draw, minimum size= 10 mm}, label distance=-3mm,  scale=.5]
 \node (A1) at (0,3.5) {};
 \node (A2) at (0,0) {$2$};
 \node (A3) at (0,-3.5) {$2$};
 \draw (A1)--(A2)--(A3);
\end{tikzpicture}
}};

\node (S3) at (-8,-10) {\scalebox{1}{
\begin{tikzpicture}[every node/.style={circle,draw, minimum size= 10 mm}, label distance=-3mm,  scale=.5]
 \node (A1) at (90:3) {};
 \node (A2) at (225:3) {};
 \node (A3) at (-45:3) {};
 \draw (A1)--(0,0);
 \draw (A2)--(0,0);
 \draw (A3)--(0,0);
  \draw[red,dashed] (-2.9,-.9) arc (90:270:1.3cm);
   \draw[red,dashed] (2.9,-.9) arc (90:-90:1.3cm);
\draw[red,dashed] (-2.9,-.9)--(2.9,-.9);
 \draw[red,dashed] (-2.9,-3.5)--(2.9,-3.5);
 \end{tikzpicture}
}};

\node (T3) at (8,-10) {\scalebox{1}{
\begin{tikzpicture}[every node/.style={circle,draw, minimum size= 10 mm}, label distance=-3mm,  scale=.5]
 \node (A1) at (90:3) {};
 \node (A2) at (225:3) {};
 \node (A3) at (-45:3) {};
 \draw (A1)--(0,0);
 \draw (A2)--(0,0);
 \draw (A3)--(0,0);
   \draw[red,dashed] (-2.9,-.9) arc (90:270:1.3cm);
   \draw[red,dashed] (2.9,-.9) arc (90:-90:1.3cm);
\draw[red,dashed] (-2.9,-.9)--(2.9,-.9);
 \draw[red,dashed] (-2.9,-3.5)--(2.9,-3.5);

\end{tikzpicture}
}};

\node (S4) at (-8,-18) {\scalebox{1}{
\begin{tikzpicture}[every node/.style={circle,draw, minimum size= 10 mm}, label distance=-3mm,  scale=.5]
 \node (A1) at (0,0) {};
 \node (A2) at (0,-5) {$2$};
 \draw (A1)--(A2);
\end{tikzpicture}
}};

\node (T4) at (8,-18) {\scalebox{1}{
\begin{tikzpicture}[every node/.style={circle,draw, minimum size= 10 mm}, label distance=-3mm,  scale=.5]
 \node (A1) at (0,0) {};
 \node (A2) at (0,-5) {$2$};
 \draw (A1)--(A2);
\end{tikzpicture}
}};

	\draw[->,>=stealth',thick=1mm]  (T1)--(T3)  node [midway,right,fill=white,text=black] { $\widetilde{a}_4=0$};
	\draw[->,>=stealth',thick=1mm]  (T3)--(T4) node [midway,right, fill=white,text=black] {$ \widetilde{a}_6=0$};
	\draw[->,>=stealth',thick=1mm]  (S1)--(S3) node [midway,left, fill=white,text=black] {$\widetilde{a_4}=0$};
	\draw[->,>=stealth',thick=1mm]  (S3)--(S4) node [midway,left, fill=white,text=black] {$\widetilde{a}_6=0$};
	\draw[->,>=stealth',thick=1mm]  (S1)--(ST1b) node [midway,left=.5cm,fill=white,text=black] {$T=0$};
		\draw[->,>=stealth',thick=1mm]  (T1)--(ST1b) node [midway,fill=white,right=.5cm,text=black] {$S=0$};

				\draw[->,>=stealth',thick=1mm]  (S3)--(ST2) node [midway,left=.3cm, fill=white,text=black] {$T=0$};
				\draw[->,>=stealth',thick=1mm]  (T3)--(ST2) node [midway,right=.3cm,fill=white,text=black] {$S=0$};
				\draw[->,>=stealth',thick=1mm]  (S4)--(ST3) node [midway,left=.3cm,fill=white,text=black] {$T=0$};
				\draw[->,>=stealth',thick=1mm]  (T4)--(ST3)  node [midway,right=.3cm,fill=white,text=black] {$S=0$};
	
				\draw[->,>=stealth',thick=1mm]  (ST1b)--(ST2) node [midway,right=.3cm, fill=white,text=black] {$\widetilde{a}_4=0$};
				\draw[->,>=stealth',thick=1mm]  (ST2)--(ST3) node [midway,right=.3cm, fill=white,text=black] {$ \widetilde{a}_6=0$};
	\end{tikzpicture}}
	\end{center}
	\caption{Fiber structure of a Spin($4$)-model with III$+$III. 
\label{Fig:Spin4IIIIII}}
	\end{figure}

\begin{figure}[htb]
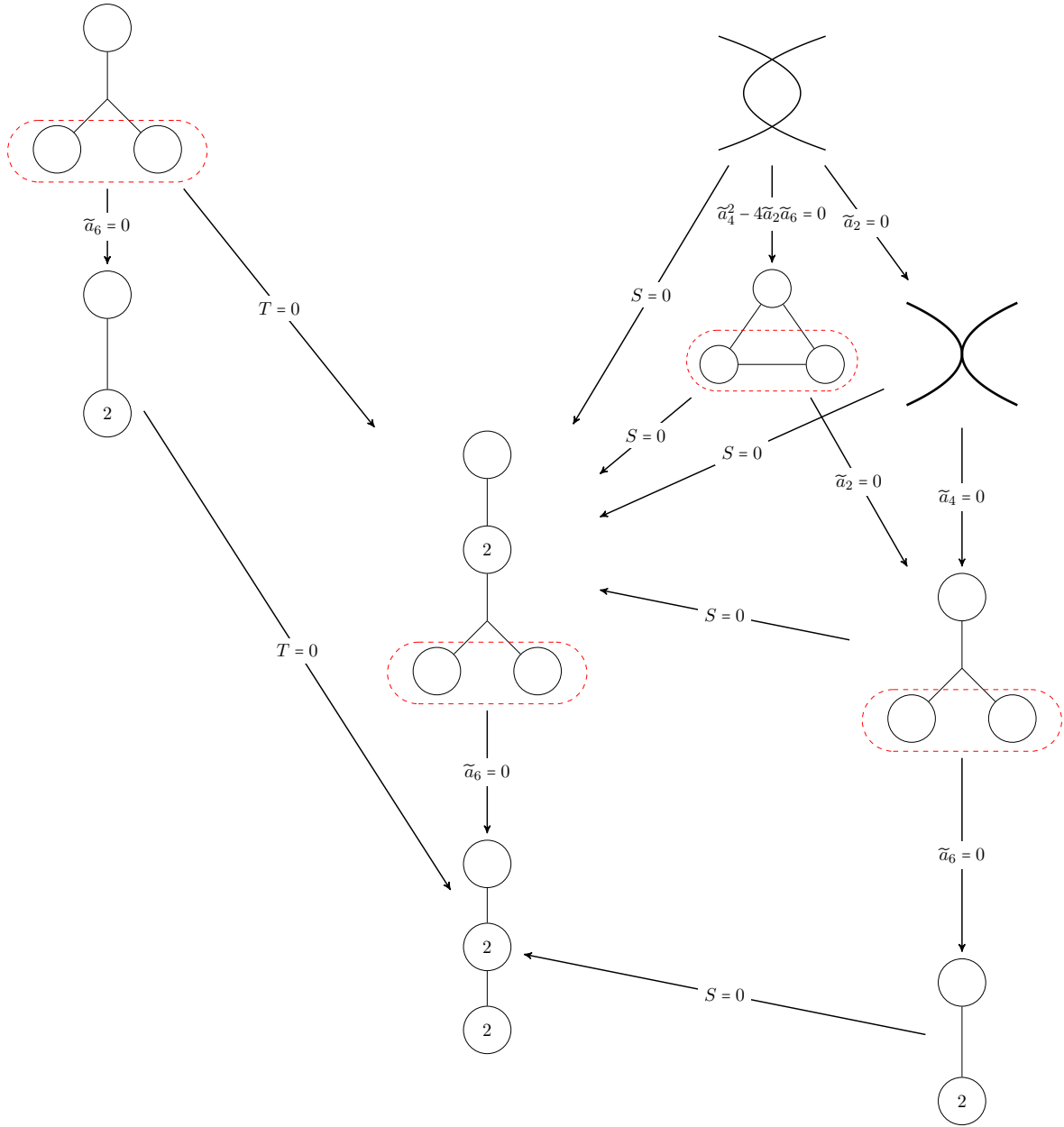

\begin{center}
\scalebox{.7}{\begin{tikzpicture}

\node (T) at (6,2) {\includegraphics[scale=1]{I2}};

\node (S) at (-8,2) {\scalebox{1}{
\begin{tikzpicture}[every node/.style={circle,draw, minimum size= 10 mm}, label distance=-3mm,  scale=.5]
 \node (A1) at (90:3) {};
 \node (A2) at (225:3) {};
 \node (A3) at (-45:3) {};
 \draw (A1)--(0,0);
 \draw (A2)--(0,0);
 \draw (A3)--(0,0);
  \draw[red,dashed] (-2.9,-.9) arc (90:270:1.3cm);
   \draw[red,dashed] (2.9,-.9) arc (90:-90:1.3cm);
\draw[red,dashed] (-2.9,-.9)--(2.9,-.9);
 \draw[red,dashed] (-2.9,-3.5)--(2.9,-3.5);

 \end{tikzpicture}
}};

\node (T1) at (10,-3.5) {\includegraphics[scale=1.8]{III}};

\node (S1) at (-8,-3.5) {\scalebox{1}{
\begin{tikzpicture}[every node/.style={circle,draw, minimum size= 10 mm}, label distance=-3mm,  scale=.5]
 \node (A1) at (0,0) {};
 \node (A2) at (0,-5) {$2$};
 \draw (A1)--(A2);
\end{tikzpicture}
}};

\node (T2) at (6,-3) {
\scalebox{.8}{
\begin{tikzpicture}[every node/.style={circle,draw, minimum size= 10mm}, scale=.5]
 \node (A1) at (0,4) {};
 \node (A2) at (-2.8,0) {};
 \node (A3) at (2.8,0) {};
 \draw (A1)--(A2)--(A3)--(A1);
 \draw[red,dashed] (-2.9,1.8) arc (90:270:1.6cm);
 \draw[red,dashed] (2.9,-1.4) arc (-90:90:1.6cm);
 \draw[red,dashed] (-2.9,-1.4)--(2.9,-1.4);
 \draw[red,dashed] (-2.9,1.8)--(2.9,1.8);
\end{tikzpicture}
}};

\node (ST) at (0,-8) {\scalebox{1}{
\begin{tikzpicture}[every node/.style={circle,draw, minimum size= 10 mm}, label distance=-3mm,  scale=.5]
 \node (A0) at (90:7) {};

\node (A1) at (90:3) {$2$};
 \node (A2) at (225:3) {};
 \node (A3) at (-45:3) {};
 \draw (A1)--(0,0);
 \draw (A2)--(0,0);
 \draw (A3)--(0,0);
 \draw (A0)--(A1);
  \draw[red,dashed] (-2.9,-.9) arc (90:270:1.3cm);
   \draw[red,dashed] (2.9,-.9) arc (90:-90:1.3cm);
\draw[red,dashed] (-2.9,-.9)--(2.9,-.9);
 \draw[red,dashed] (-2.9,-3.5)--(2.9,-3.5);

\end{tikzpicture}
}};

\node (ST1) at (0,-16) {\scalebox{1}{
\begin{tikzpicture}[every node/.style={circle,draw, minimum size= 10 mm}, label distance=-3mm,  scale=.5]
 \node (A1) at (0,3.5) {};
 \node (A2) at (0,0) {$2$};
 \node (A3) at (0,-3.5) {$2$};
 \draw (A1)--(A2)--(A3);
\end{tikzpicture}
}};

\node (T3) at (10,-10) {\scalebox{1}{
\begin{tikzpicture}[every node/.style={circle,draw, minimum size= 10 mm}, label distance=-3mm,  scale=.5]
 \node (A1) at (90:3) {};
 \node (A2) at (225:3) {};
 \node (A3) at (-45:3) {};
 \draw (A1)--(0,0);
 \draw (A2)--(0,0);
 \draw (A3)--(0,0);
   \draw[red,dashed] (-2.9,-.9) arc (90:270:1.3cm);
   \draw[red,dashed] (2.9,-.9) arc (90:-90:1.3cm);
\draw[red,dashed] (-2.9,-.9)--(2.9,-.9);
 \draw[red,dashed] (-2.9,-3.5)--(2.9,-3.5);

\end{tikzpicture}
}};

\node (T4) at (10,-18) {\scalebox{1}{
\begin{tikzpicture}[every node/.style={circle,draw, minimum size= 10 mm}, label distance=-3mm,  scale=.5]
 \node (A1) at (0,0) {};
 \node (A2) at (0,-5) {$2$};
 \draw (A1)--(A2);
\end{tikzpicture}
}};

												\draw[->,>=stealth',thick=1mm]  (T1)--(ST) node [midway, fill=white,text=black] {$S=0$};

	\draw[->,>=stealth',thick=1mm]  (T1)--(T3) node [midway, fill=white,text=black] {$\widetilde{a}_4=0$};
	\draw[->,>=stealth',thick=1mm]  (T3)--(T4)node [midway, fill=white,text=black] {$\widetilde{a}_6=0$};
	\draw[->,>=stealth',thick=1mm]  (T2)--(T3) node [midway, fill=white,text=black] {$\widetilde{a}_2=0$};
	
	\draw[->,>=stealth',thick=1mm]  (ST)--(ST1)node [midway, fill=white,text=black] {$\widetilde{a}_6=0$};
				\draw[->,>=stealth',thick=1mm]  (T2)--(ST) node [midway, fill=white,text=black] {$S=0$};
			\draw[->,>=stealth',thick=1mm]  (T3)--(ST) node [midway, fill=white,text=black] {$S=0$};
			\draw[->,>=stealth',thick=1mm]  (T4)--(ST1) node [midway, fill=white,text=black] {$S=0$};

\draw[->,>=stealth',thick=1mm]  (S)--(S1) node [midway, fill=white,text=black] {$\widetilde{a}_6=0$};
				\draw[->,>=stealth',thick=1mm]  (S1)--(ST1) node [midway, fill=white,text=black] {$T=0$};
				\draw[->,>=stealth',thick=1mm]  (S)--(ST) node [midway, fill=white,text=black] {$T=0$};

\draw[->,>=stealth',thick=1mm]  (T)--(T1) node [midway, fill=white,text=black] {$\widetilde{a}_2=0$};
				\draw[->,>=stealth',thick=1mm]  (T)--(T2) node [midway, fill=white,text=black] {$\widetilde{a}_4^2-4\widetilde{a}_2\widetilde{a}_6=0$};
				\draw[->,>=stealth',thick=1mm]  (T)--(ST) node [midway, fill=white,text=black] {$S=0$};

	\end{tikzpicture}}
	\end{center}
	\caption{Fiber structure of a Spin($4$)-model with IV$^{ns}+$I$_2^{\text{ns}}$. 
\label{Fig:Spin4IVns}}
	\end{figure}

In this paper, we add to previous work on SO($4$) and Spin($4$)-models in many ways. 
We consider other collisions than  $\text{I}_2^{\text{ns}}+\text{I}_2^{\text{ns}}$ and  before to specialize to the case of Calabi-Yau threefolds, we study these geometries over  a base  arbitrary dimension. 
Geometrically, SO($4$) and Spin($4$)-models are characterized by the collision of two Kodaira fibers whose dual graphs are of type  $\widetilde{A}_1$. 
While the Kodaira fiber of  type I$_2$ has dual graph  $\widetilde{A}_1$, it is not the only one. That means there are many more ways to realize geometrically a model of type SO($4$) or Spin($4$) than a collision of type I$_2$+I$_2$.
In fact, the dual graph $\widetilde{\text{A}}_1$ is the most versatile one as it can be realized geometrically by five distinct types of singular fibers, namely 
 \begin{equation}
\text{Fibers with dual graph}\  \widetilde{A}_1:\quad 
 \text{I}_2^{\text{ns}}, \quad \text{I}_2^{\text{s}}, \quad \text{III}, \quad \text{I}_3^{\text{ns}}, \quad \text{IV}^{\text{ns}}.
 \end{equation}
 We avoid realizing A$_1$ with fibers of  type $\text{I}^{\text{ns}}_3$  to stay away from  terminal singularities. 
The elliptic fibrations that we consider are  constructed by crepant resolutions of Weierstrass models  corresponding to the following collisions of singularities 
\begin{equation}\label{eq:Collisions}
\begin{array}{c}
 \text{I}^{\text{ns}}_2+\text{I}^{\text{ns}}_2, \quad
\text{I}^{\text{s}}_2+\text{I}^{\text{ns}}_2,
\quad 
\text{I}^{\text{s}}_2+\text{I}^{\text{s}}_2, 
 \quad
  \text{III}+\text{I}_2^{\text{ns}}, \quad
  \text{III}+\text{I}_2^{\text{s}}, \quad
   \text{III}+\text{III}, \quad
 \text{IV}^{\text{ns}}+\text{I}_2^{\text{s}}, 
 \quad \text{IV}^{\text{ns}}+\text{I}_2^{\text{ns}}.
 \end{array}
\end{equation}
These collisions define a gauge theory with Lie algebra A$_1\oplus$A$_1$. There are many  gauge groups with this Lie algebra, but as we will explain, we only get SO($4$) or Spin($4$) in this case respectively when the Mordell--Weil group is $\mathbb{Z}/2\mathbb{Z}$ or trivial. 
It is important to distinguish between I$_2^{\text{s}}$ and I$_2^{\text{ns}}$ and to fix the Mordell--Weil group as they lead to completely different fiber structures as seen by comparing Figures \ref{Fig:SO4I2nsI2ns}, \ref{Fig:SO4I2nsI2s}, \ref{Fig:SO4I2sI2s}, \ref{Fig:SO4IIII2ns}, \ref{Fig:SO4},  \ref{Fig:Spin4},  \ref{Fig:Spin4nsns},   \ref{Fig:Spin4IIIns},  \ref{Fig:Spin4IIIIII}, and  \ref{Fig:Spin4IVns}.

 The models considered in this paper are given by  Weierstrass equations listed in Table \ref{Table:Weierstrass}.
The fiber at the collisions are listed in Table \ref{Table:Collisions}.
For the SO($4$)-model, the Mordell--Weil condition forces the class of the two divisors $S$ and $T$ to satisfy the linear relation 
  $$S+T=4L,$$  
 where $L=c_1(\mathscr{L})$ is the first Chern class of the fundament line bundle $\mathscr{L}$ of the Weierstrass model (see Definition \ref{Def:Weierstrass}). 
In the Calabi-Yau case,  $L=-K$ where $K$ is the canonical class of the base $B$ of the elliptic fibration.

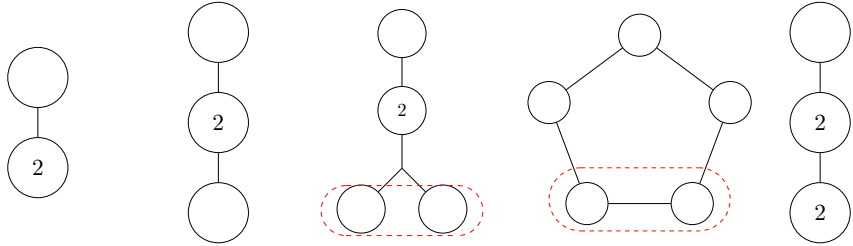
\begin{figure}[htb]
\begin{center}
\scalebox{.8}{\begin{tikzpicture}

\node (S1) at (-3,0) {\scalebox{1}{
\begin{tikzpicture}[every node/.style={circle,draw, minimum size= 10 mm}, label distance=-3mm,  scale=.5]
 \node (A1) at (0,0) {};
 \node (A2) at (0,-3) {$2$};
 \draw (A1)--(A2);
\end{tikzpicture}
}};

\node (S3)  at (3,0) 
{\scalebox{.8}{
\begin{tikzpicture}[every node/.style={circle,draw, minimum size= 10 mm}, label distance=-3mm,  scale=.4]
 \node (A0) at (90:7) {};
\node (A1) at (90:3) {$2$};
 \node (A2) at (225:3) {};
 \node (A3) at (-45:3) {};
 \draw (A1)--(0,0);
 \draw (A2)--(0,0);
 \draw (A3)--(0,0);
 \draw (A0)--(A1);
  \draw[red, dashed] (-2.9,-.9) arc (90:270:1.3cm);
   \draw[red, dashed] (2.9,-.9) arc (90:-90:1.3cm);
\draw[red, dashed] (-2.9,-.9)--(2.9,-.9);
 \draw[red, dashed] (-2.9,-3.5)--(2.9,-3.5);
\end{tikzpicture}}
};

\node (S2) at (0,0) {\scalebox{1}{
\begin{tikzpicture}[every node/.style={circle,draw, minimum size= 10 mm}, label distance=-3mm,  scale=.5]
 \node (A1) at (0,0) {};
 \node (A2) at (0,-3) {$2$};
  \node (A3) at (0,-6) {};
 \draw (A1)--(A2)--(A3);
\end{tikzpicture}
}};

\node (S5) at (10,0) {\scalebox{1}{
\begin{tikzpicture}[every node/.style={circle,draw, minimum size= 10 mm}, label distance=-3mm,  scale=.5]
 \node (A1) at (0,0) {};
 \node (A2) at (0,-3) {$2$};
  \node (A3) at (0,-6) {$2$};
 \draw (A1)--(A2)--(A3);
\end{tikzpicture}
}};

\node (S4)  at (7,0) 
{
\scalebox{.7}{
\begin{tikzpicture}[every node/.style={circle,draw, minimum size= 10 mm}, label distance=-2mm,  scale=.5]
 \node (A1) at (0,4) {};
 \node (A2) at (-4.3,0.8) {};
 \node (A3) at (-2.5,-4) {};
 \node (A4) at (2.5,-4) {};
 \node (A5) at (4.3,0.8) {};
 \draw (A1)--(A2)--(A3)--(A4)--(A5)--(A1);
 \draw[red,dashed] (-2.8,-2.3) arc (90:270:1.5cm);
 \draw[red, dashed] (2.8,-2.3) arc (90:-90:1.5cm);
 \draw[red, dashed] (-2.8,-2.3)--(2.8,-2.3);
 \draw[red, dashed] (-2.8,-5.3)--(2.8,-5.3);
\end{tikzpicture}
}};
  \end{tikzpicture}}
  \end{center}
  \caption{Non-Kodaira fibers. Nodes surrounded by a red ellipsis are obtained only after a field extension. For example, all these fibers appear in the collision I$_2^{\text{s}}+$I$_2^{\text{s}}$. 
  \label{Figure:NonKodaira}}
  \end{figure}

 Each of the Weierstrass models listed in Table \ref{Table:Weierstrass} has two crepant resolutions connected by an Atiyah flop. 
Using the main theorem of  \cite{Euler}, we determine a generating function for the Euler characteristic of SO($4$) and Spin($4$)-models (see Theorem  \ref{Thm:Euler.Char.SO4}  and Theorem \ref{Thm.Spin4.Euler}). We  study in detail the fiber degeneration of each of these elliptic fibrations and identify new non-Kodaira fibers (see Figures \ref{Figure:NonKodaira}, \ref{Fig:SO4I2nsI2ns},  \ref{Fig:SO4I2nsI2s}, \ref{Fig:SO4I2sI2s}, \ref{Fig:SO4IIII2ns}, \ref{Fig:SO4},  \ref{Fig:Spin4},  \ref{Fig:Spin4nsns},   \ref{Fig:Spin4IIIns},  \ref{Fig:Spin4IIIIII}, and  \ref{Fig:Spin4IVns}
).

When the elliptic fibration is a threefold, we  compute the triple intersection numbers of the fibral divisors (see Theorem \ref{Thm:TripleSO4}). 
Assuming the Calabi-Yau condition, we can also compute the  Hodge numbers of the elliptic fibration. 
 The five-dimensional theory has two distinct Coulomb phases separated by a wall defined by a weight of the vector representation corresponding to the difference of the fundamental weights of each A$_1$ forming the Lie algebra of type D$_2$. 
When the elliptic fibration is a Calabi-Yau threefold, the  triple intersection numbers of the fibral divisors give 
 the Chern-Simons levels of the theory in a Calabi-Yau compactification of M-theory to a five dimensional supergravity theory with eight supercharges. 
  In such a five-dimensional theory, we can constrain the number of charged hypermultiplets by comparing the triple intersection numbers of fibral divisors of the elliptic fibration with the cubic $5d$ prepotential. This point of view was presented in \cite{Grimm:2015zea} and explicitly implemented in \cite{ES,F4,G2,EKY2}. For an SO($4$)-model this is enough to completely fix the number of charged hypermultiplets. But for a Spin($4$)-model, this only gives two unresolved linear relations. The number of charged multiplets are fixed by using intersecting-brane techniques or anomaly cancellations from an uplift to a chiral six-dimensional gauged supergravity theory, coming from a Calabi-Yau compactification of F-theory. 
In all cases, we check that the matter content we obtained is compatible with an anomaly free six-dimensional supergravity theory.

\subsection{Spin groups, orthogonal groups, and half-spin groups}\label{sec:Spin}
Assuming that a Lie group $G$ is complex and connected, we only need to know its fundamental group $\pi_1(G)$ and the type of  its Lie algebra $\mathfrak{g}$ to determine $G$ up to isomorphism. 
The Lie algebra only determines the local structure of the group $G$. 
In F-theory, the  Mordell--Weil group of the elliptic fibration is isomorphic to the fundamental group of the Lie group. Assuming that the Mordell--Weil group has a trivial rank, and is therefore purely a torsion group $T$, 
we can then retrieve the group $G$ as the quotient $G=\widetilde{G}/\widetilde{T}$, where $\widetilde{G}=\exp(\mathfrak{g})$ 
is the simply connected and compact connected group  
with  Lie algebra $\mathfrak{g}$ and  $\widetilde{T}$ is a normal subgroup of 
the center $Z(\widetilde{G})$ of $\widetilde{G}$ isomorphic to $T$. Note that  different isomorphic subgroups $\widetilde{T}_1$ and $\widetilde{T}_2$  of the center $Z(\widetilde{G})$ can give non-isomorphic quotient $\widetilde{G}/\widetilde{T}_1$ and $\widetilde{G}/\widetilde{T}_2$.  
The fundamental group $\pi_1(G)$ and the center $Z(G)$ of $G$ are respectively  isomorphic to $T$ and  the quotient $Z(\widetilde{G})/\widetilde{T}$. 
\begin{equation}
G=\widetilde{G}/\widetilde{T},  \quad \widetilde{G}:=\exp(\mathfrak{g}), \quad  T\cong \widetilde{T}\subset Z(\widetilde{G}), \quad \pi_1(G)\cong T, \quad Z(G)=Z(\widetilde{G})/\widetilde{T}.
\end{equation}
In our case of interest,  we recall that the universal covering of a compact gauge group  with Lie algebra  of type D$_2\cong$A$_1\oplus$A$_1$ is  Spin($4$) 
\begin{equation}
\widetilde{G}=\text{Spin($4$)}\cong \text{SU($2$)}\times\text{SU($2$)}.
\end{equation}
The center of Spin($4+4n$) is the Klein's four-group $\mathbb{Z}/ 2\mathbb{Z}\times \mathbb{Z}/ 2\mathbb{Z}$: 
\begin{equation}
Z=\mathbb{Z}/ 2\mathbb{Z}\times \mathbb{Z}/ 2\mathbb{Z}\cong\{  \pm \mathbb{I}, \pm \Gamma_* \}.
\end{equation}
where $\mathbb{I}$ is the identity and $\Gamma_*$ is the product  of all gamma matrices and squares to the identity $\Gamma^2_*=\mathbb{I}$. 
The matrix $\Gamma_*$ is used to define Weyl spinors  of Spin($4+4n$).  
Each non-neutral element $g$ of $\mathbb{Z}/ 2\mathbb{Z}\times \mathbb{Z}/ 2\mathbb{Z}$ generates a subgroup $\langle g \rangle$ isomorphic to $\mathbb{Z}/ 2\mathbb{Z}$  and the three possibilities account for all the possible embedding of   $\mathbb{Z}/ 2\mathbb{Z}$  in  $\mathbb{Z}/ 2\mathbb{Z}\times\mathbb{Z}/ 2\mathbb{Z}$.  
 Each of the corresponding quotient  $Z/\langle g\rangle$ is  isomorphic to $\mathbb{Z}/ 2\mathbb{Z}$. But each quotient  Spin($4$)/$\langle g\rangle $ is a
  different group as expressed by the following exact sequences. 
 \begin{align}
\begin{aligned}
& 1\longrightarrow\    \langle - \mathbb{I}  \rangle \cong \mathbb{Z}/2\mathbb{Z}   \longrightarrow  
\text{Spin($4+4n$)} \longrightarrow  \text{SO($4+4n$)}
 \longrightarrow 1\\
& 1\longrightarrow \langle + \Gamma_* \rangle  \cong \mathbb{Z}/2\mathbb{Z}\longrightarrow  
\text{Spin($4+4n$)} 
\longrightarrow \text{HSpin$^+$($4+4n$)} \longrightarrow 1\\
&1\longrightarrow  \langle  -\Gamma_* \rangle \cong \mathbb{Z}/2\mathbb{Z} \longrightarrow \text{Spin($4+4n$)} \longrightarrow \text{HSpin$^-$($4+4n$)} \longrightarrow 1\\
\end{aligned}
\end{align}
The group SO($4+4n$) is the $  \mathbb{Z}/2\mathbb{Z}$ quotient of Spin($4+4n$) when $  \mathbb{Z}/2\mathbb{Z}$ is generated by  minus the identity of Spin($4+4n$).  
When  $  \mathbb{Z}/2\mathbb{Z}$ is generated by $\Gamma_*$ or $-\Gamma_*$, we get a {\em half-spin group}. 
In the case of SO($4$), we have 
\begin{equation}
 \text{HSpin$^+$($4$)}\cong \text{SO($3$)$\times$ SU($2$)}, \quad \text{HSpin$^-$($4$)}\cong \text{SU($2$)$\times$ SO($3$)}.
\end{equation}
Half-spin groups are called  {\em semi-spin group} in  Bourbaki \cite[Chap 8,\S13.4]{Bourbaki.GLA79}. 
The quotient of Spin$(4+4n)$ by its center is the adjoint group PSO($4+4n)$. Hence, there are four compact groups of type D$_{2+2n}$, namely 
 the simply connected group Spin($4+4n$), the adjoint group PSO($4+4n)$, the half-spin groups HSpin$^\pm$($4+4n$), and the orthogonal group SO($4+4n$).

\subsection{The simplest SO($4$)-model as a Miranda model}
The simplest SO($4$)-model is realized by III+III, the collision of two fibers of type III, in an elliptic fibration with Mordell--Weil group $\mathbb{Z}/2\mathbb{Z}$. Its defining equation is 
 \begin{equation}
y^2z= x(x^2 + st z^2).
\end{equation}
As an illustration, we  quickly derive this equation. First,  the general elliptic fibration with Mordell--Weil group $\mathbb{Z}/2\mathbb{Z}$ is \cite{Aspinwall:1998xj,Husemoller}
 \begin{equation}
y^2z= x(x^2 + a_2 z^2+ a_4 z^3).
\end{equation}
The generator of the Mordell--Weil group is the section $x=y=0$ and the neutral element is $x=z=0$. 
A fiber of type III over the generic point of $S=V(S)$ requires that the valuation of $a_2$ and $a_4$ be:
 $$v_S(a_2)\geq 1, \quad  v_S(a_4)=1.$$
 Hence, we should have 
 \begin{equation}
y^2z= x(x^2 + \tilde{a}_2 s^{1+m} z^2+ \tilde{a}_4 s  z^3), \quad m\in \mathbb{Z}_{\geq 0}.
\end{equation}  
The discriminant is $ \tilde{a}_4^2 s^3 (4 \tilde{a}_4 - \tilde{a}_2^2 s)$. 
We can therefore take $\tilde{a}_4=t$ to have  a collision of type III$+$I$_2^{\text{ns}}$ on the divisors $S=V(s)$ and  $T=V(t)$. 
Since $a_4=st$, we have the linear relation $S+T=4L$.  
The fiber I$_2^{\text{ns}}$ is replaced by a  fiber of type III when $v_T(\tilde{a}_2)\geq 1$. 
That would give 
 \begin{equation}
y^2z= x(x^2 + \tilde{a}_2 s^{1+m}t^{1+n} z^2+ st  z^3), \quad m, n\in \mathbb{Z}_{\geq 0}.
\end{equation}  
If we take the lowest valuations ($m=n=0$), the coefficient $\tilde{a}_2$ has to be section of $\mathscr{L}^{-\otimes 2}$ while $a_2$  is a section of $\mathscr{L}^{\otimes 2}$.  A general solution is  simply to take  $\tilde{a}_2=0$, which gives\footnote{For example, if the base is $\mathbb{P}^r$, this is the only possibility as a line bundle of $\mathbb{P}^r$ and its inverse cannot have  non-trivial sections.  }
 \begin{equation}
y^2z= x(x^2 + st z^2).
\end{equation}

The  reduced discriminant $\Delta_{red}= st$ is a normal transverse divisor and the $j$-invariant is a constant morphism taking the value $j=1728$ everywhere. After a crepant resolution defined by a sequence of two blow-ups, we get a fiber of type III over the generic point of $S=V(s)$ and the generic point of $T=V(t)$.  Over the generic  point of their intersection $S\cap T$, the fiber degenerates to
The fiber III degenerates to  a non-Kodaira fiber of type $1-2-1$, which is a contraction of a fiber of type I$_0^*$. There are no other singular fibers. 
This model satisfies all the conditions of a Miranda model. In Miranda regularization,  the collision III+III is replaced by a chain of collisions of  type III+I$_0^*$+III  (see \cite[Table 13.1]{Miranda}) by blowing up the intersection of the two divisors. 
The intersection becomes an exceptional divisor of the base over which the generic fiber  is of type I$_0^{*\text{ss}}$. 
Here, we avoid such a blowup of the base since it modifies the canonical class and introduce an additional component in the gauge algebra changing the  gauge algebra from type A$_1\oplus$A$_1$ to type A$_1\oplus$ B$_3\oplus$ A$_1$

\section{Summary of results}
In this section, we categorize all the possible collisions of the fibers that yield SO($4$) and Spin($4$)-models, and summarize the results of the paper. We first state the geometrical setup and results including Euler characteristics, Hodge numbers, and the triple intersection polynomials in section \ref{sec.sum.geo}, and describe their application to the five-dimensional supergravity theories and their six-dimensional uplifted theories in section \ref{sec.sum.5d6d}. We then list the collision of singularities in section \ref{sec.sum.col}.

\subsection{Geometry}\label{sec.sum.geo}

Weierstrass equations for SO($4$) and Spin($4$)-models with minimal valuations of the coefficients are given by Table \ref{Table:Weierstrass}.

 \begin{table}[htb]
 \begin{center}
  $
\begin{array}{|c| c | c | l |}
\hline 
 \text{Group} & \text{Collision}  &\text{Mordell--Weil}&  \text{Weierstrass model}  \\
\hline
 \multirow{5}{*}{SO($4$)}& \text{I}^{\text{ns}}_2+\text{I}^{\text{ns}}_2\to \text{I}^{\text{ns}}_4&   \multirow{5}{*}{$\mathbb{Z}/ 2\mathbb{Z}$} & y^2z=x^3+a_2x^2z+stxz^2  \\
&\text{I}^{\text{ns}}_2+\text{I}^{\text{s}}_2\to \text{I}^{\text{s}}_4 & & y^2z+a_1 xyz=x^3+\widetilde{a}_2 tx^2z+stxz^2 \\
 &\text{I}^{\text{s}}_2+\text{I}^{\text{s}}_2\to \text{I}^{\text{s}}_4 &  & y^2z+a_1 xyz=x^3+ \widetilde{a}_2 st x^2 z+stxz^2 \\
  &\text{III}+\text{I}^{\text{ns}}_2\to 1-2-1  & & y^2z=x^3+\widetilde{a}_2 s x^2 z+stxz^2 \\
 &\text{III+III}\to 1-2-1 & &  y^2z=x^3+\widetilde{a}_2 st x^2 z+stxz^2 \\
\hline
 \multirow{6}{*}{Spin($4$)} &   \text{I}^{\text{ns}}_2+\text{I}^{\text{ns}}_2\to \text{I}^{\text{ns}}_4 & \multirow{6}{*}{trivial} & y^2z=x^3+a_2x^2z+\widetilde{a}_4 stxz^2+ \widetilde{a}_6 s^2 t^2\\
 &\text{I}^{\text{ns}}_2+\text{I}^{\text{s}}_2  \to \text{I}^{\text{s}}_4& & y^2z+a_1 x y z =x^3+\widetilde{a}_2t x^2z+\widetilde{a}_4  stxz^2 + \widetilde{a}_6 s^2 t^2 \\
 &\text{I}^{\text{s}}_2+\text{I}^{\text{s}}_2 \to \text{I}^{\text{s}}_4 & &y^2z+a_1 x y z =x^3+\widetilde{a}_2st x^2z+\widetilde{a}_4  stxz^2 + \widetilde{a}_6 s^2 t^2 \\
&\text{III}+\text{I}^{\text{ns}}_2 \to 1-2-1  && y^2z =x^3+\widetilde{a}_2 s x^2z+\widetilde{a}_4  stxz^2 + \widetilde{a}_6 s^2 t^2\\
 & \text{III}+\text{III} \to 1-2-1 &&    y^2z =x^3+\widetilde{a}_2 st x^2z+\widetilde{a}_4  stxz^2 + \widetilde{a}_6 s^2 t^2 \\
 &\text{IV}^{\text{ns}}+\text{I}_2^{\text{ns}} \to1-2-(1,1) & &   y^2z=x^3+\widetilde{a}_2sx^2z+\widetilde{a}_4s^2txz^2+\widetilde{a}_6 s^2t^2z^3   \\ 
\hline
\end{array}
$
\end{center}
\caption{Weierstrass equations and collision rules \label{Table:Weierstrass}}
\end{table}

 Let $Y_0$ be one of  the Weierstrass models considered in Table \ref{Table:Weierstrass}. Then $Y_0$ has two  distinct crepant resolutions $f^\pm: Y^\pm\to Y_0$. One is given by the  sequence of two  blowups $f^+$. 
 The other crepant resolution is obtained by exchanging the order of the blowups.

\begin{equation}\label{Eq:Blowups1}
  \begin{tikzcd}[column sep=huge] 
 \scalebox{1}{$f^+$ :} \quad  X_0  \arrow[leftarrow]{r} {(x,y,s|e_1)} & \arrow[leftarrow]{r}{(x,y,t|w_1)} X_1^+ &X_2^+.
  \end{tikzcd}
\end{equation}
 \begin{equation}\label{Eq:Blowups1}
  \begin{tikzcd}[column sep=huge] 
  \scalebox{1}{$f^-$ :} \quad X_0  \arrow[leftarrow]{r} {(x,y,t|w_1)} &  \arrow[leftarrow]{r}{(x,y,s|e_1)} X_1^- &X_2^-.
  \end{tikzcd}\end{equation}
The two resolutions are connected by an Atiyah flop. These two resolutions are not isomorphic to each other as the triple intersection numbers are not symmetric under the permutation of  $S$ and $T$.

  \begin{restatable}{thm}{Pushspin}
\label{Thm.Spin4.Euler}
 The generating polynomial of the Euler characteristic of a Spin($4$)-model  obtained by a crepant resolution of a Weierstrass model given in  Table \ref{Table:Weierstrass} is
$$
\begin{aligned}
\begin{split}
\chi(Y)=\ & 2\left( \frac{S \left(-6 L^2 (4 T+3)+L (8 (T-1) T-9)+T (5 T+4)\right)}{(2 L+1) (S+1) (T+1) (-6 L+2 S+2 T-1)} \right. \\
&\quad\quad \left. +\frac{S^2 (L (8 T+6)+5 T+3)+3(2 L+1) \left(T^2-L (3 T+2)\right)}{(2 L+1) (S+1) (T+1) (-6 L+2 S+2 T-1)}\right) c(B) .
\end{split}
\end{aligned}
$$
In particular, in the case of a Calabi-Yau threefold that is also a Spin($4$)-model we have
$$
\chi(Y)=-2 \left(30 K^2+15 K (S+T)+3 S^2+3 T^2+4 S T\right).
$$
\end{restatable}

\begin{restatable}{thm}{Pushspinhodge}
\label{Thm:HodgeNumbersSpin4}
The Hodge numbers of a Spin($4$)-model 
 given by the crepant resolution of a Weierstrass model given in Table \ref{Table:Weierstrass} are
$$
h^{1,1}(Y)=13 - K^2 , \quad h^{2,1}(Y) =13+29 K^2+15 K (S+T)+3 S^2+4 S T+3 T^2.
$$
\end{restatable}

For the Euler characteristic and the Hodge numbers of the Spin($4$)-models, see their proof and detailed description in section \ref{pf:spin4.euler}.

\begin{restatable}{thm}{Pushso}
\label{Thm:Euler.Char.SO4}
The generating polynomial of the Euler characteristic of an SO($4$)-model given by the crepant resolution of a Weierstrass model given in  Table \ref{Table:Weierstrass} is
$$
\chi(Y)=\frac{4 ( 3 L + 4 T L -T^2)}{(1+T) (1+4 L-T)} c(B) .
$$
In particular, if the SO($4$)-model  is a Calabi-Yau threefold,we have  
$$
\chi(Y)=-4 (9 K^2 + 4 K T + T^2) .
$$
\end{restatable}

\begin{restatable}{thm}{Pushsohodge}
The Hodge numbers of an SO($4$)-model 
 given by the crepant resolution of a Weierstrass model given in Table \ref{Table:Weierstrass} are
$$
h^{1,1}(Y)=13 - K^2 , \quad h^{2,1}(Y)=13 + 17 K^2 + 8 K T + 2 T^2.
$$
\end{restatable}

 For the Euler characteristic and the Hodge numbers of the SO($4$)-models, see their proof and detailed description in section \ref{pf:so4.euler}.

\begin{restatable}{thm}{PushTripleSpinplus}
\label{Thm:Spin4Triple}
 Let $f^+:Y^+\to Y_0$ be the crepant resolution where $Y_0$ is any of the Spin($4$)-model  listed in Table \ref{Table:Weierstrass}. 
The triple intersection polynomial  of $Y^+$ is 
$$
\begin{aligned}
\begin{split}
\mathscr{F}_{trip}^+
=& \ \int_Y \pi_* f_*\Big[ \Big(\psi_0 D_0^s+\psi_1 D_1^s+\phi_0 D_0^t +\phi_1 D_1^t\Big)^3 \Big] \\
=& \ 2T (-2 L+S-T) \phi _1^3 -6 S T \psi _1 \phi _1^2 -2S(2 L+S) \psi _1^3 \\
& +2T (2 L-S-2 T) \phi _0^3+6T \phi _0^2 \left(\phi _1 (-2 L+S+T)-S \psi _1\right) -4S (S-L) \psi _0 \left(\psi _0-\psi _1\right)^2 \\
&-2S(2 L+S) \psi_0\psi _1\left(\psi _0-\psi_1 \right) +6T \phi _0 \left(\phi _1^2 (2 L-S) -2S \left(\psi _0-\psi _1\right)^2+2 S \psi _1 \phi _1\right) .
\end{split}
\end{aligned}
$$
The triple intersection polynomial in the fibration $Y^-$ defined by exchanging the order of the blowup is $\mathscr{F}_{trip}^-$ and is obtained from $\mathscr{F}_{trip}^-$ by the involution  $\psi\leftrightarrow \phi$.
\end{restatable}

 The triple intersection polynomial for the Spin($4$)-models are derived in section \ref{Thm:TripleSO4}. The triple intersection for an SO($4$)-model is then derived from the one of a Spin($4$)-model by the specialization $S\to -4K-T$.
 
\begin{restatable}{thm}{PushTripleSOplus}
\label{Thm:SO4Triple}
 Let $f^+:Y^+\to Y_0$ be the crepant resolution where $Y_0$ is any of the SO($4$)-model  listed in Table \ref{Table:Weierstrass}. 
The triple intersection polynomial  of $Y^+$ is 
\begin{align} \nonumber
\begin{split}
\mathscr{F}_{trip}^+=&
\pi_* f_* \Big(\psi_0 D_0^s+\psi_1 D_1^s+\phi_0 D_0^t +\phi_1 D_1^t\Big)^3\\ 
=&-2 T \phi _0^3 (2 L+T)-2 (4 L-T) \left(\psi _0-\psi _1\right)^2 \left(\psi _0 (6 L-2 T)+\psi _1 (6 L-T)\right)\\
&+6 T \phi _0 \left(-\left(\psi _0-\psi _1\right)^2 (4 L-T)+2 \psi _1 \phi _1 (4 L-T)+\phi _1^2 (T-2 L)\right)
+6 T \phi _0^2 \left(\psi _1 (T-4 L)+2 L \phi _1\right)
\\
&-2  \left(24 L^2-10 L T+T^2\right) \psi_1^3+6 T (T-4 L) \psi_1 \phi_1^2+4 T (L-T)  \phi_1^3.
\end{split}
\end{align}
\end{restatable}
 This triple intersection model of an SO($4$)-model is described in detail in section \ref{sec:TripSO4}.

\subsection{Applications to $5d$ and $6d$ supergravity theories}\label{sec.sum.5d6d}

The $\mathbb{Z}/2\mathbb{Z}$ quotients of  Spin($4$) can be  SO($4$),  or the half-spin groups SO($3$)$\times$SU($2$) and SU($2$)$\times$ SO($3$). But which one is realized by the collisions presented in Table \ref{Table:Weierstrass}? This question is answered by scrutinizing the representation $\mathbf{R}$ associated with the elliptic fibration. 
 
We recall that a representation of A$_1\oplus$A$_1$ is given by two spins $(j_1, j_2)$ where $j_1$ is in $\frac{1}{2}\mathbb{Z}_{\geq 0}$. It we name the representation by two numbers indicating the dimension of each projection, $(j_1, j_2)$ is the same as  $(\bf{2j_1+1},\bf{2j_2+1})$.   Each representation $(j_1, j_2)$ is a valid representation of Spin($4$). But is only a projective representation of the three possible $\mathbb{Z}/2\mathbb{Z}$ quotient of Spin($4$). 
  More explicitly, the representations of the semi-spin group SU($2$)$\times$ SO($3$) are those with spin $(j_1, j_2)$ where  $j_2$ an integer, while the representations of the semi-spin group SO($3$)$\times$ SU($2$) have spin $(j_1, j_2)$ where $j_1$ an integer, and the representations of SO($4$) have spin $(j_1, j_2)$ such that  $j_1+j_2$ an integer. 
\begin{table}[H]
\begin{center}
$$
\begin{tabular}{| r  |  l  |}
\hline 
Group & Representation  $(j_1, j_2)$ or $(\bf{2j_1+1}, \bf{2j_2+1})$\\ 
\hline
Spin($4$)  &  $j_1, j_2\in \frac{1}{2}\mathbb{Z}_{\geq 0}$  \\
Spin$^+$($4$)  &$j_1\in\mathbb{Z}_{\geq 0}, \    j_2\in \frac{1}{2}\mathbb{Z}_{\geq 0}$ \\
Spin$^-$($4$)  &$ j_1\in \frac{1}{2}\mathbb{Z}_{\geq 0}, \  j_2\in\mathbb{Z}_{\geq 0}$ \\
SO($4$)  &  $j_1, j_2\in \frac{1}{2}\mathbb{Z}_{\geq 0}$,  $j_1+j_2\in\mathbb{Z}_{\geq 0}$\\
\hline
\end{tabular}
$$
\end{center}
\caption{Representations for SO($4$) and Spin($4$) groups.  \label{Table:Representations}}
\end{table}
 The determination of the representation $\mathbf{R}$ associated to an elliptic fibration is explained in section \ref{Sec:Rep}. 
In the present case, we find that
\footnote{We denote a representation by the dimensions $(\bf{d_1}, \bf{d_2})$ of its two projections.}: 
\begin{center}
\begin{tabular}{rl}
SO($4$)-model: &\quad $\mathbf{R}=(\bf{3},\bf{1})\oplus(\bf{1},\bf{3})\oplus(\bf{2},\bf{2})$ \\
Spin($4$)-model: &\quad $\mathbf{R}=(\bf{3},\bf{1})\oplus(\bf{1},\bf{3})\oplus(\bf{2},\bf{2})\oplus (\bf{2},\bf{1})\oplus(\bf{1},\bf{2})$ \\
\end{tabular}
\end{center}
The representation $(\bf{3},\bf{1})\oplus(\bf{1},\bf{3})$ is the adjoint representation of the Lie algebra of type  D$_2\cong$ A$_1\oplus$ A$_1$ and the representation $(\bf{2},\bf{1})\oplus(\bf{1},\bf{2})$ is the spin representation of D$_2$.
The representation $(\bf{2},\bf{2})$ is the vector representation of D$_2$. 
The collisions listed  in Table \ref{Table:Weierstrass} always produce  the bifundamental representation $(\mathbf{2},\mathbf{2})$ of spin ($\frac{1}{2}, \frac{1}{2})$ of the Lie algebra of type D$_2$. 
Such a representation is incompatible with the half-spin groups and left only SO($4$) and Spin($4$) as possible options. \footnote{The bifundamental representation $(\mathbf{2},\mathbf{2})$ of  A$_1\oplus$ A$_1$  is the vector representation of SO($4$) of dimension $4$, which, in terms of spins of the two SU($2$) forming a Spin($4$), is the representation ($\frac{1}{2},\frac{1}{2}$). 
The bifundamental representation  rules out the groups  SU($2$)$\times$ SO($3$) and SO($3$)$\times$ SU($2$), but is a valid representation of both SO($4$) and Spin($4$).  }

From the point of view of the elliptic fibration, the group Spin($4$) requires a trivial Mordell--Weil group while the group SO($4$) requires  a Mordell--Weil group $\mathbb{Z}/ 2\mathbb{Z}$.  
We are not aware of any  F-theory construction  of a gauge theory with the semi-spin group  SU($2$)$\times$SO($3$).

We denote by $g_S$ and $g_T$ the genus of $S$ and $T$. 
The SO($4$)-model only has adjoint representations ($\mathbf{Adj}^-=(\bf{3},\bf{1})$ and $\mathbf{Adj}^+=(\bf{1},\bf{3})$) and the vector representation $\bf{V}=(\bf{2},\bf{2})$. 
 For the  Spin($4$)-model, there are additional hypermultiplets transforming  in the two semi-spin representations ($\mathbf{Spin}^-$ $(\bf{2},\bf{1})$ and $\mathbf{Spin}^+$ $(\bf{1},\bf{2})$) of D$_2\cong \mathfrak{so}(4)$. 
 If we denote by $\Delta'$ the third component of the discriminant locus, then the number of hypermultiplets transforming in the semi-spin representations $\mathbf{Spin^\pm}$ of D$_2\cong \mathfrak{so}(4)$ are given by the intersection numbers 
 $S\cdot \Delta'$ and $T\cdot \Delta'$. The class of $\Delta'$ is $-2(4K+T+S)$. In particular, it is zero when we specialize to the SO($4$)-model. 

We show that for an SO($4$)-model  and a Spin($4$)-model with the matter content discussed above  and summarized on  Table \ref{Table:Data}, all anomalies are canceled in a six-dimensional ${\cal N}=(1,0)$ supergravity theory. 
Comparing the triple intersection numbers of an SO($4$)-model with the prepotential of a five-dimensional ${\cal N}=1$ theory with matter charged under the same representation $\mathbf{R}$ completely fixes the numbers of charged hypermultiplets.
However this is not the case for a Spin($4$)-model, since there are more representations involved; the comparison with the triple intersection numbers only give two linear relations. Using additional information from intersecting branes, we check that the six-dimensional uplifted theory can be anomaly-free by the Green-Schwarz mechanism.

\clearpage

\begin{figure}[bth]
\begin{center}
\begin{tabular}{rl}
\begin{minipage}[b]{0.35\textwidth}
$\begin{tikzcd}[column sep=huge, scale=.8] 
Y^- \arrow[rightarrow]{rdd}{ \scalebox{1.1}{$f^-$ }} \arrow[leftrightarrow,dashed]{rr} {\scalebox{1.2}{flop}}&  & Y^+ \arrow[rightarrow,swap]{ldd} { \scalebox{1.1}{$f^+$}}\\
& & \\
& Y_0 & 
  \end{tikzcd}
$\end{minipage}
&
\begin{minipage}{0.35\textwidth}
\begin{tikzpicture}[scale=.3]
				
			\draw (90:7)--(210:7)--(330:7)--(90:7);
			\draw (90:7)--(90:-4.5);
									\node at (-87:5.7) {\scalebox{1}{$[1;-1]$}};	
					\node at (200:-2) {\scalebox{2}{$+$}};
					\node at (160:2) {\scalebox{2}{$-$}};													
			\end{tikzpicture}
			\end{minipage}
			\end{tabular}
			\end{center}
\caption{
Coulomb phases of an SO($4$)-model or a Spin($4$)-model with matter in the representation $(\mathbf{2},\mathbf{2})$. 
The addition of the representation $(\bf{3},\bf{1})$, $(\bf{1},\bf{3})$, $(\bf{2},\bf{1})$, or $(\bf{1},\bf{2})$ do not change the chamber structure.
The only weight defining an interior wall is the weight $[1;-1]$ of the representation $(\mathbf{2},\mathbf{2})$. 
  Geometrically, the chambers are identified by the presence of a curve with weights $\pm[1,-1]$ with respect to the fibral divisors $D_1^s$ and   $D_1^t$ that project respectively to the divisors $S$ and $T$ and do not touch the zero section of the elliptic fibration. 
 \label{fig:SO4Phases} }
\end{figure}
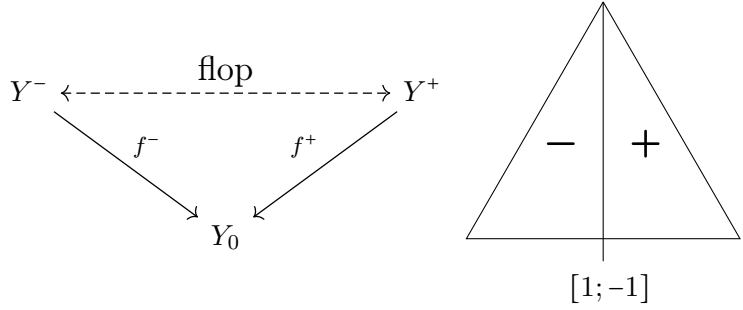

  \begin{table}[H]
  \begin{center}
  \scalebox{.94}{
$
  \begin{array}{|c|l|}
  \hline
\multirow{5}{*}{SO($4$)} \quad& \mathbf{R}=(\bf{3},\bf{1})\oplus(\bf{1},\bf{3})\oplus(\bf{2},\bf{2})\\
&\chi(CY_3)=-4 (9 K^2 + 4 K T + T^2) \\
& h^{1,1}(CY_3)=13 - K^2 , \quad h^{2,1}(CY_3)=13 + 17 K^2 + 8 K T + 2 T^2\\
& \mathscr{F}^{+}_{triple}=
-2  \left(24 L^2-10 L T+T^2\right) \psi_1^3+6 T (T-4 L) \psi_1 \phi_1^2+4 T (L-T)  \phi_1^3
\\
& 
 n_{\bf{1,3}}=g_T,\quad
n_{\bf{3,1}}=g_S ,\quad n_{\bf{2,2}}=S\cdot T\\
\hline
 \multirow{5}{*}{ Spin($4$)}\quad & \mathbf{R}=(\bf{3},\bf{1})\oplus(\bf{1},\bf{3})\oplus(\bf{2},\bf{2})\oplus (\bf{2},\bf{1})\oplus(\bf{1},\bf{2})\\
 & \chi(CY_3)=-2 \left(30 K^2+15 K S+15 K T+3 S^2+4 S T+3 T^2\right)\\
 &\mathscr{F}_{triple}^+=2T (-2 L+S-T) \phi _1^3 -6 S T \psi _1 \phi _1^2 -2S(2 L+S) \psi _1^3\\
 & h^{1,1}(CY_3)=13 - K^2 , \quad h^{2,1}(CY_3)=13+29 K^2+15 K S+15 K T+3 S^2+4 S T+3 T^2\\
& n_{\bf{3,1}}=g_S,\quad  n_{\bf{1,3}}=g_T, \quad  n_{\bf{2,2}}=S\cdot T , \quad n_{\bf{2,1}}=S\cdot V(\tilde{b}_8),\quad n_{\bf{1,2}}=T\cdot V(\tilde{b}_8)\\
\hline
\end{array}
$
}
\end{center}
\caption{Data of the low energy effective theory of a  SO($4$) and  a Spin($4$)-model compactified on a Calabi-Yau threefolds. 
$\mathbf{R}$ is the representation, $\chi$ is the Euler characteristic of the Calabi-Yau threefold, $S$ and $T$ are the two divisors supporting the reducible fibers, 
$g_S$ and $g_T$ are the genus of these divisors. The number of representations in the irreducible representation $\mathbf{R}_i$ is denoted $n_{\mathbf{R}_i}$. 
 $\mathscr{F}_{triple}^+$ is the triple intersection polynomial for the variety $Y^+$ defined by the crepant resolution of equation \eqref{Eq:Blowups1}.  
In the resolution $Y^-$, we have a distinct $\mathscr{F}_{triple}^-$ obtained by exchanging $\psi_1$ and $\phi_1$. 
\label{Table:Data}
}
  \end{table}

  \clearpage
 \subsection{Collision of Singularities} \label{sec.sum.col}
 \begin{table}[H]
 \begin{center}
 \scalebox{.8}{
 \begin{tabular}{|c|}
 \hline
 \begin{tikzpicture}
\node at (-9,0) {\scalebox{1.5}{I$_2^{ns}+$I$_2^{ns}$}};
\node (S) at (-6,0) {\includegraphics[scale=1]{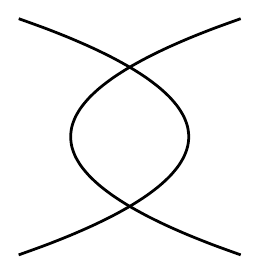}};
\node (T) at (-2,0) {\includegraphics[scale=1]{I2}};

\node (ST) at (5,0) {
\begin{tikzpicture}[every node/.style={circle,draw, minimum size= 10 mm}, scale=.5]
 \node (A1) at (0,3) {};
 \node (A2) at (-3,0) {};
 \node (A3) at (3,0) {};
 \node (A4) at (0,-3) {};
 \draw (A1)--(A2)--(A4)--(A3)--(A1);
 \draw[red,dashed] (-3.3,1.7) arc (90:270:1.5cm);
 \draw[red,dashed] (3.3,1.7) arc (90:-90:1.5cm);
 \draw[red,dashed]  (-3.3,1.7)--(3.3,1.7);
  \draw[red,dashed] (-3.3,-1.3)--(3.3,-1.3);
\end{tikzpicture}
};

			\draw[->,>=stealth',thick=1mm]  (0,0)--(2,0);
			
			\node at (-4,0) {\scalebox{3}{$+$}}	;	
	\end{tikzpicture}
	
	\\
	\hline
	\begin{tikzpicture}
	\node at (-9,0) {\scalebox{1.5}{I$_2^{ns}+$I$_2^{s}$}};
		\node at (-9.3,-.8) {\scalebox{1.5}{or\   I$_2^{s}+$I$_2^{s}$}};
\node (S) at (-6,0) {\includegraphics[scale=1]{I2}};
\node (T) at (-2,0) {\includegraphics[scale=1]{I2}};

\node (ST) at (5,0) {
\begin{tikzpicture}[every node/.style={circle,draw, minimum size= 10 mm}, scale=.5]
 \node (A1) at (0,3) {};
 \node (A2) at (-3,0) {};
 \node (A3) at (3,0) {};
 \node (A4) at (0,-3) {};
 \draw (A1)--(A2)--(A4)--(A3)--(A1);
\end{tikzpicture}
};

			\draw[->,>=stealth',thick=1mm]  (0,0)--(2,0);
			
			\node at (-4,0) {\scalebox{3}{$+$}}	;	
	\end{tikzpicture}
	\\
	\hline

	\begin{tikzpicture}
	\node at (-15,0) {\scalebox{1.5}{III+III}};
\node (S) at (-12,0) {\includegraphics[scale=1.5]{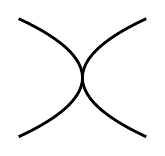}};
\node at (-10,0) {\scalebox{3}{$+$}}	;	
\node (T) at (-8,0) {\includegraphics[scale=1.5]{III}};
\draw[->,>=stealth',thick=1mm]  (-5.5,0)--+(2,0);
\node (ST) at (-2,0) {
\begin{tikzpicture}[every node/.style={circle,draw, minimum size= 10 mm}, scale=.5]
 \node (A1) at (0,3) {};
 \node (A2) at (0,0) {$2$};
 \node (A3) at (0,-3) {};
 \draw (A1)--(A2)--(A3);
\end{tikzpicture}
};

	\end{tikzpicture}
\\
\hline
	\begin{tikzpicture}
	\node at (-11,0) {\scalebox{1.5}{III+I$_2^{\text{ns}}$}};
		\node at (-11.5,-1) {\scalebox{1.5}{or \ III+I$_2^{\text{s}}$}};
\node (S) at (-8,0) {\includegraphics[scale=1.5]{III}};
\node (T) at (-4,0) {\includegraphics[scale=1]{I2}};

\node (ST) at (2,0) {
\begin{tikzpicture}[every node/.style={circle,draw, minimum size= 10 mm}, scale=.5]
 \node (A1) at (0,3) {};
 \node (A2) at (0,0) {$2$};
 \node (A3) at (0,-3) {};
 \draw (A1)--(A2)--(A3);
\end{tikzpicture}
};

			\draw[->,>=stealth',thick=1mm]  (-2,0)--(0,0);
			
			\node at (-6,0) {\scalebox{3}{$+$}}	;	
	\end{tikzpicture}
\\
\hline

	 \begin{tikzpicture}
\node at (-9,0) {\scalebox{1.5}{IV$^{ns}+$I$_2^{ns}$}};
\node (S) at (-6,0) {
\begin{tikzpicture}[every node/.style={circle,draw, minimum size= 10 mm}, label distance=-3mm,  scale=.4]
 \node (A1) at (90:3) {};
 \node (A2) at (225:3) {};
 \node (A3) at (-45:3) {};
 \draw (A1)--(0,0);
 \draw (A2)--(0,0);
 \draw (A3)--(0,0);
  \draw[red,dashed] (-2.9,-.9) arc (90:270:1.3cm);
   \draw[red,dashed] (2.9,-.9) arc (90:-90:1.3cm);
\draw[red,dashed] (-2.9,-.9)--(2.9,-.9);
 \draw[red,dashed] (-2.9,-3.5)--(2.9,-3.5);
  \end{tikzpicture}};
\node (T) at (-2,0) {\includegraphics[scale=1]{I2}};

\node (ST) at (5,0) {\scalebox{1}{
\begin{tikzpicture}[every node/.style={circle,draw, minimum size= 10 mm}, label distance=-3mm,  scale=.4]
 \node (A0) at (90:7) {};

\node (A1) at (90:3) {$2$};
 \node (A2) at (225:3) {};
 \node (A3) at (-45:3) {};
 \draw (A1)--(0,0);
 \draw (A2)--(0,0);
 \draw (A3)--(0,0);
 \draw (A0)--(A1);
  \draw[red,dashed] (-2.9,-.9) arc (90:270:1.3cm);
   \draw[red,dashed] (2.9,-.9) arc (90:-90:1.3cm);
\draw[red,dashed] (-2.9,-.9)--(2.9,-.9);
 \draw[red,dashed] (-2.9,-3.5)--(2.9,-3.5);

\end{tikzpicture}
}};

			\draw[->,>=stealth',thick=1mm]  (0,0)--(2,0);
			
			\node at (-4,0) {\scalebox{3}{$+$}}	;	
	\end{tikzpicture}\\
	\hline
	\end{tabular}
	}
	\end{center}
	\caption{Collisions at the intersections of two Kodaira fibers with dual graph $\widetilde{\text{A}}_1$. The representation produced is always the bifundamental representation $(\mathbf{2},\mathbf{2})$ of A$_1\oplus$A$_1$. 
	Nodes surrounded by a  red ellipse form a non-split node and are obtained individually only after a field extension (they are ``related by monodromies''). 
		\label{Table:Collisions} }
	\end{table}
\clearpage

\clearpage

\clearpage

\begin{figure}[htb]

\begin{center}
\scalebox{1}{\begin{tikzpicture}
\node (S) at (-6,0) {\includegraphics[scale=1.5]{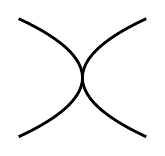}};
\node (T) at (6,0) {\includegraphics[scale=1]{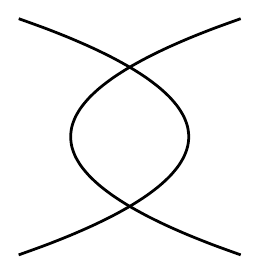}};
\node (TT) at (6,-4) {\includegraphics[scale=1.5]{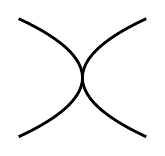}};
\node (ST) at (0,-3) {
\begin{tikzpicture}[every node/.style={circle,draw, minimum size= 10 mm}, scale=.5]
 \node (A1) at (0,3) {};
 \node (A2) at (0,0) {$2$};
 \node (A3) at (0,-3) {};
 \draw (A1)--(A2)--(A3);
\end{tikzpicture}
};

\draw[->,>=stealth',thick=1mm]  (T)--($(TT)+(0,1)$);
\draw[->,>=stealth',thick=1mm]  (TT)--($(ST)+(1,-1)$);
			\draw[->,>=stealth',thick=1mm]  (S)--($(ST)+(-2,2)$);
			\draw[->,>=stealth',thick=1mm]  (T)--($(ST)+(2,2)$);
												
												\node  at (-2.4,-.5) {$\displaystyle{T=0}$}; 
																							\node  at (2.3,-.5) {$\displaystyle{S=0}$};
																							\node  at (2,-3.5) {$\displaystyle{S=0}$};	
																							\node  at (5,-2.5) {$\displaystyle{\widetilde{a}_2=0}$};	
	\end{tikzpicture}}
	\end{center}
	\caption{Fiber structure of an SO($4$)-model with III+I$_2^{\text{ns}}$. 
	The Weierstrass equation is $y^2z=x^3 +\text{a}_2 st^2 z+ s t x z^2$. 
		\label{Fig:SO4IIII2ns}}
	
\begin{center}
\scalebox{1}{\begin{tikzpicture}
\node (S) at (-6,0) {\includegraphics[scale=1.5]{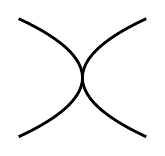}};
\node (T) at (6,0) {\includegraphics[scale=1.5]{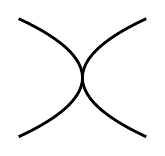}};

\node (ST) at (0,-3) {
\begin{tikzpicture}[every node/.style={circle,draw, minimum size= 10 mm}, scale=.5]
 \node (A1) at (0,3) {};
 \node (A2) at (0,0) {$2$};
 \node (A3) at (0,-3) {};
 \draw (A1)--(A2)--(A3);
\end{tikzpicture}
};

			\draw[->,>=stealth',thick=1mm]  (S)--($(ST)+(-2,2)$);
			\draw[->,>=stealth',thick=1mm]  (T)--($(ST)+(2,2)$);
												
												\node  at (-3.3,-1) {$\displaystyle{T=0}$}; 
																							\node  at (3.3,-1) {$\displaystyle{S=0}$};	
	\end{tikzpicture}}
	\end{center}
	\caption{Fiber structure of an SO($4$)-model with III+III. 
	The Weierstrass equation is $y^2z=x^3 + s t x z^2$ and there are moduli coming from the Weierstrass coefficients since we necessary have $S+T=4L$ and $a_2=0$. 
	The discriminant $\Delta\propto s^3 t^3$ is a divisor with simple normal crossing and the $j$-invariant is constant and equal to $1728$. 
	\label{Fig:SO4}}
\vspace{1cm}

	\end{figure}

\clearpage

\clearpage

\begin{figure}[htb]
\begin{center}
\scalebox{.7}{\begin{tikzpicture}

\node (S) at (-5,2) {\includegraphics[scale=1]{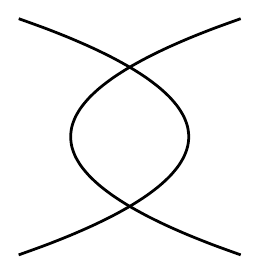}};
\node (T) at (6,2) {\includegraphics[scale=1]{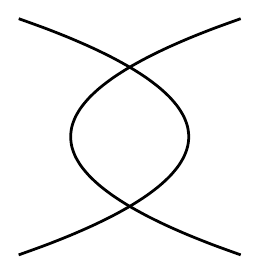}};

\node (ST) at (0,-3) {
\begin{tikzpicture}[every node/.style={circle,draw, minimum size= 10 mm}, scale=.5]
 \node (A1) at (0,3) {};
 \node (A2) at (-3,0) {};
 \node (A3) at (3,0) {};
 \node (A4) at (0,-3) {};
 \draw (A1)--(A2)--(A4)--(A3)--(A1);
  \draw[red,dashed] (-2.8,1.4) arc (90:270:1.5cm);
 \draw[red,dashed] (2.8,1.4) arc (90:-90:1.5cm);
 \draw[red,dashed] (-2.8,1.4)--(2.8,1.4);
 \draw[red,dashed] (-2.8,-1.6)--(2.8,-1.6);

\end{tikzpicture}
};

\node (S1) at (-10,-4) {\includegraphics[scale=1.8]{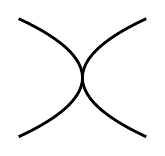}};
\node (T1) at (10,-3.5) {\includegraphics[scale=1.8]{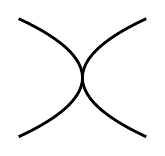}};

\node (S2) at (-5,-3.5) {
\scalebox{.8}{
\begin{tikzpicture}[every node/.style={circle,draw, minimum size= 10mm}, scale=.5]
 \node (A1) at (0,4) {};
 \node (A2) at (-2.8,0) {};
 \node (A3) at (2.8,0) {};
 \draw (A1)--(A2)--(A3)--(A1);
 \draw[red,dashed] (-2.9,1.8) arc (90:270:1.6cm);
 \draw[red,dashed] (2.9,-1.4) arc (-90:90:1.6cm);
 \draw[red,dashed] (-2.9,-1.4)--(2.9,-1.4);
 \draw[red,dashed] (-2.9,1.8)--(2.9,1.8);
\end{tikzpicture}
}};

\node (T2) at (6,-3) {
\scalebox{.8}{
\begin{tikzpicture}[every node/.style={circle,draw, minimum size= 10mm}, scale=.5]
 \node (A1) at (0,4) {};
 \node (A2) at (-2.8,0) {};
 \node (A3) at (2.8,0) {};
 \draw (A1)--(A2)--(A3)--(A1);
 \draw[red,dashed] (-2.9,1.8) arc (90:270:1.6cm);
 \draw[red,dashed] (2.9,-1.4) arc (-90:90:1.6cm);
 \draw[red,dashed] (-2.9,-1.4)--(2.9,-1.4);
 \draw[red,dashed] (-2.9,1.8)--(2.9,1.8);
\end{tikzpicture}
}};

\node (ST1)  at (3.5,-11) {
\scalebox{.8}{
\begin{tikzpicture}[every node/.style={circle,draw, minimum size= 10 mm}, label distance=-2mm,  scale=.5]
 \node (A1) at (0,4) {};
 \node (A2) at (-4.3,0.8) {};
 \node (A3) at (-2.5,-4) {};
 \node (A4) at (2.5,-4) {};
 \node (A5) at (4.3,0.8) {};
 \draw (A1)--(A2)--(A3)--(A4)--(A5)--(A1);
 \draw[red,dashed] (-4.6,2.4) arc (90:270:1.5cm);
 \draw[red,dashed] (4.6,2.4) arc (90:-90:1.5cm);
 \draw[red,dashed] (-4.6,2.4)--(4.6,2.4);
 \draw[red,dashed] (-4.6,-0.6)--(4.6,-0.6);
 \draw[red,dashed] (-2.8,-2.3) arc (90:270:1.5cm);
 \draw[red,dashed] (2.8,-2.3) arc (90:-90:1.5cm);
 \draw[red,dashed] (-2.8,-2.3)--(2.8,-2.3);
 \draw[red,dashed] (-2.8,-5.3)--(2.8,-5.3);
\end{tikzpicture}
}};

\node (ST1b) at (-3,-10.5)
{\scalebox{.8}{
\begin{tikzpicture}[every node/.style={circle,draw, minimum size= 10 mm}, label distance=-3mm,  scale=.5]
 \node (A1) at (0,0) {};
 \node[label=right:2] (A2) at (0,-4) {};
 \node (A3) at (0,-8) {};
 \draw (A1)--(A2)--(A3);
\end{tikzpicture}
}};

\node (ST2) at (0,-17.5) {\scalebox{.9}{
\begin{tikzpicture}[every node/.style={circle,draw, minimum size= 10 mm}, label distance=-3mm,  scale=.5]
 \node (A0) at (90:7) {};
 \node (A1) at (90:3) {$2$};
 \node (A2) at (225:3) {};
 \node (A3) at (-45:3) {};
 \draw (A1)--(0,0);
 \draw (A2)--(0,0);
 \draw (A3)--(0,0);
 \draw (A0)--(A1);
 \draw[red,dashed] (-2.9,-.9) arc (90:270:1.3cm);
 \draw[red,dashed] (2.9,-.9) arc (90:-90:1.3cm);
 \draw[red,dashed] (-2.9,-.9)--(2.9,-.9);
 \draw[red,dashed] (-2.9,-3.5)--(2.9,-3.5);
\end{tikzpicture}
}};

\node (ST3) at (0,-24) {\scalebox{.9}{
\begin{tikzpicture}[every node/.style={circle,draw, minimum size= 10 mm}, label distance=-3mm,  scale=.5]
 \node (A1) at (0,3.5) {};
 \node (A2) at (0,0) {$2$};
 \node (A3) at (0,-3.5) {$2$};
 \draw (A1)--(A2)--(A3);
\end{tikzpicture}
}};

\node (S3) at (-10,-11) {\scalebox{1}{
\begin{tikzpicture}[every node/.style={circle,draw, minimum size= 10 mm}, label distance=-3mm,  scale=.5]
 \node (A1) at (90:3) {};
 \node (A2) at (225:3) {};
 \node (A3) at (-45:3) {};
 \draw (A1)--(0,0);
 \draw (A2)--(0,0);
 \draw (A3)--(0,0);
  \draw[red,dashed] (-2.9,-.9) arc (90:270:1.3cm);
   \draw[red,dashed] (2.9,-.9) arc (90:-90:1.3cm);
\draw[red,dashed] (-2.9,-.9)--(2.9,-.9);
 \draw[red,dashed] (-2.9,-3.5)--(2.9,-3.5);

 \end{tikzpicture}
}};

\node (T3) at (10,-10) {\scalebox{1}{
\begin{tikzpicture}[every node/.style={circle,draw, minimum size= 10 mm}, label distance=-3mm,  scale=.5]
 \node (A1) at (90:3) {};
 \node (A2) at (225:3) {};
 \node (A3) at (-45:3) {};
 \draw (A1)--(0,0);
 \draw (A2)--(0,0);
 \draw (A3)--(0,0);
   \draw[red,dashed] (-2.9,-.9) arc (90:270:1.3cm);
   \draw[red,dashed] (2.9,-.9) arc (90:-90:1.3cm);
\draw[red,dashed] (-2.9,-.9)--(2.9,-.9);
 \draw[red,dashed] (-2.9,-3.5)--(2.9,-3.5);

\end{tikzpicture}
}};

\node (S4) at (-10,-18) {\scalebox{1}{
\begin{tikzpicture}[every node/.style={circle,draw, minimum size= 10 mm}, label distance=-3mm,  scale=.5]
 \node (A1) at (0,0) {};
 \node (A2) at (0,-5) {$2$};
 \draw (A1)--(A2);
\end{tikzpicture}
}};

\node (T4) at (10,-18) {\scalebox{1}{
\begin{tikzpicture}[every node/.style={circle,draw, minimum size= 10 mm}, label distance=-3mm,  scale=.5]
 \node (A1) at (0,0) {};
 \node (A2) at (0,-5) {$2$};
 \draw (A1)--(A2);
\end{tikzpicture}
}};

												\draw[->,>=stealth',thick=1mm]  (S1)--(S3) node [midway, fill=white,text=black] {$ \widetilde{a}_4=0$};
												\draw[->,>=stealth',thick=1mm]  (S3)--(S4) node [midway, fill=white,text=black] {$ \widetilde{a}_6=0$};
												\draw[->,>=stealth',thick=1mm]  (S2)--(S3);
												\node [fill=white,text=black]  at ($(S2)+(-1,-2)$)  {$a_2=0$};

	\draw[->,>=stealth',thick=1mm]  (T1)--(T3) node [midway, fill=white,text=black] {$ \widetilde{a}_4=0$};
	\draw[->,>=stealth',thick=1mm]  (T3)--(T4) node [midway, fill=white,text=black] {$ \widetilde{a}_6=0$};
	\draw[->,>=stealth',thick=1mm]  (T2)--(T3) node [midway, fill=white,text=black] {$ a_2=0$};
	
	\draw[->,>=stealth',thick=1mm]  (ST)--(ST1);
	\node [fill=white,text=black] at ($(ST1)+(-2,5)$) {$\widetilde{a}_4^2-4a_2\widetilde{a}_6=0$};
	\draw[->,>=stealth',thick=1mm]  (ST)--(ST1b);
		\draw[->,>=stealth',thick=1mm]  (ST1)--(ST2) node [midway, fill=white,text=black] {$a_2=0$};
	\draw[->,>=stealth',thick=1mm]  (ST2)--(ST3) node [midway, left=.3cm,fill=white,text=black] {$ \widetilde{a}_6=0$};
		\draw[->,>=stealth',thick=1mm]  (S2)--(ST1);
		\node [fill=white,text=black]  at ($(ST)+(-1,-3)$)  {$a_2=0$};
				\draw[->,>=stealth',thick=1mm]  (T2)--(ST1);
				\node [fill=white,text=black] at  ($(T2)+(-1.2,-4.2)$)  {$S=0$};
				\draw[->,>=stealth',thick=1mm]  (S3)--(ST2) node [midway, fill=white,text=black] {$T=0$};
				\draw[->,>=stealth',thick=1mm]  (T3)--(ST2) node [midway, fill=white,text=black] {$S=0$};
				\draw[->,>=stealth',thick=1mm]  (S4)--(ST3) node [midway, fill=white,text=black] {$T=0$};
				\draw[->,>=stealth',thick=1mm]  (T4)--(ST3) node [midway, fill=white,text=black] {$S=0$};
	
\node [fill=white,text=black] at ($(ST)+(-0.1,-4.6)$)  {$T=0$};
				\draw[->,>=stealth',thick=1mm]  (S1)--(ST1b) node [midway, fill=white,text=black] {$T=0$};
				\draw[->,>=stealth',thick=1mm]  (T1)--(ST1b) node [midway, fill=white,text=black] {$S=0$};
				\draw[->,>=stealth',thick=1mm]  (ST1b)--(ST2) node [midway, fill=white,text=black] {$ \widetilde{a}_4=0$};

\draw[->,>=stealth',thick=1mm]  (S)--(S1) node [midway, fill=white,text=black] {$a_2=0$};
				\draw[->,>=stealth',thick=1mm]  (S)--(S2) node [midway, fill=white,text=black] {$\widetilde{a}_4^2-4a_2 \widetilde{a}_6=0$};
				\draw[->,>=stealth',thick=1mm]  (S)--(ST) node [midway, fill=white,text=black] {$ T=0$};

\draw[->,>=stealth',thick=1mm]  (T)--(T1) node [midway, fill=white,text=black] {$a_2=0$}; 
				\draw[->,>=stealth',thick=1mm]  (T)--(T2) node [midway, fill=white,text=black] {$\widetilde{a}_4^2-4a_2\widetilde{a}_6=0$};
				\draw[->,>=stealth',thick=1mm]  (T)--(ST)node [midway, fill=white,text=black] {$S=0$};

	\end{tikzpicture}}
	\end{center}
	\caption{Fiber structure of a Spin($4$)-model with I$_2^{\text{ns}}+$I$_2^{\text{ns}}$. 
\label{Fig:Spin4nsns}}
	\end{figure}
\clearpage

\clearpage

\clearpage

\clearpage

\clearpage

\section{Preliminaries}

\begin{defn}
An elliptic fibration is a proper surjective morphism  $\varphi:Y\to B$ between algebraic varieties such that the generic fiber of $\varphi$ is a smooth algebraic curve of genus one 
and $\varphi$ is endowed with a rational section.  
\end{defn}
We work over the complex numbers and assume that the base $B$ is a smooth projective variety. 
The locus of points of $B$ over which the fiber is singular is called the discriminant locus of $\varphi$. 
Under mild assumptions, the discriminant is a Cartier divisor \cite{Dolgavcev.Purity}.

\subsection{Singular fibers over generic points}\label{Sec:Kodaira}

Kodaira has classified the possible types of  singular fibers of a minimal elliptic surface. N\'eron arrived to the same classification in an arithmetic setting based on Weierstrass models. 
For higher dimensional elliptic fibrations, one can define a notion of minimality and Kodaira's classification continues to hold for geometric fibers over  generic points of the discriminant locus.

We recall that a fiber over a point $p$ is scheme defined with respect to the  residue field of the point $p$. 
If the point $p$ is a closed point, its residue field is just the ground field and the fiber is always geometric when the  ground field is algebraically closed. 
But if the point $p$ is the generic point of an irreducible variety, its residue field might not be algebraically closed. 
In that case, some components of the singular fiber that are irreducible with respect to the residue field of $p$ might become reducible after a field extension. 
Such a component is said to not be geometrically irreducible. 

The fibers over generic points are classified by decorated Kodaira fibers where the decoration (``split'', ``non-split", ``semi-split") keeps track of the minimal field extension needed to make all component of the singular fiber (and the divisors defined by intersection of components) geometrically irreducible. 
It is enough to consider only quadratic and cubic field extension. The split case corresponds to the case in which the fiber has components that are geometrically irreducible. 
Kodaira fibers of  type I$_1$, II, III, III$^*$, and II$^*$ are always geometric. 
Kodaira fibers of type I$_{n\geq 2}$, IV, IV$^*$, I$_{n\geq 0}^*$ are said to be split when they have geometrically irreducible component or non-split otherwise. 
For these fibers,  a quadratic field extension is required to make them geometrically irreducible. 
For the fiber I$_0^*$, the non-split and split cases require respectively a cubic and quadratic field extension. 

The fiber of type I$_2$ is very special in the sense that its non-split and split type have the same number of irreducible components,  but  distinguished themselves by the splitting properties of their intersection points. 
The fiber of type I$_2$ is composed of two rational curves intersecting transversally at two distinct geometric points. 
Let $\kappa$ be the residue field of the point $p$ over which the fiber I$_2$ is considered. 
When the two points of intersection form   a $\kappa$-irreducible divisor on each component, the curve (with respect to the residue field of $p$), the fiber is said to be  non-split, 
otherwise, the divisor is the sum of two points that are rational with respect to the residue field of $p$ and the fiber is said to be split. 
At the collision of two I$_2$ fibers, the generic fiber can be I$_4^{\text{ns}}$ or I$_4^{\text{s}}$. The difference between I$_4^{\text{ns}}$ or I$_4^{\text{s}}$ over a codimension-two loci evaporates when the base is a surface since 
then these fibers are located over closed points with residue fields that are algebraically closed. That means that the fibers over codimension-two points of an elliptic threefolds are always geometric.

While geometric fibers have dual graphs that are always ADE affine Dynkin diagrams, the fiber type over generic points are twisted affine Dynkin diagram. 
We denoted an affine Dynkin diagram by $\widetilde{g}$ and its Langlands dual by $\widetilde{\mathfrak{g}}^t$, this is the unique twisted Dynkin diagram whose  Cartan matrix is the transpose of the Cartan matrix of $\tilde{\mathfrak{g}}$.

If $K_i$ are decorated Kodaira fibers and $S_i$ are irreducible divisors of the base, a model of type $K_1+K_2+\cdots +K_n$ is an elliptic fibration for which there exists two irreducible components $S_1$ and $S_2$ in the  discriminant locus such that 
 the type of the fiber over the generic point of $S_i$ is $K_i$ and the  fiber over the generic point of  any other component of the discriminant locus is irreducible (Kodaira type II or I$_1$).

\subsection{Lie algebra, representations, and Lie groups from  an elliptic fibrations}\label{Sec:Rep}
If the dual fibers of $K_1$ and $K_2$ are affine Dynkin diagrams of type $\widetilde{\mathfrak{g}}^t_1$ and $\widetilde{\mathfrak{g}}^t_2$, then the Lie algebra associated with the $K_1+K_2$-model is  
$$\mathfrak{g}=\mathfrak{g}_1\oplus \mathfrak{g}_2.$$ 

Let $C$ be a vertical curve, i.e.  a curve contained in a fiber of the elliptic fibration.
Let $S$ be an irreducible component of the reduced discriminant of the elliptic fibration $\varphi: Y\to B$. 
The pullback of  $\varphi^* S$ has irreducible components $D_0, D_1, \ldots, D_n$, where $D_0$ is the component touching the section of the elliptic fibration.   
The divisors $D_a$ are called fibral divisors. 
The {\em weight vector} of $C$ over $S$ is  by definition the vector ${\varpi}_S(C)=(-D_1\cdot C, \ldots, -D_n\cdot C)$ of intersection numbers $D_i\cdot C$ for $i=1,\ldots, n$.

The irreducible curves of the degenerations over codimension-two loci give weights of a representation $\mathbf{R}$. However, they only give a subset of weights. Hence, we need an algorithm that retrieves the full representation $\mathbf{R}$ given only a few of its weights.  
This problem can be addressed systematically using the notion of a saturated set of weights introduced by  Bourbaki  \cite[Chap.VIII.\S 7. Sect. 2]{Bourbaki.GLA79}.

\begin{defn}[Saturated set of weights]
A set $\Pi$ of integral weights is {\em saturated} if for any weight $\varpi\in\Pi$ and any simple root $\alpha$,  the weight $\varpi-i\alpha$ is also in $\Pi$ for any $i$ such that $0\leq i\leq \langle \varpi,\alpha\rangle$. 
A saturated set has {\em highest weight} $\lambda$ if $\lambda\in\Lambda^+$ and $\mu\prec\lambda$ for any $\mu\in \Pi$. 
\end{defn}
\begin{defn}[Saturation of a subset]
Any subsets $\Pi$ of weights is contained in a unique smallest saturated subset. We call it the saturation of $\Pi$. 
\end{defn}
\begin{prop}\quad 
\begin{enumerate}[label=(\alph*)]
\item A saturated set of weights is invariant under the action of the Weyl group.  
\item The saturation of a set of weights $\Pi$ is finite if and only if  the set $\Pi$ is finite.
\item A saturated set with highest weight $\lambda$  consists of all dominant weights lower than or equal to $\lambda$ and their conjugates under the Weyl group.  
\end{enumerate}
\end{prop}
\begin{proof}
See   \cite[Chap. III \S 13.4]{Humphreys}.
\end{proof}

\begin{thm}[{(\cite[Chap.VIII.\S 7. Sect. 2, Corollary to Prop. 5]{Bourbaki.GLA79}.)}]\label{Thm:R-Saturation}
Let $\Pi$ be a finite saturated  set of weights. Then there exists a finite dimensional  $\mathfrak{g}$-module  whose set of weights is $\Pi$. 
\end{thm}

\begin{defn}
To a $G$-model, we associate a representation $\mathbf{R}$ of the Lie algebra $\mathfrak{g}$ as follows. 
 The weight vectors of the irreducible vertical   rational curves of the fibers over codimension-two points form  a set $\Pi$  whose  saturation defines uniquely a representation  $\mathbf{R}$ 
  by  Theorem \ref{Thm:R-Saturation}.   We call this representation  $\mathbf{R}$  the representation of the $G$-model.\footnote{
   This definition is a formalization of the method of Aspinwall and Gross \cite[\S 4]{Aspinwall:1996nk}. See also \cite{Marsano}. 
Note that we always get the adjoint representation as a summand of $\mathbf{R}$.
  } 
  \end{defn}

The  unique compact, connected, and simply connected Lie group with Lie algebra $\mathfrak{g}$ is 
$$\widetilde{G}=\exp(\mathfrak{g}).
$$
Assuming that the Mordell--Weil group has  rank $r$ and torsion subgroup $H$, the gauge group attached to the elliptic fibration 
requires the specification of an embedding of $H$ in the center of $\widetilde{G}$
$$
H\cong  \widetilde{H}\subset Z(\widetilde{G}).
$$
 Then 
$$G= U(1)^r \times \widetilde{G}/\widetilde{H}, \quad H\cong \widetilde{H}, \quad \widetilde{H}\mathrel{\unlhd} Z(\widetilde{G}).$$ 
Two different  isomorphic subgroups of $Z(\widetilde{G})$ can  give two different quotients $\widetilde{G}/\widetilde{H}$. The choice of the correct embedded is restricted by the representation $\mathbf{R}$ attached to the elliptic fibration since not all representation of  Lie algebra of a group $G$ is a representation of  $G$. 

\subsection{Hyperplane arrangement I($\mathfrak{g},\mathbf{R}$)}

Let  $\mathfrak{g}$ be a semi-simple Lie algebra and $ \mathbf{R}$ a representation of $\mathfrak{g}$.
The kernel of each  weight $\varpi$ of $\mathbf{R}$ defines a  hyperplane $\varpi^\perp$ through the origin of the Cartan sub-algebra of $\mathfrak{g}$. 
\begin{defn}[hyperplane arrangement I($\mathfrak{g},\mathbf{R}$)]
The hyperplane arrangement I($\mathfrak{g},\mathbf{R}$) is defined inside the dual fundamental Weyl chamber of $\mathfrak{g}$, i.e. the  dual cone of the fundamental Weyl chamber of $\mathfrak{g}$, and its hyperplanes are the set of kernels of  the weights of $\mathbf{R}$. 
\end{defn}
For each $G$-model, we associate the hyperplane arrangement $\mathrm{I}(\mathfrak{g},  \mathbf{R})$ using the representation $\mathbf{R}$ induced by the weights of vertical rational curves produced by degenerations of the generic fiber over codimension-two points of the base. We then study the incidence structure of the hyperplane arrangement I$(\mathfrak{g}, \mathbf{R})$ \cite{EJJN1,EJJN2,G2,F4,Hayashi:2014kca}.

\subsection{Non-simply-connected simple groups in F-theory}

The classification of connected Lie groups over the complex or real numbers is based on the following theorem whose content can be traced back to Cartan and Lie. 
In what follows, all groups and Lie algebras are defined over the complex numbers or the real numbers. We refer to  \cite[Chap III, \S 6]{Bourbaki.GLA13} for more information.

\newpage
\begin{thm}\label{Thm:Lie}  \quad
\begin{enumerate}
\item Any connected  Lie group $G$ defined over the complex numbers or the real numbers is isomorphic to a quotient $\tilde{G}/K$ of its universal covering group $\tilde{G}$ by  a discrete  central subgroup $K$   of $\tilde{G}$ isomorphic to the fundamental group of $G$. 

\item Two connected  Lie groups having the same Lie algebra are locally isomorphic. Two  simply connected Lie groups having the same Lie algebra are isomorphic. 

\item (Cartan--Lie theorem)  If $\mathfrak{g}$ is a finite dimensional Lie algebra, there exists a simply connected Lie group whose associated Lie algebra is isomorphic to $\mathfrak{g}$.
\end{enumerate}
\end{thm}
\begin{proof}  See \cite[Chap III, \S 6.3, Theorem 3]{Bourbaki.GLA13}.  \end{proof}

The center of  $G$ is isomorphic to the quotient $Z(\tilde{G})/K$. In particular, this implies that the fundamental group of any connected group is Abelian. The third  assertion is the Cartan--Lie theorem (usually called Lie's  third fundamental theorem).\footnote{ Lie proved the local existence of a Lie group. Cartan proved the global existence and the simply connected property. } 
The second assertion (without the simply connected specialization) is Lie's second fundamental theorem. 
The following definition is inspired by the Cartan--Lie theorem (second assertion of Theorem \ref{Thm:Lie}).
\begin{defn}[simply connected Lie group associated with a Lie algebra]
Given a  finite dimensional Lie algebra $\mathfrak{g}$, the 
 simply connected Lie group whose Lie algebra is isomorphic to $\mathfrak{g}$ is called the {\em simply connected Lie group  associated with} $\mathfrak{g}$ and is denoted by 
 $\exp(\mathfrak{g})$. 
\end{defn}
A direct consequence of Theorem \ref{Thm:Lie} is that the  classification of connected simple Lie groups reduces to the classification of simply connected simple Lie groups and of the subgroups of their centers. 
The centers of simply connected simple  complex Lie groups are given in Table \ref{Table:Centers}.

If $\mathfrak{g}$ is G$_2$, F$_4$, or E$_8$, there is  a unique connected, simple, complex compact Lie group with Lie algebra  $\mathfrak{g}$.
If  $\mathfrak{g}$ is A$_{p-1}$ (with $p$ a prime number), B$_{3+n}$, C$_{2+n}$, D$_{5+2n}$, E$_6$,  E$_7$, there are two compact connected Lie groups with Lie algebra $\mathfrak{g}$, namely the simply connected group $G=\exp(\mathfrak{g})$ and the centerless group 
$G_{ad}:=G/Z(G)$. For symplectic groups, we denote by Sp($2n$) for the compact and connected simple complex Lie group with Lie algebra C$_n$, that is 
Sp($2n$)$\cong $ USp($2n$).

The case of A$_{n-1}$  involves the classification of subgroups of $\mathbb{Z}/n\mathbb{Z}$. 
The subgroups of $\mathbb{Z}/n\mathbb{Z}$ are the cyclic groups $\mathbb{Z}/r\mathbb{Z}$ such that $r$ is a divisor of $n$. 
There are three compact connected Lie groups with Lie algebra D$_{5+2n}$ and four with Lie algebra D$_{4+2n}$. \footnote{The Lie algebra D$_{5+2n}$ has three distinct compact groups: Spin($10+4n$), its $\mathbb{Z}/2\mathbb{Z}$ quotient SO($10+4n$), and the centerless group PSO($10+4n$). 
 The center of Spin(${8+4n}$) is $(\mathbb{Z}/2\mathbb{Z} )^2=\{\pm 1, \pm \Gamma_*\}$ with $\Gamma^2_*=1$. We have 4 distinct  subgroups:  the trivial group, the full group $(\mathbb{Z}/2\mathbb{Z} )^2$, and three proper subgroups --each generated by a non-neutral element  of 
  $(\mathbb{Z}/2\mathbb{Z} )^2$--each isomorphic to 
 $\mathbb{Z}/2\mathbb{Z}$. As a result there are give compact groups with Lie algebra D$_{4+2n}$: the simply connected group Spin($8+4n$), the orthogonal group SO($8+4n$),  the center less group PSO($8+4n$), and two groups HSpin($8+4n$) that are isomorphic to each other. One is a quotient of Spin($8+4n$) by $\{1, \Gamma_*\}$ and the other is a quotient by $\{1,-\Gamma_*\}$.
} Namely, 
 the simply connected group Spin($8+2n$), its $\mathbb{Z}/2\mathbb{Z}$ quotient SO($8+2n$), and its centerless quotient PSO($8+2n$).

\begin{table}
\begin{center}
\begin{tabular}{|c|c|c|c|c|p{.5cm}|}
\hline 
$\mathfrak{g}$ & $G=\exp(\mathfrak{g})$ & Center $Z(G)$ &   $G_{Ad}:=G/Z(G)$ & Quotient with non-trivial center \\
\hline
A$_{n-1}$ & SU($n$) & $\mathbb{Z}/n\mathbb{Z}$ &  PSU($n$) &  SU($n$)/ ($\mathbb{Z}/r\mathbb{Z}$) with $r|n$ and $r\neq 1, n$\\
\hline 
B$_{3+n}$ & Spin($7+2n$)& $\mathbb{Z}/2\mathbb{Z}$ & SO($7+2n$) &- \\
\hline
C$_{2+n}$ & Sp($4+2n$) &  $\mathbb{Z}/2\mathbb{Z}$ &  PSp($4+2n$) & -\\
\hline
D$_{4+2n}$ & Spin($8+4n$) &$(\mathbb{Z}/2\mathbb{Z})^2$ & PSO($8+4n$) & SO($8+4n$),  HSpin$^\pm$($8+4n$) \\
\hline
D$_{5+2n}$ &Spin($10+4n$) &$\mathbb{Z}/4\mathbb{Z}$ &PSO($10+4n$)  & SO($10+4n$)\\
\hline
E$_6$ & E$_6$ & $\mathbb{Z}/3\mathbb{Z}$ &E$_6$/ ($\mathbb{Z}/3\mathbb{Z}$) & -\\
\hline
E$_7$ & E$_7$ &$\mathbb{Z}/2\mathbb{Z}$ & E$_7$/($\mathbb{Z}/2\mathbb{Z}$)  & -\\
\hline
E$_8$ & E$_8$ & trivial& E$_8$ & -\\
\hline
F$_4$ & F$_4$ &trivial& F$_4$  & -\\
\hline
G$_2$ & G$_2$ &trivial& G$_2$  &- \\
\hline
\end{tabular}
\end{center}
\caption{Classification of connected compact simple  complex Lie groups. 
The group $G=\exp(\mathfrak{g})$ is the unique simply connected compact group with Lie algebra $\mathfrak{g}$. We denote its center by $Z$. 
The group $G_{Ad}=G/Z$ is the unique centerless connected simple group with Lie algebra $\mathfrak{g}$. The group $G$ and $G_{Ad}$ are isomorphic when $Z(G)$ is the trivial group, that is for $\mathfrak{g}=$ G$_2$, F$_4$, or E$_8$. 
$G$ and $G_{Ad}$ are the only connected compact simple groups with Lie algebra $\mathfrak{g}$ except in the case of $\mathfrak{g}=$A$_{n-1}$ with $n$ a non-prime number and  $\mathfrak{g}=$D$_{4+2n}$ where the orthogonal group SO($8+4n$) and the 
half spin group HSpin($8+4n$) are  
non-simply connected with center $\mathbb{Z}/2\mathbb{Z}$.  
The orthogonal group SO($2m$) is the quotient of Spin$(2n)$ by $\pm 1$. 
The groups HSpin$^\pm$($8+4m$) are the  $\mathbb{Z}/2\mathbb{Z}$ quotient of Spin($8+4m$) for which one of the half-spin representations is faithful. 
The group HSpin$^\pm$($8+4m$) are isomorphic to each  other. The half-spin groups HSpin$^\pm$($8+4m$) are isomorphic to  SO($8+4m$) if and only if $m=0$. 
\label{Table:Centers}}
\end{table}

\subsection{Weierstrass models and Deligne's formulaire}
\begin{defn}[Weierstrass model {\cite{Esole.Elliptic}}]\label{Def:Weierstrass}
Consider a variety $B$ endowed with a line bundle $\mathscr{L}\rightarrow B$.
A Weierstrass model $\mathscr{E} \rightarrow B$ over $B$ is a hypersurface  cut out by the zero locus of a section of the line bundle of $\mathscr{O}(3)\otimes\pi^* \mathscr{L}^{\otimes 6}$ in the projective bundle  $\mathbb{P}(\mathscr{O}_B \oplus\mathscr{L}^{\otimes 2} \oplus\mathscr{L}^{\otimes 3})\rightarrow B$. 
We denote by $\mathscr{O}(1)$  the dual of the tautological line bundle of the projective bundle, and denote by $\mathscr{O}(n)$ ($n>0$) its $n$th-tensor product.  
The relative projective coordinates of the $\mathbb{P}^2$ bundle are denoted by $[x:y:z]$. In particular,  $x$ is a section of $\mathscr{O}(1)\otimes\pi^* \mathscr{L}^{\otimes 2}$, $y$ is a section of $\mathscr{O}(1)\otimes\pi^* \mathscr{L}^{\otimes 3}$, and $z$ is a section of $\mathscr{O}(1)$. 
 Following Tate and Deligne's notation, the defining equation of a  Weierstrass model is
$$
\mathscr{E}: \quad  zy(y +a_1 x+ a_3 z) -(x^3 +a_2 x^2 z+a_4 xz^2 +a_6 z^3)=0,
$$
 where the coefficient $a_i$ ($i=1,2,3,4,6$) is a section of $\mathscr{L}^{\otimes i}$ on $B$. Such a hypersurface is an elliptic fibration since over the generic point of the base, the fiber is a nonsingular  cubic planar curve with a rational point ($x=z=0$). 
    We use the convention of  Deligne's formulaire \cite{Deligne.Formulaire}: 
$$
\begin{aligned}
b_2 &= a_1^2+ 4 a_2,\quad
b_4 = a_1 a_3 + 2 a_4 ,\quad
b_6  = a_3^2 + 4 a_6 , \quad
b_8  =b_2 a_6 -a_1 a_3 a_4 + a_2 a_3^2-a_4^2,\\
c_4 & = b_2^2 -24 b_4, \quad
c_6 = -b_2^3+ 36 b_2 b_4 -216 b_6,\\
\Delta &= -b_2^2 b_8 -8 b_4^3 -27 b_6^2 + 9 b_2 b_4 b_6,\quad  j =\frac{c_4^3}{\Delta}.
\end{aligned}
$$
These quantities are related by the  relations
$$
4 b_8 =b_2 b_6 -b_4^2 \quad \text{and}\quad 1728 \Delta=c_4^3 -c_6^2. 
$$ 
\end{defn}

 The discriminant locus is the subvariety of $B$ cut out by the equation $\Delta=0$, and is the locus of points  $p$ of the base $B$ such that the fiber over $p$ (i.e. $Y_p$)  is singular. 
 Over a generic point of $\Delta$, the fiber is a nodal cubic that  degenerates to a cuspidal  cubic over the codimension-two locus $c_4=c_6=0$. Up to isomorphisms, the
 $j$-invariant $j=c_4^3/ \Delta$ uniquely characterizes nonsingular elliptic curves.

\subsection{Pushforward formulas}\label{Sec:Intersection}

Many of the  intersection theory computations performed on elliptic fibrations come down to  the following three theorems. 
The first one is a theorem of Aluffi which gives the Chern class after a blowup along a local complete intersection. 
The second theorem is a pushforward theorem that provides a user-friendly method to compute invariant of the blowup space in terms of the original space. 
The last theorem is a direct consequence of functorial properties of the Segre class and gives a simple method to pushforward analytic expressions in the Chow ring of a projective bundle to  the Chow ring of its base.

\begin{defn}[Resolution of singularities]
A resolution of singularities of a variety $Y$ is a proper birational morphism $\varphi:\widetilde{Y}\longrightarrow Y$  such that  
$\widetilde{Y}$ is nonsingular
and  $\varphi$ is an isomorphism away  from the singular  locus of $Y$. In other words, $\widetilde{Y}$ is nonsingular and  if $U$ is the singular locus of $Y$, $\varphi$ maps $\varphi^{-1}(Y\setminus U)$isomorphically  onto $Y\setminus U$.  
\end{defn}

\begin{defn}[Crepant birational map]
A  birational map $\varphi:\widetilde{Y}\to Y$ between two algebraic varieties with  $\mathbb{Q}$-Cartier canonical classes is said to be {\em crepant} if it preserves the canonical class, i.e.  
$
K_{\widetilde{Y}}=\varphi^\ast K_Y.
$
\end{defn}

\begin{thm}[Aluffi, {
{\cite[Lemma 1.3]{Aluffi_CBU}}}]
\label{Thm:AluffiCBU}
Let $Z\subset X$ be the  complete intersection  of $d$ nonsingular hypersurfaces $Z_1$, \ldots, $Z_d$ meeting transversally in $X$.  Let  $f: \widetilde{X}\longrightarrow X$ be the blowup of $X$ centered at $Z$. We denote the exceptional divisor of $f$  by $E$. The total Chern class of $\widetilde{X}$ is then:
$$
c( T{\widetilde{X}})=(1+E) \left(\prod_{i=1}^d  \frac{1+f^* Z_i-E}{1+ f^* Z_i}\right)  f^* c(TX).
$$
\end{thm}

\begin{thm}[Esole--Jefferson--Kang,  see  {\cite{Euler}}] \label{Thm:Push}
    Let the nonsingular variety $Z\subset X$ be a complete intersection of $d$ nonsingular hypersurfaces $Z_1$, \ldots, $Z_d$ meeting transversally in $X$. Let $E$ be the class of the exceptional divisor of the blowup $f:\widetilde{X}\longrightarrow X$ centered 
at $Z$.
 Let $\widetilde{Q}(t)=\sum_a f^* Q_a t^a$ be a formal power series with $Q_a\in A_*(X)$.
 We define the associated formal power series  ${Q}(t)=\sum_a Q_a t^a$, whose coefficients pullback to the coefficients of $\widetilde{Q}(t)$. 
 Then the pushforward $f_*\widetilde{Q}(E)$ is
 $$
  f_*  \widetilde{Q}(E) =  \sum_{\ell=1}^d {Q}(Z_\ell) M_\ell, \quad \text{where} \quad  M_\ell=\prod_{\substack{m=1\\
 m\neq \ell}}^d  \frac{Z_m}{ Z_m-Z_\ell }.
 $$ 
\end{thm}

\begin{thm}[{See  \cite{Euler} and  \cite{AE1,AE2,Fullwood:SVW,EKY1}}]\label{Thm:PushH}
Let $\mathscr{L}$ be a line bundle over a variety $B$ and $\pi: X_0=\mathbb{P}[\mathscr{O}_B\oplus\mathscr{L}^{\otimes 2} \oplus \mathscr{L}^{\otimes 3}]\longrightarrow B$ a projective bundle over $B$. 
 Let $\widetilde{Q}(t)=\sum_a \pi^* Q_a t^a$ be a formal power series in  $t$ such that $Q_a\in A_*(B)$. Define the auxiliary power series $Q(t)=\sum_a Q_a t^a$. 
Then 
$$
\pi_* \widetilde{Q}(H)=-2\left. \frac{{Q}(H)}{H^2}\right|_{H=-2L}+3\left. \frac{{Q}(H)}{H^2}\right|_{H=-3L}  +\frac{Q(0)}{6 L^2},
$$
 where  $L=c_1(\mathscr{L})$ and $H=c_1(\mathscr{O}_{X_0}(1))$ is the first Chern class of the dual of the tautological line bundle of  $ \pi:X_0=\mathbb{P}(\mathscr{O}_B \oplus\mathscr{L}^{\otimes 2} \oplus\mathscr{L}^{\otimes 3})\rightarrow B$.
\end{thm}

\subsection{Euler characteristic of crepant resolutions}

Using $p$-adic integration and the Weil conjecture, Batyrev proved the following theorem.

\begin{thm}[Batyrev, \cite{Batyrev.Betti}]
\label{thm:Batyrev}
Let $X$ and $Y$ be irreducible birational smooth $n$-dimensional projective algebraic varieties 
over $\mathbb{C}$. Assume that there exists a birational rational map $\varphi: X   - \rightarrow Y$ that does not 
change the canonical class. Then $X$ and $Y$ have the same Betti numbers. 
\end{thm}
 Batyrev's result was strongly inspired by string dualities, in particular by the work of Dixon, Harvey, Vafa, and Witten \cite{Dixon:1986jc}. 
As a direct consequence of Batyrev's theorem, the Euler characteristic of a crepant resolution of a variety with Gorenstein canonical singularities is independent on the choice of resolution. 
We identify the Euler characteristic as the degree  of the total  (homological) Chern class of a crepant resolution $f: \widetilde{Y }\longrightarrow Y$ of a Weierstrass model $Y\longrightarrow B$:
$$
\chi(\widetilde{Y})=\int c(\widetilde{Y}).
$$
We then use the birational invariance of the degree under the pushfoward to express the Euler characteristic as a class in the Chow ring of the projective bundle $X_0$. We subsequently push this class forward to the base to obtain a rational function depending upon only the total Chern class of the base $c(B)$, the first Chern class $c_1(\mathscr L)$, and the class $S$ of the divisor in $B$:
$$
\chi(\widetilde{Y})=\int_B \pi_* f_* c(\widetilde{Y}).
$$
In view of Theorem \ref{thm:Batyrev}, this Euler characteristic is independent of the choice of a crepant resolution. 

\subsection{Hodge numbers for Calabi-Yau elliptic threefolds}
 Using motivitic integration, Kontsevich shows in his famous ``String Cohomology'' Lecture at Orsay that birational equivalent Calabi-Yau varieties have the same class in the completed Grothendieck ring \cite{Kontsevich.Orsay}. 
Hence, birational equivalent Calabi-Yau varieties have the same  Hodge-Deligne polynomial, Hodge numbers, and Euler characteristic. 
 In this section, we compute the Hodge numbers of crepant resolutions of Weierstrass models in the case of  Calabi-Yau threefolds.

\begin{thm}[Kontsevich, (see \cite{Kontsevich.Orsay})]
Let $X$ and $Y$ be birational equivalent Calabi-Yau varieties over the complex numbers. Then $X$ and $Y$ have the same Hodge numbers. 
\end{thm}
\begin{rem}
In Kontsevich's theorem, a Calabi-Yau variety is a nonsingular complete projective variety of dimension $d$ with a trivial canonical divisor. 
To compute Hodge numbers in this section, we use the following stronger definition of a Calabi-Yau variety.
\end{rem}

\begin{defn}\label{defn:CY}
A \emph{Calabi-Yau variety} is a smooth compact projective variety $Y$ of dimension $n$ with a trivial canonical class and such that  $H^i(Y,\mathscr{O}_X)=0$ for $1\leq i\leq n-1$.
\end{defn}

 We first recall some basic definitions and relevant classical theorems.  
\begin{thm}[Noether's formula]
 If $B$ is a smooth compact, connected, complex surface with canonical class $K_B$ and Euler number  $c_2$, then
$$
\chi(\mathscr{O}_B)=1-h^{0,1}(B)+h^{0,2}(B),\quad    \chi(\mathscr{O}_B)=\frac{1}{12}(K^2+c_2).
$$
\end{thm}
When $B$ is a smooth compact rational surface, we have a simple  expression of   $h^{1,1}(B)$ as a function of $K^2$ using the following lemma. 
\begin{lem}\label{lem:NoetherRational}
Let $B$ be a smooth compact rational surface with  canonical class $K$. Then 
\begin{equation}
h^{1,1}(B)=10-K^2.
\end{equation}
\end{lem}
\begin{proof}
Since $B$ is a rational surface, $h^{0,1}(B)=h^{0,2}(B)=0$. Hence $c_2=2+h^{1,1}(B)$ and the  lemma follows from  Noether's formula. 
\end{proof}

We now compute  $h^{1,1}(Y)$ using the Shioda--Tate--Wazir theorem  \cite[Corollary 4.1]{Wazir}. 
\begin{thm}\label{Thm:STW2} Let $Y$ be a smooth Calabi-Yau threefold  elliptically fibered over a smooth variety $B$ with Mordell-Weil group of rank zero. Then,
\begin{equation}\nonumber
h^{1,1}(Y)=h^{1,1}(B)+f+1, \quad h^{2,1}(Y)=h^{1,1}(Y)-\frac{1}{2}\chi(Y),
\end{equation}
where $f$ is the number of geometrically irreducible fibral divisors not touching the zero section. In particular, if $Y$ is a $G$-model with $G$ being a semi-simple group, then $f$ is the rank of $G$. 
\end{thm}

\subsection{Prepotential of a $5d$ ${\cal N}=1$ supergravity theory}\label{sec:5dsugra}
The compactification of M-theory on a Calabi-Yau threefold $Y$ yields a  five dimensional supergravity theory with eight supercharges coupled to  $h^{1,1}(Y)$  vector multiplets and $h^{2,1}(Y)+1$ neutral hypermultiplets \cite{Cadavid:1995bk}. 
The gravity multiplet also contains a gauge field called the graviphoton. The kinetic terms of the vector multiplets and the graviphoton,  together with the coefficients of the Chern Simons terms, are derived from the prepotential $\mathcal{F}(\phi)$, which is a real function of the scalar fields of the vector multiplets. 
After integrating out massive charged vector and matter fields, the prepotential receives  a one-loop quantum correction protected from additional quantum corrections by supersymmetry. The vector multiplets transform in the adjoint representation of the gauge group while the  hypermultiplets transform in  representation $\mathbf{R}=\bigoplus_i \mathbf{R}_i$ of the gauge group, where $\mathbf{R}_i$ are irreducible components of $\mathbf{R}$. 

The Coulomb branches of the theory correspond to the chambers of the hyperplane arrangement I($\mathfrak{g},\mathbf{R}$). 
By matching the crepant resolutions with the chambers of I($\mathfrak{g},\mathbf{R}$), we determine which resolutions correspond to which phases of the Coulomb branch. 
 The triple intersection numbers of the fibral divisors correspond to the coefficient of the Chern-Simons couplings of the five dimensional gauge theory and  
  can be compared with the  Intrilligator--Morrison--Seiberg (IMS) prepotential, which is the  one-loop quantum contribution to the prepotential of the five-dimensional gauge theory
. Since in field theory the Chern-Simons couplings are linear in the numbers $n_{\mathbf{R}_i}$ of hypermultiplets transforming in the  irreducible representation $\mathbf{R}_i$  such that $\mathbf{R}=\bigoplus_i \mathbf{R}_i$, computing the triple intersection numbers provides a way to determine the numbers  $n_{\mathbf{R}_i}$ from the topology of the elliptic fibration. 
We observe by direct computation in each chamber that the  numbers we find do not depend on the choice of the crepant resolution. 
This  idea of using the triple intersection numbers to determine the number of multiplets transforming in a given representation was used previously in \cite{ES} for SU($n)$-models  and most recently for F$_4$-models  in \cite{F4}. 
This technique has been advocated by Grimm and Hayashi in \cite{Grimm:2011fx}.

\begin{table}[htb]
\begin{center}
\begin{tabular}{|c|c|}
\hline
Multiplet & Fields \\
\hline 
Graviton &  $(g_{\mu\nu}, A_{\mu},\psi_\mu)$\\\hline
Vector  &  $(A_\mu,\varphi, \lambda)$\\\hline
Hyper &  $(q, \zeta)$  \\\hline
\end{tabular}
\end{center}
\caption{Supermultiplets for $\mathcal{N}=1$ five dimensional supergravity.
The indices $\mu$ and $\nu$ refer to the five dimensional spacetime coordinates. 
The tensor $g_{\mu\nu}$ is the metric of the five dimensional spacetime.  The fields $\psi_\mu,\lambda,\zeta$  are symplectic Majorana spinors.
The field $\psi_\mu$ is the gravitino and $A_\mu$ is the graviphoton.   
The hyperscalar $q$ is a quaternion composed of four real scalar fields. 
   \label{Table:5DMatter}}
\end{table}

The Intrilligator--Morrison--Seiberg (IMS)  prepotential is the one-loop quantum contribution to the prepotential of a five-dimensional gauge theory with the matter fields in the representations $\mathbf{R}_i$ of the gauge group. Let   $\phi$ denote an element of the  Cartan subalgebra of the Lie algebra $\mathfrak{g}$, $\alpha_\ell$ the fundamental roots, $\varpi$ a weight, and $\langle \varpi,\phi \rangle$ the evaluation of a  weight $\varpi$ on an element $\phi$ of the Cartan subalgebra. 
 The Intrilligator--Morrison--Seiberg (IMS) prepotential is \cite{IMS}
\begin{align}
6\mathscr{F}_{\text{IMS}} (\phi)=&\frac{1}{2} \left(
\sum_{\ell} |\langle \alpha_\ell, \phi \rangle|^3-\sum_{i} \sum_{\varpi\in \mathbf{R}_i} n_{\mathbf{R}_i} |\langle \varpi, \phi\rangle|^3 
\right).\nonumber
\end{align}
The full prepotential also contains a contribution proportional to the third Casimir invariant of the Lie algebra $\mathfrak{g}$; for simple groups, it is only nonzero for SU($N$) groups with $N\geq 3$. 

 For a given choice of a Lie algebra $\mathfrak{g}$, choosing a  dual  fundamental  Weyl chamber resolves the  absolute values in the sum over the roots. 
 We then consider the arrangement of hyperplanes $\langle \varpi, \phi\rangle=0$, where $\varpi$ runs through all the weights of all the representations $\mathbf{R}_i$. 
 They define the hyperplane arrangement I($\mathfrak{g},\mathbf{R}=\bigoplus_i \mathbf{R}_i)$ restricted to the dual fundamental Weyl chamber. 
 If none of these hyperplanes intersect the interior of the dual fundamental Weyl chamber, we can safely remove the absolute values in the sum over the weights. 
 Otherwise, we have hyperplanes partitioning the fundamental Weyl chamber into subchambers. Each of these subchambers is defined by the signs of the linear forms $\langle \varpi, \phi\rangle$. 
 Two such subchambers are adjacent when they differ by the sign of a unique linear form.

Each of the subchambers is called a {\em Coulomb phase} of the gauge theory. 
 The transition from one chamber to an adjacent chamber is a {\em phase transition} that geometrically corresponds to a flop between different crepant resolutions of the same singular Weierstrass model.  
   The number of chambers of such a hyperplane arrangement is physically the number of phases of the Coulomb branch of the gauge theory.

\subsection{Anomaly Cancellations in $6d$ ${\cal N}=(1,0)$ supergravity theories} \label{sec:anomaly}

The matter content of the six-dimensional ${\cal N}=(1,0)$ supergravity theory are given by \cite{Green:1984bx} and reviewed in Table \ref{Table:6d}. 
The scalar manifold of the tensor multiplets is the symmetric space $\text{SO}(1, n_T)/\text{SO}(n_T)$ where $n_T$ is the number of tensor multiplets. 
The scalar manifold of the hypermultiplet is a quaternionic-K\"ahler manifold of quaternionic dimension $n_H$, where $n_H$ is the number of hypermultiplets.

\begin{table}[htb]
\begin{center}
\begin{tabular}{r c l}
supergravity multiplets: & \hspace{2cm}& $(g_{\mu\nu}, B^-_{\mu\nu}, \psi_\mu{}^{A-})$\\
tensor multiplets: & & $(B_{\mu\nu}^+,\chi^{A+},\sigma)$\\
vector multiplets: & & $(A_\mu, \lambda^{A-})$\\
hypermultiplets: & & $(4\phi, \zeta^+)$ 
\end{tabular}\\
\end{center}
\caption{The indices $\mu,\nu=0,\ldots, 5$ label spacetime indices, $A=1,2$ labels the fundamental representation 
of the $R$-symmetry SU($2$), and $\pm$ denotes the chirality of Weyl spinors or the self-duality ($+$)  or anti-self-duality ($-$) of the field strength of antisymmetric two-forms. 
The gravitini $\psi_\mu{}^{A-}$, the tensorini $\chi^{A+}$, and gaugini $ \lambda^{A-}$ are symplectic Majorana Weyl spinors. 
The hyperino $\zeta^+$ is a Weyl spinor invariant under the $R$-symmetry group SU($2$)$_R$. \label{Table:6d}}
\end{table}

Consider a gauged six-dimensional $\mathcal{N}=(1,0)$ supergravity theory  with a semi-simple gauge group $G=\prod_a G_a$, $n_V^{(6)}$ vector multiplets, $n_T$ tensor multiplets, and $n_H$ hypermultiplets consisting of $n_H^0$ neutral hypermultiplets and $n_H^{ch}$ charged under a representation $\bigoplus_i\mathbf{R}_i$ of the gauge group with $\mathbf{R}_i=\bigotimes_{a} \mathbf{R}_{i,a}$, where $\mathbf{R}_{i,a}$ is an irreducible representation of the simple component  $G_a$ of the semi-simple group $G$. The vector multiplets belongs to the adjoint of the gauge group (hence $n_V=\dim G$). 
 CPT invariance requires the representation  $\bigoplus_i\mathbf{R}_i$ to be quaternionic.

By denoting the zero weights of a representation $\bf{R}_i$ as $\bf{R}^{(0)}_i$, the charged dimension of the hypermultiplets in representation $\bf{R}_i$ is given by $\dim{\bf{R}_i}-\dim{\bf{R}_{i}^{(0)}}$, as the hypermultiplets of zero weights are considered neutral. For a representation $\bf{R}_i$, $n_{\bf{R}_i}$ denotes the multiplicity of the representation $\bf{R}_i$. Then the number of charged hypermultiplets is simply \cite{GM1,GM2}
\begin{equation}
n_H^{ch}=\sum_{i} n_{\bf{R}_i} \left( \dim{\bf{R}_i}-\dim{\bf{R_{i}^{(0)}}} \right).
\end{equation}
The total number of hypermultiplets is the sum of the neutral hypermultiplets and the charged hypermultiplets. 
For a compactification on a Calabi-Yau threefold $Y$, the  number of  neutral hypermultiplets is  $n_H^0=h^{2,1}(Y)+1$ \cite{Cadavid:1995bk}. The number of each multiplet is
\begin{align}\label{eqn:N-Multiplets}
n_V^{(6)}&=\dim{G}, \quad n_T=h^{1,1}(B)-1=9-K^2 , \\
n_H&=n_H^0+n_H^{ch}=h^{2,1}(Y)+1+\sum_{i} n_{\bf{R}_i} \left( \dim{\bf{R}_i}-\dim{\bf{R_{i}^{(0)}}} \right),
\end{align}
where the (elliptically fibered) base $B$ is a rational surface.

The anomaly polynomial I$_8$ has a pure gravitational contribution of the form  $\tr R^4$ where $R$ is the Riemann tensor thought of as a $6\times 6$ real matrix of two-form values. 
To apply the Green-Schwarz mechanism, the coefficient of $\tr R^4$ in the anomaly polynomial is required to vanish \cite{RandjbarDaemi:1985wc}
\begin{equation}
n_H-n_V^{(6)}+29n_T-273=0.
\end{equation}
In order to define the remainder terms of the anomaly polynomial I$_8$, 
we define 
\begin{equation}
X^{(n)}_{a}=\tr_{\bf{adj}}F^n_a -\sum_{i}n_{\bf{R}_{i,a}}\tr_{\bf{R}_{i,a}}F^n_a, \quad 
Y_{ab}=\sum_{i,j} n_{\bf{R}_{i,a}, \bf{R}_{j,b}} \tr_{\bf{R}_{i,a}}F^2_a \tr_{\bf{R}_{j,b}}F^2_b,
\end{equation}
where $n_{\bf{R}_{i,a}, \bf{R}_{j,b}}$ is the number of hypermultiplets transforming in the representation $(\mathbf{R}_{i,a},\mathbf{R}_{j,b})$ of $G_a\times G_b$.
The trace identities for a representation $\mathbf{R}_{i,a}$ of a simple group $G_a$ are
\begin{equation}
\tr_{\bf{R}_{i,a}} F^2_a=A_{\bf{R}_{i,a}} \tr_{\bf{F}_a} F^2_a , \quad \tr_{\bf{R}_{i,a}} F^4_a=B_{\bf{R}_{i,a}} \tr_{\bf{F}_a} F^4_a+C_{\bf{R}_{i,a}} (\tr_{\bf{F}_a} F^2_a )^2
\end{equation}
with respect to a  reference representation $\bf{F}_a$ for each simple component $G_a$. To define the anomaly polynomial I$_8$, we introduce the following expressions \cite{Schwarz:1995zw}: 
\begin{align}
X^{(2)}_a&=\overbrace{\left(A_{a,\bf{adj}}-\sum_{i}n_{\bf{R}_{i,a}}A_{\bf{R}_{i,a}}\right)}^{A_a}   \tr_{\bf{F}_a} F^2_a, \\
X^{(4)}_a&=\overbrace{\left(B_{a,\bf{adj}}-\sum_{i}n_{\bf{R}_{i,a}}B_{\bf{R}_{i,a}}\right)}^{B_a}\tr_{\bf{F}_a}F^4_a
+\overbrace{\left(C_{a,\bf{adj}}-\sum_{i}n_{\bf{R}_{i,a}}C_{\bf{R}_{i,a}}\right)}^{C_a}(\tr_{\bf{F}_a}F^2_a)^2 , \\
Y_{ab}&=\sum_{i,j} n_{\bf{R}_{i,a},\bf{R}_{j,b}} A_{\bf{R}_{i,a}} A_{\bf{R}_{j,b}} \tr_{\bf{F}_a}F^2_a \tr_{\bf{F}_b}F^2_b.
\end{align}
For each simple component $G_a$, the anomaly polynomial  I$_8$ has a pure gauge contribution proportional to the quartic term $\tr F^4_a$ that is required to vanish in order to factorize I$_8$: 
\begin{equation}
B_{a,\bf{adj}}-\sum_{i}n_{\bf{R}_{i,a}}B_{\bf{R}_{i,a}}=0.
\end{equation}
When the coefficients of all quartic terms  ($\tr R^4$ and $\tr F^4_a$) vanish,  the remaining part of the anomaly polynomial I$_8$ is \cite{Sadov:1996zm}
\begin{equation}
I_8=\frac{K^2}{8} (\tr R^2)^2 +\frac{1}{6}\sum_{a} X^{(2)}_{a} \tr R^2-\frac{2}{3}\sum_{a} X^{(4)}_{a}+4\sum_{a<b}Y_{ab}.
\end{equation}
The anomalies are canceled by the Green-Schwarz mechanism when I$_8$ factorizes \cite{Green:1984bx,Sagnotti:1992qw,Schwarz:1995zw}.

There is a subtlety on the representations that are charged on more than a simple component of the group, as it affects not only  $Y_{ab}$ but also $X^{(2)}_a$ and $X^{(4)}_a$. Consider a representation $(\bf{R_1},\bf{R_2})$ for of a semi-simple group with two simple components $G=G_1\times G_2$, where $\bf{R_a}$ is a representation of $G_a$. Then this representation contributes to $n_{\bf{R_1}}$ $\dim{\bf{R_2}}$ times, and contributes to $n_{\bf{R_2}}$ $\dim{\bf{R_1}}$ times:
\begin{equation}
n_{\bf{R_1}}=\dim{\bf{R_2}} \ n_{\bf{R_1},\bf{R_2}}, \quad n_{\bf{R_2}}=\dim{\bf{R_1}} \ n_{\bf{R_1},\bf{R_2}}.
\end{equation}

If the coefficients of $\tr R^4$ and $\tr F_a^4$ vanish and $G=G_1\times G_2$,  the remainder of the anomaly polynomial is given by
\begin{equation}
I_8 =\frac{K^2}{8} (\tr R^2)^2 +\frac{1}{6} (X^{(2)}_{1} +X^{(2)}_{2}) \tr R^2-\frac{2}{3} (X^{(4)}_{1}+X^{(4)}_{2})+4Y_{12} .
\end{equation}
If I$_8$ factors as $\frac{1}{2}\Omega_{ij} X^{(4)}_i X^{(4)}_j$, then the anomaly is cancelled by adding the counter term 
$\Omega_{ij} B_i \wedge   X_j^{(4)}$ to the Lagrangian. 
The modification of the field strength $H^{(i)}$ of  the antisymmetric tensor $B^{(i)}$ are $H^{(i)}=dB^{(i)} +\omega^{(i)}$, where $\omega^{(i)}$ is a proper combination of Yang-Mills and gravitational Chern-Simons terms.

\section{SO($4$)-model}

The generic SO($4$)-model is defined as an $\text{I}_2^{\text{ns}}+\text{I}_2^{\text{ns}}$-model with a $\mathbb{Z}/ 2\mathbb{Z}$ Mordell--Weil group. 
We assume that the first and second I$_2^{\text{ns}}$ are the generic fibers of two smooth divisors $S=V(s)$ and $T=V(t)$ intersecting transversally. 
  Using the work of Miranda or Nakayama, we expect  a geometric fiber of type I$_4$ at the collision of $S$ and $T$. 
 All the other SO($4$)-model are then obtained as degenerations of this one.  
\subsection{Weierstrass equation, crepant resolutions and flops}
We recall that the generic Weierstrass model with a $\mathbb{Z}/ 2\mathbb{Z}$ Mordell--Weil group is 
\begin{equation}
y^2z=x^3+a_2x^2z+a_4xz^2,
\end{equation}
where $a_2$ and $a_4$ are respectively sections of  $\mathscr{L}^{\otimes 2}$ and 
of  $\mathscr{L}^{\otimes 4}$.
The discriminant locus of this elliptic fibration is $\Delta=16 a_4^2 (a_2^2-4 a_4)$. The  generic fiber  over $V(a_4)$ is of type I$_2^{\text{ns}}$ while the generic fiber over $V(a_2^2-4 a_4)$ is of  type I$_1$. 
An SO($4$)-model is obtained by requiring $a_4$ to factor into two irreducible components  
\begin{equation}\label{eq:a4}
a_4=st.
\end{equation} 
We assume that $S=V(s)$ and $T=V(t)$ are two smooth divisors intersecting transversally.  
Since $a_4$ is a section of $\mathscr{L}^{\otimes 4}$, the equation $a_4=st$ imposes the following linear constraint on the class of $S$ and $T$:
\begin{equation}
S+T=4L,
\end{equation}
where we denote the first Chern class of $\mathscr{L}$ as $L=c_1(\mathscr{L})$. 
The Weierstrass model of this SO($4$)-model is then  
\begin{equation}
y^2z=x^3+a_2x^2z+stxz^2.
\end{equation}
Its discriminant locus is 
\begin{equation}
\Delta=16 s^2 t^2 \left(a_2^2-4 s t\right),
\end{equation}
which consists of three components
\begin{equation}
 S=V(s), \quad T=V(t),\   \text{and}  \  \Delta'=a_2^2-4 s t. 
\end{equation}
When the base is of dimension three or higher, the divisor $\Delta'$ has double point singularities at $V(a_2,s,t)$. 
The divisors $S$ and $T$ do not intersect  $\Delta'$ transversally as the intersection is non-reduced.

We can easily deform the elliptic fibration to produce other fibers still giving an SO($4$)-model. 
\begin{align}
\text{I}_2^{\text{ns}}+\text{I}_2^{\text{s}}:&\quad y^2z+a_1 y x z=x^3+\widetilde{a}_2 tx^2z+s t xz^2\\
\text{I}_2^{\text{s}}+\text{I}_2^{\text{s}}:&\quad y^2z+a_1 y x z=x^3+ \widetilde{a}_2 s t x^2 z +s t xz^2\\
\text{III}+\text{I}_2^{\text{s}}:&\quad y^2z=x^3+ \widetilde{a}_2 s  x^2 z +s t xz^2\\
\text{III}+\text{I}_2^{\text{s}}:&\quad y^2z+\widetilde{a}_1s x y z=x^3+\widetilde{a}_2 s  x^2 z +s t xz^2\\
\text{III}+\text{III}:&\quad y^2z=x^3+ \widetilde{a}_2 s t  x^2 z +s t xz^2.
\end{align}      
 We give a crepant resolution via following sequence of two blowups
\begin{equation}
  \begin{tikzcd}[column sep=huge] 
  X_0  \arrow[leftarrow]{r} {(x,y,s|e_1)} & \arrow[leftarrow]{r}{(x,y,t|w_1)} X_1 &X_2.
  \end{tikzcd}
\end{equation}
The other crepant resolution connected by a flop is obtained by exchanging $S$ and $T$. For that reason, we work with the current resolution as the other one is described by a simple exchange of $S$ and $T$. 
The relative ``projective coordinates" of $X_2$ are  then given by
\begin{equation}
[e_1w_1x : e_1w_1y : z=1][w_1x : w_1y : s][x : y : t].
\end{equation}
The proper transform of the original Weierstrass model is 
\begin{equation}
y^2z=e_1w_1x^3+a_2x^2z+stxz^2 .
\end{equation}
By the Jacobian criterion, assuming that $S$ and $T$ are transverse is enough to prove that this proper transform is smooth since $(x,y,st)$ is an empty ideal after the two blowups.

In the SO($4$)-model, the Atiyah flop is easily seen by considering a partial resolution that is a common blowdown for both resolutions. 
\begin{equation}
  \begin{tikzcd}[column sep=huge] 
& X_0  \arrow[rightarrow]{d} {(x,y|e_1)}   & \\
  & X_{1} \arrow[rightarrow]{dl} {(e_1,s|e_2)} \arrow[rightarrow]{dr} {(e_1,t|w_2)}  & \\
  X_2^{-} \arrow[leftrightarrow,dashed]{rr}{} & &   X_2^{+}
  \end{tikzcd}
\end{equation}
After the first blow-up, we get the partial resolution 
\begin{equation}
\mathscr{E}_{1} :\quad  e_1(y^2-e_1x^3-a_2x^2)=stx.
\end{equation}
Since $(x,y)$ cannot vanish at the same time, the previous equation is of the type $u_1 u_2 -u_3 u_4=0$, which is exactly the equation of the singularity whose resolutions determine the  Atiyah's flop.
The two crepant resolutions are then obtained by blowing up $(e_1,s)$ or $(e_1,t)$.

\subsection{Fiber structure}
We can now study the fiber structure of this elliptic fibration.  We assume that the base is of arbitrary dimension. 
We have four fibral divisors: 
\begin{equation}
\begin{cases}
 D_0^s &:\quad s=y^2-e_1w_1x^3-a_2x^2=0,  \\
 D_1^s &:\quad e_1=y^2-a_2x^2-stx=0, \\
 D_0^t &:\quad t=y^2-e_1w_1x^3-a_2x^2=0, \\
 D_1^t &:\quad w_1=y^2-a_2x^2-stx=0 .
 \end{cases}
\end{equation}
The divisors $D_0^s$ and $D_1^s$  (resp. $D_0^t$ and $D_1^t$) project to $S$ (resp. $T$). All these divisors are conic bundles. 
  Over the generic point of $S$, we have two fibers $C_0^s$ and $C_1^s$ intersecting at 
\begin{equation}
C_0^s \cap C_1^s: s=e_1=y^2-a_2x^2=0.
\end{equation}
The intersection points are the two roots of $y^2-a_2 x^2=0$. 
Thus, this is does represent an I$_2^{ns}$ since computing the roots requires taking the square root of $a_2$,
  the two points coincide when $a_2=0$ yielding a fiber of type III. 
The same is true over $T$, where the generic fiber I$_2^{\text{ns}}$ degenerates to a fiber III over $T\cap V(a_2)$.
 At the intersection of $S$ and $T$,  we get 
\begin{align}
\begin{cases}
D_0^s\cap D_0^t &: s=t=y^2-e_1w_1x^3-a_2x^2=0 \rightarrow \eta^{00} , \\
D_1^s\cap D_0^t &: e_1=t=y^2-a_2x^2=0 \rightarrow \eta^{10\pm} , \\
D_1^s\cap D_1^t &: e_1=w_1=y^2-a_2x^2-stx=0 \rightarrow \eta^{11},
\end{cases}
\end{align} 
where the intersections of the new fibers are given by
\begin{align}
\eta^{00} \cap \eta^{10\pm}&: s=e_1=t=y^2-a_2x^2=0,\\
\eta^{10\pm} \cap \eta^{11}&: e_1=t=w_1=y^2-a_2x^2=0.
\end{align}
This gives a fiber of type  I$_4^{\text{ns}}$ (see Figure  \ref{Fig:SO4}). 
 Over  $V(S,T,a_2)$, the fiber I$_4^{\text{ns}}$ degenerates further to a non-Kodaira fiber of type 
$1-2-1$ as the  two rational curves  $\eta^{10\pm }$ coincide. There are no other degenerations.

At the collision $S\cap T$, the splitting of the curves $C_0^s$, $C_1^s$, $C_0^t$, and $C_1^t$ are given by 
\begin{align}
\begin{cases}
&C_0^s \rightarrow \eta^{00} , \\
& C_1^s \rightarrow \eta^{10+}+\eta^{10-}+\eta^{11} , \\
&C_0^t \rightarrow \eta^{00}+\eta^{10+}+\eta^{10-} , \\
&  C_1^t \rightarrow \eta^{11} ,
\end{cases} 
\end{align}
This induces linear relations that we exploit to compute the intersection numbers between the curves of the collision and the  fibral divisors
\begin{equation}
\begin{array}{c|cccc}
 & D_0^s & D_1^s & D_0^t & D_1^t \\
 \hline
\eta^{00} & -2 & 2 & 0 & 0 \\
\eta^{10+}+\eta^{10-} & 2 & -2 & -2 & 2 \\
\eta^{10\pm} & 1 & -1 & -1 & 1 \\
\eta^{11} & 0 & 0 & 2 & -2  .
\end{array} 
\end{equation}
Hence, we see that we get the weight $[1;-1]$ from the curves $\eta^{10\pm}$. Since both $[1]$ and $[-1]$ are from the representation $\bf{2}$ of $\frak{su}(2)$, we can deduce that the matter content we get is a bifundamental, $(\bf{2},\bf{2})$, corresponding to the vector representation of SO($4$).
Hence the matter contents of the I$_2^{ns}+$I$_2^{ns}$-model with a $\mathbb{Z}/ 2\mathbb{Z}$ Mordell--Weil group is given by
\begin{equation}
\bf{R}=(\bf{3},\bf{1})\oplus(\bf{1},\bf{3})\oplus(\bf{2},\bf{2}).
\end{equation}

\subsection{Euler characteristic} \label{pf:so4.euler}

Now we compute the Euler characteristic  using the blowup maps for getting this crepant resolution. 
 The following theorems are direct specializations of  Theorem \ref{Thm.Spin4.Euler} and  \ref{Thm:HodgeNumbersSpin4} for  Spin($4$)-models after imposing the condition $S+T=4L$. 
  \Pushso*

\Pushsohodge*

\subsection{Triple Intersection and the Prepotential}\label{sec:TripSO4}
In this subsection, we compute the triple intersection polynomial of the SO($4$)-model. 
\begin{align}
\begin{split}
\mathscr{F}_{trip}^+=&
\pi_* f_* \Big(\psi_0 D_0^s+\psi_1 D_1^s+\phi_0 D_0^t +\phi_1 D_1^t\Big)^3\\ 
=&-2 T \phi _0^3 (2 L+T)-2 (4 L-T) \left(\psi _0-\psi _1\right)^2 \left(\psi _0 (6 L-2 T)+\psi _1 (6 L-T)\right)\\
&+6 T \phi _0 \left(-\left(\psi _0-\psi _1\right)^2 (4 L-T)+2 \psi _1 \phi _1 (4 L-T)+\phi _1^2 (T-2 L)\right)
+6 T \phi _0^2 \left(\psi _1 (T-4 L)+2 L \phi _1\right)
\\
&-2  \left(24 L^2-10 L T+T^2\right) \psi_1^3+6 T (T-4 L) \psi_1 \phi_1^2+4 T (L-T)  \phi_1^3.
\end{split}
\end{align}
In the flop, $\mathscr{F}_{trip}^-$ is obtained by the involution  $\psi\leftrightarrow \phi$.
Using the Intrilligator-Morrison-Seiberg approach, we compute the prepotential of the five-dimensional theory to be
\begin{equation}
6\mathscr{F}_{\text{IMS}}=-8 (n_{\bf{1},\bf{3}}-1)\left| \phi _1\right|^3-n_{\bf{2},\bf{2}}\Big(\left| \phi _1-\psi _1\right|^3-\left| \phi _1+\psi _1\right|^3\Big)-8 (n_{\bf{3},\bf{1}}-1)\left| \psi _1\right|^3 .
\end{equation}
In the chamber $\phi_1-\psi_1>0$, the prepotential is  
\begin{equation}
6\mathscr{F}^+_{\text{IMS}}=-8 (n_{\bf{1,3}}-1) \phi _1^3-2 (n_{\bf{2,2}}+4n_{\bf{3,1}}-4) \psi _1^3-6n_{\bf{2,2}} \phi _1^2  \psi _1.
\label{IMS.I2nsI2ns}
\end{equation}
In the chamber $\phi_1-\psi_1<0$, the prepotential is  
\begin{equation}
6\mathscr{F}^-_{\text{IMS}}=-8 (n_{\bf{1,3}}-1) \psi_1^3-2 (n_{\bf{2,2}}+4n_{\bf{3,1}}-4) \phi_1^3-6n_{\bf{2,2}} \psi _1^2  \phi_1.
\label{IMS.SO4.m}
\end{equation}
\subsection{Counting charged hypermultiplets in $5d$}
Matching the $5d$ prepotential with the triple intersection numbers, we can determine the number of representations. We find a perfect match between the triple intersection number and the prepotential when
\begin{align}
n_{\bf{2,2}}&=-T (-4 L + T), \quad n_{\bf{1,3}}=\frac{1}{2} \left(-L T+T^2+2\right) ,\quad 
n_{\bf{3,1}}=\frac{1}{2} \left(12 L^2-7 L T+T^2+2\right)\\
\intertext{Using the relation $T+S=4L$, we can rewrite these numbers as }
n_{\bf{2,2}}&=T S, \quad n_{\bf{1,3}}=g_T ,\quad 
n_{\bf{3,1}}=g_S,
\label{nR.I2nsI2ns}
\end{align}
where $g_T$ and $g_S$ are respectively the genus of $T$ and $S$. 
\subsection{Anomaly cancellations}
The number of vector multiplets $n_V^{(6)}$, tensor multiplets $n_T$, and hypermultiplets $n_H$ are  (see equations \eqref{eqn:N-Multiplets}):
\begin{align}
\begin{split}
n_V^{(6)}&=6, \quad n_T=9-K^2, \\
n_H&=h^{2,1}(Y)+1+n_{\bf{3},\bf{1}}(3-1)+n_{\bf{2},\bf{2}} (4-0)+n_{\bf{1},\bf{3}}(3-1)=18 + 29 K^2.
\end{split}
\end{align}
Thus, we can conclude that the purely gravitational anomaly is canceled:
\begin{equation}
n_H-n_V^{(6)}+29n_T-273=0.
\end{equation}

Now consider the remaining parts of the anomaly polynomial
\begin{align}
I_8&=\frac{9-n_T}{8} (\tr R^2)^2 +\frac{1}{6} (X^{(2)}_{1} +X^{(2)}_{2}) \tr R^2-\frac{2}{3} (X^{(4)}_{1}+X^{(4)}_{2})+4Y_{12} ,
\end{align}
where 
\begin{align}
X^{(2)}_{1}&=\left(A_{\bf{3}}(1-n_{\bf{3}})-n_{\bf{2}}A_{\bf{2}}\right)\tr_{\bf{2}}F^2_1 =-6 K (4 K + T)\tr_{\bf{2}}F^2_1\\
\begin{split}
X^{(4)}_{1}&=\left(B_{\bf{3}}(1-n_{\bf{3}})-n_{\bf{2}}B_{\bf{2}}\right)\tr_{\bf{2}}F^4_1+\left(C_{\bf{3}}(1-n_{\bf{3}})-n_{\bf{2}}C_{\bf{2}}\right) (\tr_{\bf{2}}F^2_1)^2 \\
&=-3 (4 K + T)^2  (\tr_{\bf{2}}F^2_1)^2
\end{split} \\
X^{(2)}_{2}&=\left(A_{\bf{3}}(1-n_{\bf{3}})-n_{\bf{2}}A_{\bf{2}}\right)\tr_{\bf{2}}F^2_1 =6 K T\tr_{\bf{2}}F^2_1\\
\begin{split}
X^{(4)}_{2}&=\left(B_{\bf{3}}(1-n_{\bf{3}})-n_{\bf{2}}B_{\bf{2}}\right)\tr_{\bf{2}}F^4_1+\left(C_{\bf{3}}(1-n_{\bf{3}})-n_{\bf{2}}C_{\bf{2}}\right) (\tr_{\bf{2}}F^2_1)^2 \\
&=-3 T^2 (\tr_{\bf{2}}F^2_1)^2
\end{split} \\
Y_{12}&=(n_{\bf{2},\bf{4}}+n_{\bf{2},\bf{\bar{4}}}) \tr_{\bf{2}}F^2_1 \tr_{\bf{4}}F^2_2=-T (4 K + T)\tr_{\bf{2}}F^2_1 \tr_{\bf{4}}F^2_2 .
\end{align}
Hence the anomaly polynomial becomes simply
\begin{equation}
I_8=\frac{1}{2} \left(\frac{1}{2}K \tr R^2+2S \tr_{\bf{2}}F^2_1+2 T\tr_{\bf{2}}F^2_2\right)^2,
\label{I8Factor.SO4}
\end{equation}
where $S+T=-4K$. Since the anomaly polynomial is a perfect square, we can conclude that the anomalies are all canceled.

\section{Spin($4$)-model }
 A Weierstrass equation for Spin($4$)  is given by the collision of type I$_2^{ns}+$I$_2^{ns}$  in an elliptic fibration with a trivial Mordell--Weil group. 
\begin{equation}
Y_0: \quad y^2z=x^3+{a}_2x^2z+\widetilde{a}_4stxz^2+\widetilde{a}_6s^2t^2z^3.
\end{equation}
This is a good starting point as the other possibilities  (see Table \ref{Table:Weierstrass}) can be described as degenerations of it. 
The discriminant locus of this model is
\begin{equation}
\Delta=-s^2 t^2 \left(4 a_2^3 \widetilde{a}_6-a_2^2 \widetilde{a}_4^2-18 a_2 \widetilde{a}_4 \widetilde{a}_6 s t+4 \widetilde{a}_4^3 s t+27 \widetilde{a}_6^2 s^2 t^2\right).
\end{equation}

\begin{thm}\label{Thm:ResSpin4}
Assuming that $S=V(s)$ and $T=V(t)$ intersect transversally, the  Weierstrass model 
$$
Y_0:\quad y^2z=x^3+{a}_2x^2z+\widetilde{a}_4stxz^2+\widetilde{a}_6s^2t^2z^3.
$$
has two distinct crepant resolutions $Y^\pm$. One is  given by the sequence of blowups 
\begin{equation}\label{Eq:Blowups11}
  \begin{tikzcd}[column sep=huge] 
X_0  \arrow[leftarrow]{r} {(x,y,s|e_1)} & \arrow[leftarrow]{r}{(x,y,t|w_1)} X_1^+ &X_2^+.
  \end{tikzcd}
\end{equation}
The other crepant resolution is obtained by exchanging the order of the blowup. 
 \begin{equation}\label{Eq:Blowups12}
  \begin{tikzcd}[column sep=huge] 
 X_0  \arrow[leftarrow]{r} {(x,y,t|w_1)} & \arrow[leftarrow]{r}{(x,y,s|e_1)} X_1^- &X_2^-.
  \end{tikzcd}\end{equation}
The proper transforms of  the  Weierstrass model $Y_0$ is then  a smooth elliptic fibration $Y^+$ or $Y^-$ are crepant resolutions of $Y_0$ connected by an Atiyah flop.
\end{thm}
The proper transforms are 
\begin{equation}
Y^\pm:\quad y^2z=e_1w_1x^3+a_2x^2z+\widetilde{a}_4stxz^2+\widetilde{a}_6 s^2 t^2 z^3 .
\end{equation}
The relative projective coordinates for $X_2^\pm$ are  then given by
\begin{equation}
X_2^+\quad [e_1w_1x : e_1w_1y : z=1][w_1x : w_1y : s][x : y : t].
\end{equation}
and for  the second sequence of blowups:
\begin{equation}
X_2^-\quad [e_1w_1x : e_1w_1y : z=1][e_1x : e_1y : t][x : y : s].
\end{equation}
The same blowups are also giving crepant resolutions connected by an Atiyah flop.

\subsection{Fiber structure}
We have four fibral divisors where two  ($D_0^s$ and  $D_1^s$) are over $S$ and the other two ($D_0^t$ and $D_1^t$) are over $T$:
\begin{align}
 D_0^s &: s=y^2-e_1w_1x^3-a_2x^2=0 , \\
 D_1^s &: e_1=y^2-a_2x^2-\widetilde{a}_4stx-\widetilde{a}_6s^2t^2=0 \\
 D_0^t &: t=y^2-e_1w_1x^3-a_2x^2=0 , \\
 D_1^t &: w_1=y^2-a_2x^2-\widetilde{a}_4stx-\widetilde{a}_6s^2t^2=0 .
\end{align}
First we can understand the fiber structure of this chamber.  Over $S$, we have two fibers $C_0^s$ and $C_1^s$, where they intersect at two points given by
\begin{equation}
C_0^s \cap C_1^s: s=e_1=y^2-a_2x^2=0.
\end{equation}
This is the same intersection from the I$_2^{ns}+$I$_2^{ns}$-model with $\mathbb{Z}/ 2\mathbb{Z}$. This enhances to a type III fiber when $a_2=0$ as well. 
 This has another enhancement over $S$ when $\widetilde{a}_4^2-4a_2\widetilde{a}_6=0$, because the conic curve $D_1^s$ splits into two curves
\begin{equation}
C_1^s \rightarrow C_{1+}^s +C_{1-}^s ,
\end{equation}
which yields the fiber I$_3^{ns}$. The intersection numbers between the new curves and the fibral divisors are given by
\begin{equation}
\begin{tabular}{c|cccc}
 & $D_0^s$ & $D_1^s$ & $D_0^t$ & $D_1^t$ \\
 \hline
$C_0^s$ & $-2$ & $2$ & 0 & 0 \\
$C_{1+}^s$ & $1$ & $-1$ & 0 & 0 \\
$C_{1-}^s$ & $1$ & $-1$ & 0 & 0 
\end{tabular}
\quad \raisebox{-25pt}{.}
\end{equation}
From the intersection numbers, we get the weight $[1;0]$ from both curves $C_{1\pm}^s$. Since $[1]$ corresponds to the representation $\bf{2}$ of $\frak{su}(2)$ and it is only charged over $S$, we get the matter content in the representation $(\bf{2},\bf{1})$.

Over $T$, the same enhancements I$_2 \rightarrow$ I$_3$ and I$_2 \rightarrow$ III exist as the two divisors are the same when $s \leftrightarrow t$ and $e_1 \leftrightarrow w_1$. For the former enhancement, the intersection numbers between the new curves via the splitting
\begin{equation}
C_1^t \rightarrow C_{1+}^t +C_{1-}^t ,
\end{equation}
and the fibral divisors are given by
\begin{equation}
\begin{tabular}{c|cccc}
 & $D_0^s$ & $D_1^s$ & $D_0^t$ & $D_1^t$ \\
 \hline
$C_0^t$ & $0$ & $0$ & $-2$ & $2$ \\
$C_{1+}^t$ & $0$ & $0$ & $1$ & $-1$ \\
$C_{1-}^t$ & $0$ & $0$ & $1$ & $-1$
\end{tabular}
\quad \raisebox{-25pt}{.}
\end{equation}
From these intersection numbers of the curves $C_{1\pm}^t$, we get the weight $[0;1]$. Since $[1]$ corresponds to the representation $\bf{2}$ of $\frak{su}(2)$ and it is only charged over $T$, we get the matter content in the representation $(\bf{1},\bf{2})$.

Over both $S$ and $T$, we get four following curves:
\begin{align}
\begin{cases}
D_0^s \cap D_0^t &: s=t=y^2-e_1w_1x^3-a_2x^2=0\rightarrow \eta^{00} , \\
D_1^s \cap D_0^t &: e_1=t=y^2-a_2x^2=0\rightarrow \eta^{10}, \\
D_1^s \cap D_1^t &: e_1=w_1=y^2-a_2x^2-\widetilde{a}_4stx-\widetilde{a}_6s^2t^2=0 \rightarrow \eta^{11}.
\end{cases}
\end{align}
The curve $\eta^{10}$ is not geometrically irreducible and consists of two geometrically irreducible curves $\eta^{10\pm}$ that requires taking the square-root  of $a_2$. 
Clearly, if in the Weierstrass equation, $a_2$ was a perfect square modulo $t$ or $s$, the fiber $\eta^{10}$ will factorize into $\eta^{10\pm}$ without the need of a field extension. 
The intersections of the new curves are 
\begin{align}
\eta^{00} \cap \eta^{10\pm}&: s=e_1=t=y^2-a_2x^2=0,\\
\eta^{10\pm} \cap \eta^{11}&: e_1=t=w_1=y^2-a_2x^2=0.
\end{align}
This gives a fiber of type  I$_4^{\text{ns}}$. When $a_2=0$, this specializes into a new fiber, as represented in Figure \ref{Fig:Spin4nsns}, where the nonsplit curves $\eta^{10\pm}$ become a single curve $\eta^{10y}$ with degeneracy two, where
\begin{equation}
\eta^{10y}: e_1=t=y=0.
\end{equation}
From these splittings of the curve,
\begin{align}
\begin{cases}
&C_0^s \rightarrow \eta^{00} , \\
& C_1^s \rightarrow \eta^{10+}+\eta^{10-}+\eta^{11}, \\
&C_0^t \rightarrow \eta^{00}+\eta^{10+}+\eta^{10-} , \\
& C_1^t \rightarrow \eta^{11},
\end{cases}
\end{align}
we compute the intersection numbers between the curves on the collision with the divisors of two I$_2^{ns}$ fibers to be
\begin{equation}
\begin{tabular}{c|cccc}
 & $D_0^s$ & $D_1^s$ & $D_0^t$ & $D_1^t$ \\
 \hline
$\eta^{00}$ & $-2$ & $2$ & $0$ & $0$ \\
$\eta^{10+}+\eta^{10-}$ & $2$ & $-2$ & $-2$ & $2$ \\
$\eta^{10\pm}$ & $1$ & $-1$ & $-1$ & $1$ \\
$\eta^{11}$ & $0$ & $0$ & $2$ & $-2$
\end{tabular}
\end{equation}
Hence we see that we get the weight $[1;-1]$ from the curves $\eta^{10\pm}$. Since both $[1]$ and $[-1]$ is from the representation $\bf{2}$ of $\frak{su}(2)$, we can deduce that the matter content we get is a bifundamental, $(\bf{2},\bf{2})$.

Hence, the matter content of the I$_2^{ns}+$I$_2^{ns}$-model with a trivial Mordell--Weil group is
\begin{equation}
\bf{R}=(\bf{3},\bf{1})\oplus(\bf{1},\bf{3})\oplus(\bf{2},\bf{1})\oplus(\bf{2},\bf{2})\oplus(\bf{1},\bf{2}).
\end{equation}

\subsection{Euler characteristic and Hodge numbers} \label{pf:spin4.euler}
We compute the Euler characteristic using the blowup maps  and the pushforward theorems of \cite{Euler}. 
Let $S=V(s)$, $T=V(t)$ be the two smooth divisors that support the fibers with dual graph $\widetilde{\text{A}}_1$.  We assume that $S$ and $T$ intersect transversally. Let  us recall that $L=c_1(\mathscr{L})$ is  the first Chern class of the fundamental line bundle of the Weierstrass model.

\Pushspin*
\begin{proof}
The  Euler characteristic of  a variety $Y$ is the degree of its homological total Chern class. 
The homological  Chern class of a Weierstrass model is 
$$
\frac{(1+H)(1+H+2\pi^*L)(1+H+3\pi^*L)}{1+3H+6\pi^*L} \pi^* c(TB).
$$
The ambient space for the Weierstrass model is the projective bundle is $\pi: X_0\to B$. 
We denote by $f_1:X_1^+\to X_0$ the first blowup with center $(x,y,s)$ and by $f_2:X^+_2\to X_1^+$ the second blowup with center $(x,y,t)$.
The center of the first and second blowups have centered of class:
$$
\begin{array}{lll}
Z^{(1)}_1= H+2\pi^*L, &\quad Z^{(1)}_2=H+3\pi^* L, & \quad Z^{(1)}_3= S\\
Z^{(2)}_1=f_1^* H+2f^*_1\pi^*L-E_1,& \quad Z^{(2)}_2=f^*_1H+3f^*_1\pi^*L-E_1,& \quad Z^{(2)}_3= T
\end{array}
$$

The total Chern class of $X_0$, $X_1$, and $X_2$ are 
$$
\begin{aligned}
c(TX_0) &=(1+H)(1+H+2\pi^*L)(1+H+3\pi^*L) \pi^* c(TB), \\
c(TX_1^+) &=(1+E_1)\frac{(1+Z^{(1)}_1)(1+Z^{(1)}_2)(1+Z^{(1)}_3)}{(1+Z^{(1)}_1)(1+Z^{(1)}_2)(1+Z^{(1)}_3)}f_1^*c(TX_0), \\
c(TX_2^+) &=(1+W_1)\frac{(1+Z^{(2)}_1)(1+Z^{(2)}_2)(1+Z^{(2)}_3)}{(1+Z^{(2)}_1)(1+Z^{(2)}_2)(1+Z^{(2)}_3)}f_2^*c(TX_1^+).
\end{aligned}
$$
After each blowup, we can pull out two powers of the exceptional divisor. It follows that the defining equation is a section of a line bundle whose first Chern class is 
$$
3f_1^*H+6f_1^*\pi^*L-2E_1.$$ 
After the second blowup, the proper transform of the defining equation is a section of a line bundle whose first Chern class is 
$$3f_2^* f_1^*H+6f_2^* f_1^*\pi^*L-2f_2^* E_1-2W_1$$ 
It follows that the total Chern class of the proper transform of the elliptic fibration is 
$$
c(TY_2^+)=\frac{c(TX_2^+)}{1+3f_2^* f_1^*H+6f_2^* f_1^*\pi^*L-2f_2^* E_1-2W_1}.
$$
The Euler characteristic is then 
$$
\chi(Y_2^+)=\int \frac{ (3f_2^* f_1^*H+6f_2^* f_1^*\pi^*L-2f_2^* E_1-2W_1)}{1+3f_2^* f_1^*H+6f_2^* f_1^*\pi^*L-2f_2^* E_1-2W_1} c(TX_1^+),
$$
where $\int A$ is the degree of $A$. 
Since the degree is invariant under a pushforward, we get 
$$
\chi(Y_2^+)=\int\pi_* f_{1*}f_{2*}\frac{ (3f_2^* f_1^*H+6f_2^* f_1^*\pi^*L-2f_2^* E_1-2W_1)}{1+3f_2^* f_1^*H+6f_2^* f_1^*\pi^*L-2f_2^* E_1-2W_1} c(TX_2^+).
$$
The pushforwards $f_{1*}$ and $f_{2*}$ are computed using Theorem \ref{Thm:Push}. 
The final pushforward $\pi_*$ is computed using Theorem \ref{Thm:PushH}. 
 These three pushforwards are purely algebraic computations.

In the Calabi--Yau case, we have $L=-K$. 
For a Calabi-Yau threefold, we just keep the terms of degree two in the base. 
\end{proof}
\Pushspinhodge*
\begin{proof}
For a Calabi-Yau threefold, we have $h^{1,0}=0$ and $\chi(Y)=2\Big(h^{1,1}(Y)-h^{2,1}(Y)\Big)$. 
Since the variety is elliptically fibered, we can compute $h^{1,1}(Y)$ by the  Shioda--Tate--Wazir theorem (see Theorem \ref{Thm:STW2}) and 
Noether's formula.  We then have 
$h^{1,1}(Y)=h^{1,1}(B)+f+1$ with $f=2$ and $h^{1,1}(B)=10-K^2$ (see Lemma \ref{lem:NoetherRational}). Finally,  $h^{2,1}(Y)=h^{1,1}(Y)-\chi(Y)/2$, where $\chi(Y)$ is given in Theorem \ref{Thm.Spin4.Euler}. 
\end{proof}

\subsection{Triple Intersection numbers}\label{Thm:TripleSO4}
In this subsection, we compute the triple intersection polynomial of the Spin($4$)-model. 
 In contrast to the Euler characteristic, the triple intersection polynomial does depend on the choice of a crepant resolution.  In particular, the two varieties $Y^\pm$ have distinct topologies as they have distinct triple intersection numbers. 

\PushTripleSpinplus*
\begin{proof}
The ambient space for the Weierstrass model is the projective bundle is $\pi: X_0\to B$. 
We denote by $f_1:X_1^+\to X_0$ the first blowup with center $(x,y,s)$ and by $f_2:X^+_2\to X_1^+$ the second blowup with center $(x,y,t)$.

The class of the fibral divisors $D_0^s$, $D_1^s$, $D_0^t$, $D_1^t$ are
$$
[D_0^s]=f_2^* f_1^* \pi^* S-f_2^*E_1, \quad [D_1^s]=f_2^* E_1, \quad [D_0^t]=f_2^* f_1^* \pi^*T-W_1, \quad [D_1^s]= W_1.
$$
$$
\begin{aligned}
\int_{Y} (f_2^*E_1 a + W_1 b + f_2^*f_1^*H b)^3
&=\int_{X_2} (f_2^* E_1 a + W_1 b + f_2^*f_1^*H c)^3(3f_2^*f_1^* H + 6 f_2^*f_1^*\pi^*L - 2f_2^* E_1 - 2 W_1)\\
&=
-2 S (S + 2 L) a^3 - 6 S T a b^2 + 2 (S - 2 L - T) T b^3 + 
 27 L^2 c^3
 \end{aligned}
$$
which gives the following non-vanishing triple intersection numbers 
$$
\int_Y  f_2^* E_1^3 =-2 S (S + 2 L), \quad \int_Y f_2^* E_1 W_1^2=-2 ST , \quad \int_Y W_1^3=2 T(S - 2 L - T) , \quad 
\int_Y f_2^* f_1^* H^3=27 L^2.
$$
The triple intersection numbers of the fibral divisors are then computed from these equations.
\end{proof}

\subsection{Counting charged hypermultiplets in $5d$}\label{Sec:CountingHypersSpin4}

 M-theory compactified on a Calabi-Yau threefold yields a five-dimensional theory with eight supercharges. 
This is reviewed in section \ref{sec:5dsugra}. 
The dynamics of the vector fields and the scalar fields of the vector multiplets depends on the prepotential $\mathscr{F}$, which gets a one-loop correction protected from additional corrections by supersymmetry. 
In the present case,  the Lie algebra is $\text{D}_2=\text{A}_1\oplus\text{A}_1$ and charged hypermultiplets transform in the representation $\mathbf{R}=(\bf{3},\bf{1})\oplus(\bf{1},\bf{3})\oplus (\bf{2},\bf{2})\oplus(\bf{2},\bf{1})\oplus (\bf{1},\bf{2})$  of D$_2$. 
In the Coulomb phase, all massive particles have been integrated out; in particular, the prepotential depends only on  fields in the Cartan sub algebra of the Lie algebra.  

 Let  $(\psi_1,\phi_1)$ be  a parametrization of the coroots written in the basis of simple coroots of the Lie algebra $\text{D}_2=\text{A}_1\oplus\text{A}_1$.  The prepotential of the five-dimensional supergravity theory with Lie algebra $\text{D}_2$ and matter in the representation 
$\mathbf{R}$ defined above  
 is \cite{IMS}
\begin{equation}
6\mathscr{F}_{\text{IMS}}=-(8n_{\bf{1},\bf{3}}+n_{\bf{1},\bf{2}}-8)\left| \phi _1\right|^3-n_{\bf{2},\bf{2}}\Big(\left| \phi _1-\psi _1\right|^3-\left| \phi _1+\psi _1\right|^3\Big)-(8n_{\bf{3},\bf{1}}+n_{\bf{2},\bf{1}}-8)\left| \psi _1\right|^3,
\end{equation}
where $n_{\mathbf{R_1},\mathbf{R}_2}$ is the number of hypermultiplets transforming in the irreducible representation $(\mathbf{R}_1,\mathbf{R}_2)$ of D$_2$. 
The dual fundamental Weyl chamber is $\psi_1>0$ and  $\phi_1>0$. 
The only weights of the representation  $\mathbf{R}$ are the weights $\pm[1;-1]$. 
It follows that there are two Coulomb chambers, each characterized by the  sign of $\phi_1-\psi_1$. 

In the chamber $\phi_1-\psi_1>0$, the prepotential is  
\begin{equation}
6\mathscr{F}^+_{\text{IMS}}=\phi _1^3 (-n_{\bf{1,2}}-8 n_{\bf{1,3}}-2 n_{\bf{2,2}}+8)+\psi _1^3 (-n_{\bf{2,1}}-8 n_{\bf{3,1}}+8)-6 n_{\bf{2,2}} \psi _1^2 \phi _1 .
\label{IMS.I2nsI2ns.NoZ2}
\end{equation}
In the chamber $\phi_1-\psi_1<0$, the prepotential is  
\begin{equation}
6\mathscr{F}^-_{\text{IMS}}=\phi _1^3 (-n_{\bf{1,2}}-8 n_{\bf{1,3}}+8)+\psi _1^3 (-n_{\bf{2,1}}-2 n_{\bf{2,2}}-8 n_{\bf{3,1}}+8)-6 n_{\bf{2,2}} \psi _1 \phi _1^2 .
\label{IMS.Spin4.m}
\end{equation}

Matching the $5d$ prepotential with the triple intersection numbers gives constraints on the number of representations that are sometimes enough to completely fix them. However, in the present case, they do determine  only the number of bifundamental $(\bf{2},\bf{2})$ while giving two linear relations for the four other numbers:
\begin{align}
n_{\bf{1,2}}+8n_{\bf{1,3}}=T (4 L - 2 S + 2 T) + 8 , \quad n_{\bf{2,1}}+8n_{\bf{3,1}}=S (4 L - 2 T + 2 S) + 8 .
\label{Rel.nR.Spin4}
\end{align}
This can be fixed using two  different methods. 

 Firstly, by using the intersecting brane picture, we compute the number of  fundamentals as the number of intersection points between appropriate components of the reduced discriminant. Since the split curve that gives the representation $(\bf{2},\bf{1})$ is from $s=a_4^2-4a_2a_6=0$, whose class is $[s](2 [a_4])=2S(4L-S-T)$, we deduce that $n_{\bf{2,1}}=2S(4L-S-T)$. Likely, for the representation $(\bf{1},\bf{2})$ is from $t=a_4^2-4a_2a_6=0$, whose class is $[t](2 [a_4])=2T(4L-S-T)$, we can also deduce that $n_{\bf{1,2}}=T(4L-S-T)$. This fixes all the number of representations to be
\begin{equation}\label{eqn:Spin4.matter}
\begin{array}{lll}
n_{\bf{1,2}}=2T(4L-S-T),\quad &n_{\bf{2,1}}=2S(4L-S-T), \quad  &n_{\bf{2,2}}=ST,\\
 n_{\bf{1,3}}=\frac{1}{2} \left(-L T+T^2+2\right)=g_T ,\quad &n_{\bf{3,1}}=\frac{1}{2} \left(-L S+S^2+2\right)=g_S, &
\end{array}
\end{equation}
where $g_C=(K\cdot C+C^2+2)/2$ is the arithmetic genus of a curve $C$. 

 Secondly,   using that Witten's genus formula holds in this context,  the number of adjoint representations $n_{\bf{3,1}}$ and $n_{\bf{1,3}}$ are given respectively  by the genus of the supporting curve $S$ and $T$ as in the last two equations 
of \eqref{eqn:Spin4.matter}. We then determine 
$n_{\bf{2,1}}$ and $n_{\bf{1,2}}$ using the linear relations \eqref{Rel.nR.Spin4} and  thereby reproducing \eqref{eqn:Spin4.matter} after imposing the Calabi-Yau condition  $-K=L$.

The  vanishing of the coefficients of $\tr R^4$ and $\tr F_{i}^4$ are necessary conditions for ensuring that the six-dimensional supergravity theory is  anomaly free.   Since here we are dealing with two SU($2$), we never have a fourth Casimir. 

\subsection{Counting hypermultiplets in $6d$ and anomaly cancellations}
Sadov's F-theory geometric interpretation of the Green-Schwarz anomaly conditions gives the following system of five equations
\begin{align}
\begin{cases}
 A_3(1-n_{\bf 3,1})-A_2( n_{\bf 2,1}+2 n_{\bf 2,2}) =6 K\cdot S,\quad 
 C_3(1-n_{\bf3,1})-C_2(n_{\bf 2,1}+2n_{\bf 2,2}) =-3 S^2,\\
 A_3(1-n_{\bf1,3})-A_2(n_{\bf 1,2}+2 n_{\bf 2,2})=6 K \cdot T, \quad
 C_3(1-n_{\bf 1,3})-C_2(n_{\bf 1,2}+2n_{\bf 2,2})=-3 T^2,\\
 A_1 A_1 n_{\bf 2,2}=  S\cdot T,
\end{cases}
\end{align}
which give\footnote{For the Lie algebra of  SU($2$), we recall that    $A_{\bf{3}}=4,\quad B_{\mathbf{3}}=B_{\mathbf{2}}=0,\quad A_{\bf{2}}=1,\quad  C_{\bf{3}}=8, \quad C_{\bf{2}}=\frac{1}{2}$. See \cite{Avramis:2005hc,Schwarz:1995zw,Erler:1993zy}.}
\begin{align}
 4(1-n_{\bf{3},\bf{1}})-( n_{\bf{2},\bf{1}}+2 n_{\bf{2},\bf{2}})=&\   6 K\cdot S, \quad 
 16(1-n_{\bf{3},\bf{1}})-(n_{\bf{2},\bf{1}}+2n_{\bf{2},\bf{2}})=-6 S^2,\label{SadovS}\\
 4(1-n_{\bf{1},\bf{3}})-(n_{\bf{1},2}+2 n_{\bf{2},\bf{2}})=&\   6 K \cdot T, \quad
 16(1-n_{\bf{1},\bf{3}})-(n_{\bf{1},\bf{2}}+2n_{\bf{2},\bf{2}})=-6 T^2, \label{SadovT}\\
  n_{\mathbf{2},\mathbf{2}} =&\   S\cdot T,\label{SadovST}
\end{align}
The solutions of these linear equations provide a direct computation of the numbers of representations (see  equation \eqref{eqn:Spin4.matter}) from a purely six-dimensional perspective.

Witten's genus formula for $S$ and  $T$ states that the number of hypermultiplets transforming in the adjoint representation is the arithmetic genus of the curve supporting the corresponding fiber. Here it becomes a direct consequence of equations \eqref{SadovS} and  \eqref{SadovT}. 
Equation \eqref{SadovST} is equivalent to the condition obtained by comparing the  triple intersection numbers and the one-loop prepotential  (the coefficient of  $\psi^2_1\phi_1$ of  $\phi_1^2\psi_1$) depending on the chamber) or by counting the number of intersection points of $S$ and $T$.

Since we have geometrically computed the Hodge numbers and the number of representations, we can check if the anomalies are canceled in the uplifted six-dimensional supergravity theory. The number of vector multiplets $n_V^{(6)}$, tensor multiplets $n_T$, and hypermultiplets $n_H$ are
\begin{align}
\begin{split}
n_V^{(6)}&=6, \quad n_T=9-K^2, \\
n_H&=h^{2,1}(Y)+1+n_{\bf{3},\bf{1}}(3-1)+n_{\bf{2},\bf{1}}(2-0)+n_{\bf{2},\bf{2}} (4-0)+n_{\bf{1},\bf{2}}(2-0)+n_{\bf{1},\bf{3}}(3-1) \\
&=29 K^2+15 K S+15 K T+3 S^2+4 S T+3 T^2+13.
\end{split}
\end{align}
Thus, a direct computation shows that the purely gravitational anomalies are also canceled: 
\begin{equation}
n_H-n_V^{(6)}+29n_T-273=0.
\end{equation}
The remaining parts of the anomaly polynomial is 
\begin{align}
I_8&=\frac{K^2}{8} (\tr R^2)^2 +\frac{1}{6} (X^{(2)}_{1} +X^{(2)}_{2}) \tr R^2-\frac{2}{3} (X^{(4)}_{1}+X^{(4)}_{2})+4Y_{12} .
\end{align}
Since A$_1$ does not have quartic Casimir invariants, the coefficients of $\tr F_1^4$ and $\tr F_2^4$ vanish. We also  have  
\begin{align}
X^{(2)}_{1}&=\left(A_{\bf{3}}(1-n_{\bf{3},\bf{1} })-(n_{\bf{2},\bf{1}}+ 2 n_{\bf{2},\bf{2} }) A_{\bf{2}} \right)\tr_{\bf{2}}F^2_1 =6 K S\tr_{\bf{2}}F^2_1,\\
X^{(2)}_{2}&=\left(A_{\bf{3}}(1-n_{\bf{1},\bf{3} })-(n_{\bf{1},\bf{2}}+ 2 n_{\bf{2},\bf{2} }) A_{\bf{2}} \right)\tr_{\bf{2}}F^2_2 =6 K T\tr_{\bf{2}}F^2_2,\\
X^{(4)}_{1}&=\left(C_{\bf{3}}(1-n_{\bf{3}})-(n_{\bf{2},\bf{1}}+ 2 n_{\bf{2},\bf{2} })C_{\bf{2}}\right) (\tr_{\bf{2}}F^2_1)^2=-3 S^2  (\tr_{\bf{2}}F^2_1)^2,\\
X^{(4)}_{2}&=\left(C_{\bf{3}}(1-n_{\bf{3}})-(n_{\bf{1},\bf{2}}+ 2 n_{\bf{2},\bf{2} })C_{\bf{2}}\right) (\tr_{\bf{2}}F^2_1)^2 =-3 T^2 (\tr_{\bf{2}}F^2_1)^2,\\
Y_{12}&=n_{\bf{2},\bf{2}} \tr_{\bf{2}}F^2_1 \tr_{\bf{4}}F^2_2= S T \tr_{\bf{2}}F^2_1 \tr_{\bf{2}}F^2_2 .
\end{align}
It follows that  the anomaly polynomial becomes
\begin{align}
I_8&=\frac{K^2}{8} (\tr R^2)^2 +K(S \tr_{\bf{2}} F_1^2+T \tr_{\bf{2}} F_2^2) \tr R^2+2 S^2 (\tr_{\bf{1}}F^2_1)^2  (\tr_{\bf{2}}F^2_1)^2+4 S T    \tr_{\bf{2}}F^2_1 \tr_{\bf{2}}F^2_2,
\end{align}
which is a perfect square
\begin{equation}
I_8=\frac{1}{2} \left(\frac{1}{2}K \tr R^2+2S \tr_{\bf{2}}F^2_1+2 T\tr_{\bf{2}}F^2_2\right)^2.
\end{equation}
Since the anomaly polynomial is a perfect square, we can safely deduce that all the anomalies are canceled by the Green--Schwarz mechanism.

\section*{Acknowledgements}

The authors are grateful to  Lara Anderson, Richard Derryberry, Ravi Jagadeesan,  Patrick Jefferson,  Julian Salazar, and Shing-Tung Yau for useful discussions. 
This paper answers a question  that was raised during the 2017 workshop on {\em Singular Geometry and Higgs Bundles in String Theory} organized by the 
American Institute of Mathematics  (AIM). 
The authors would like to thank the support and hospitality of AIM and all the participants of the workshop. The authors are particularly thankful to  those who attended the focus group on the mathematics of F-theory: 
 Richard Derryberry, William Donovan, Olivia Dumitrescu,
Matthew Woolf,  and Laura Schaposnik.  
M.E. is supported in part by the National Science Foundation (NSF) grant DMS-1701635  ``Elliptic Fibrations and String Theory''.
M.J.K. would like to acknowledge a partial support from NSF grant PHY-1352084. 
M.J.K. is thankful to Daniel Jafferis for his support and guidance.


\begin{thebibliography}{10}

\bibitem{Aluffi_CBU}
P.~Aluffi.
\newblock Chern classes of blow-ups.
\newblock {\em Math. Proc. Cambridge Philos. Soc.}, 148(2):227--242, 2010.

\bibitem{AE1}
P.~Aluffi and M.~Esole.
\newblock {Chern class identities from tadpole matching in type IIB and
  F-theory}.
\newblock {\em JHEP}, 03:032, 2009.

\bibitem{AE2}
P.~Aluffi and M.~Esole.
\newblock {New Orientifold Weak Coupling Limits in F-theory}.
\newblock {\em JHEP}, 02:020, 2010.

\bibitem{Anderson:2017zfm} 
  L.~B.~Anderson,  M.~Esole, L.~Fredrickson, and L.~P.~Schaposnik,
  Singular geometry and Higgs bundles in string theory,
  arXiv:1710.08453 [math.DG].


\bibitem{Anderson:2017rpr} 
  L.~B.~Anderson, J.~J.~Heckman, S.~Katz and L.~Schaposnik,
  T-Branes at the Limits of Geometry,
  JHEP {\bf 1710}, 058 (2017)
%

\bibitem{Aspinwall:1996nk} 
  P.~S.~Aspinwall and M.~Gross, 
  The SO(32) heterotic string on a K3 surface,
  Phys.\ Lett.\ B {\bf 387}, 735 (1996)
  doi:10.1016/0370-2693(96)01095-7
  [hep-th/9605131].
  
  \bibitem{Aspinwall:1998xj} 
  P.~S.~Aspinwall and D.~R.~Morrison,
   Nonsimply connected gauge groups and rational points on elliptic curves, 
  JHEP {\bf 9807}, 012 (1998)
  [hep-th/9805206].
%



  
  \bibitem{Avramis:2005hc} 
  S.~D.~Avramis and A.~Kehagias,
  A Systematic search for anomaly-free supergravities in six dimensions, 
  JHEP {\bf 0510}, 052 (2005)



\bibitem{Batyrev.Betti}
V.~V. Batyrev.
\newblock Birational {C}alabi-{Y}au {$n$}-folds have equal {B}etti numbers.
\newblock In {\em New trends in algebraic geometry ({W}arwick, 1996)}, volume
  264 of {\em London Math. Soc. Lecture Note Ser.}, pages 1--11. Cambridge
  Univ. Press, Cambridge, 1999.
%
%

\bibitem{Bershadsky:1996nh} 
  M.~Bershadsky, K.~A.~Intriligator, S.~Kachru, D.~R.~Morrison, V.~Sadov and C.~Vafa,
   Geometric singularities and enhanced gauge symmetries,
  Nucl.\ Phys.\ B {\bf 481}, 215 (1996)
  doi:10.1016/S0550-3213(96)90131-5
  [hep-th/9605200].
\bibitem{Bershadsky:1996nu} 
  M.~Bershadsky and A.~Johansen,
  Colliding singularities in F theory and phase transitions,
  Nucl.\ Phys.\ B {\bf 489}, 122 (1997)
  doi:10.1016/S0550-3213(97)00027-8
  [hep-th/9610111].
  
  \bibitem{Bershadsky:1998vn} 
  M.~Bershadsky, T.~Pantev and V.~Sadov,
  F theory with quantized fluxes,
  Adv.\ Theor.\ Math.\ Phys.\  {\bf 3}, 727 (1999)
  doi:10.4310/ATMP.1999.v3.n3.a9
  
  \bibitem{Bonetti:2011mw} 
  F.~Bonetti and T.~W.~Grimm,
  ``Six-dimensional (1,0) effective action of F-theory via M-theory on Calabi-Yau threefolds,''
  JHEP {\bf 1205}, 019 (2012)



\bibitem{Bourbaki.GLA13} N.~Bourbaki, {\it Groups and Lie Algebras. Chap. 1--3}, 
Translated from the 1971 French original. Reprint of the 1989 English translation.
Elements of Mathematics (Berlin). 
Springer-Verlag, Berlin,   1989.   


\bibitem{Bourbaki.GLA79} N.~Bourbaki, {\it Groups and Lie Algebras. Chap. 7-9.}, 
Translated from the 1975 and 1982 French originals by A.~Pressley. Elements of Mathematics (Berlin).
Springer-Verlag, Berlin, 2005.   


\bibitem{Cadavid:1995bk} 
  A.~C.~Cadavid, A.~Ceresole, R.~D'Auria and S.~Ferrara,
  Eleven-dimensional supergravity compactified on Calabi-Yau threefolds,
  Phys.\ Lett.\ B {\bf 357}, 76 (1995)
  

%
\bibitem{Cremmer:1978km} 
  E.~Cremmer, B.~Julia and J.~Scherk,
   Supergravity Theory in Eleven-Dimensions,
  Phys.\ Lett.\  {\bf 76B}, 409 (1978).

	
\bibitem{Deligne.Formulaire}
P.~Deligne.
\newblock Courbes elliptiques: formulaire d'apr{\`e}s {J}. {T}ate.
\newblock In {\em Modular functions of one variable, {IV} ({P}roc. {I}nternat.
  {\text{s}}ummer {\text{s}}chool, {U}niv. {A}ntwerp, {A}ntwerp, 1972)}, pages 53--73.
  Lecture Notes in Math., Vol. 476. Springer, Berlin, 1975.
  
  
    
  \bibitem{Dixon:1986jc}
L.~J. Dixon, J.~A. Harvey, C.~Vafa, and E.~Witten.
\newblock {Strings on Orbifolds. 2.}
\newblock {\em Nucl. Phys.}, B274:285--314, 1986.

  
  \bibitem{MR1242006}
I.~Dolgachev and M.~Gross.
\newblock Elliptic threefolds. {I}. {O}gg-{S}hafarevich theory.
\newblock {\em J. Algebraic Geom.}, 3(1):39--80, 1994.

\bibitem{Dolgavcev.Purity}
I.~V. Dolgachev.
\newblock On the purity of the degeneration loci of families of curves.
\newblock {\em Invent. Math.}, 8:34--54, 1969.


  \bibitem{Erler:1993zy} 
  J.~Erler,
   Anomaly cancellation in six-dimensions, 
  J.\ Math.\ Phys.\  {\bf 35}, 1819 (1994).

  \bibitem{Esole.Elliptic} 
  M.~Esole,
   Introduction to Elliptic Fibrations, 
  Math.\ Phys.\ Stud.\  {\bf 9783319654270}, 247 (2017).

  \bibitem{EJJN1}
M.~Esole, S.~G. Jackson, R.~Jagadeesan, and A.~G. No{\"e}l.
\newblock {Incidence Geometry in a Weyl Chamber I: GL$_n$}, 
\newblock arXiv:1508.03038 [math.RT].



\bibitem{EJJN2} 
  M.~Esole, S.~G.~Jackson, R.~Jagadeesan and A.~G.~No{\"e}l,
  Incidence Geometry in a Weyl Chamber II: $SL_n$,
  arXiv:1601.05070 [math.RT].
  
  
  
\bibitem{G2} 
  M.~Esole, R.~Jagadeesan and M.~J.~Kang,
  The Geometry of G$_2$, Spin(7), and Spin(8)-models,
  arXiv:1709.04913 [hep-th].


\bibitem{Euler} 
  M.~Esole, P.~Jefferson and M.~J.~Kang,
  Euler Characteristics of Crepant Resolutions of Weierstrass Models,
  arXiv:1703.00905 [math.AG].

\bibitem{EKY2} 
  M.~Esole, M.~J.~Kang and S.~T.~Yau,
  ``Mordell--Weil Torsion, Anomalies, and Phase Transitions,''
  arXiv:1712.02337 [hep-th].


\bibitem{EKY1} 
  M.~Esole, M.~J.~Kang and S.~T.~Yau,
  A New Model for Elliptic Fibrations with a Rank One Mordell--Weil Group: I. Singular Fibers and Semi-Stable Degenerations,
  arXiv:1410.0003 [hep-th].

\bibitem{F4} 
  M.~Esole, P.~Jefferson and M.~J.~Kang,
  The Geometry of F$_4$-Models,
  arXiv:1704.08251 [hep-th].
  
  
   
  
  
  \bibitem{Esole:2012tf} 
  M.~Esole and R.~Savelli,
  Tate Form and Weak Coupling Limits in F-theory,
  JHEP {\bf 1306}, 027 (2013)

\bibitem{ES} 
  M.~Esole and S.~H.~Shao,
   M-theory on Elliptic Calabi-Yau Threefolds and $6d$ Anomalies, 
  arXiv:1504.01387 [hep-th].


\bibitem{ESY1}
M.~Esole, S.-H. Shao, and S.-T. Yau.
\newblock {Singularities and Gauge Theory Phases}.
\newblock {\em Adv. Theor. Math. Phys.}, 19:1183--1247, 2015.

\bibitem{ESY2} 
  M.~Esole, S.~H.~Shao and S.~T.~Yau,
  Singularities and Gauge Theory Phases II,
  Adv.\ Theor.\ Math.\ Phys.\  {20}, 683 (2016)



\bibitem{EY} 
  M.~Esole and S.~T.~Yau,
  Small resolutions of SU(5)-models in F-theory,
  Adv.\ Theor.\ Math.\ Phys.\   {17}, no. 6, 1195 (2013)
  
  

\bibitem{EFY}
M.~Esole, J.~Fullwood, and S.-T. Yau.
\newblock {$D_5$ elliptic fibrations: non-Kodaira fibers and new orientifold
  limits of F-theory}.
\newblock Commun.\ Num.\ Theor.\ Phys.\  {\bf 09}, no. 3, 583 (2015).
%


\bibitem{Fullwood:SVW}
J.~Fullwood.
\newblock {On generalized Sethi-Vafa-Witten formulas}.
\newblock {\em J. Math. Phys.}, 52:082304, 2011.
%


\bibitem{GM1}
A.~Grassi and D.~R. Morrison.
\newblock Group representations and the Euler characteristic of elliptically
  fibered Calabi-Yau threefolds.
\newblock {\em J. Algebraic Geom.}, 12(2):321--356, 2003.


\bibitem{GM2} 
  A.~Grassi and D.~R.~Morrison,
  ``Anomalies and the Euler characteristic of elliptic Calabi-Yau threefolds,''
  Commun.\ Num.\ Theor.\ Phys.\  {\bf 6}, 51 (2012)
  doi:10.4310/CNTP.2012.v6.n1.a2
%

\bibitem{Green:1984bx} 
  M.~B.~Green, J.~H.~Schwarz and P.~C.~West,
   Anomaly Free Chiral Theories in Six-Dimensions, 
  Nucl.\ Phys.\ B {\bf 254}, 327 (1985).
  
  
  
\bibitem{Grimm:2011fx} 
  T.~W.~Grimm and H.~Hayashi,
  F-theory fluxes, Chirality and Chern-Simons theories,
  JHEP {\bf 1203}, 027 (2012)

  
  \bibitem{Grimm:2015zea} 
  T.~W.~Grimm and A.~Kapfer,
  Anomaly Cancelation in Field Theory and F-theory on a Circle,
  JHEP {\bf 1605}, 102 (2016)
  doi:10.1007/JHEP05(2016)102
  [arXiv:1502.05398 [hep-th]].
  

\bibitem{Hayashi:2014kca}
H.~Hayashi, C.~Lawrie, D.~R. Morrison, and S.~Sch\"afer-Nameki.
\newblock {Box Graphs and Singular Fibers}.
\newblock {\em JHEP}, 1405:048, 2014.


\bibitem{Humphreys}
J.~Humphreys, {\it Introduction to Lie Algebras and Representation Theory}, Graduate Texts in Mathematics 9, Springer-Verlag, New York, 1972. 

\bibitem{Husemoller} D.~Husem\"oller, {\it Elliptic curves}. Second edition. With appendices by Otto Forster, Ruth Lawrence and Stefan Theisen. Graduate Texts in Mathematics, 111. Springer-Verlag, New York, 2004. 


\bibitem{IMS}
K.~A. Intriligator, D.~R. Morrison, and N.~Seiberg.
\newblock {Five-dimensional supersymmetric gauge theories and degenerations of
  Calabi-Yau spaces}.
\newblock {\em Nucl.Phys.}, B497:56--100, 1997.



\bibitem{Kodaira}
K.~Kodaira.
\newblock On compact analytic surfaces. {II}, {III}.
\newblock {\em Ann. of Math. (2) 77 (1963), 563--626; ibid.}, 78:1--40, 1963.

\bibitem{Kontsevich.Orsay}
M.~Kontsevich.
\newblock String cohomology, December 1995.
\newblock Lecture at Orsay.



  


\bibitem{Marsano}
J.~Marsano and S.~Sch\"afer-Nameki.
\newblock {Yukawas, G-flux, and Spectral Covers from Resolved Calabi-Yau's}.
\newblock {\em JHEP}, 11:098, 2011.


\bibitem{Matsuki.Weyl}
K.~Matsuki,  Weyl groups and birational transformations among minimal
  models, \href{http://dx.doi.org/10.1090/memo/0557}{{\em Mem. Amer. Math.
  Soc.} {\bfseries 116} no.~557, (1995) vi+133}.
  
  \bibitem{Mayrhofer:2014opa} 
  C.~Mayrhofer, D.~R.~Morrison, O.~Till and T.~Weigand,
   Mordell--Weil Torsion and the Global Structure of Gauge Groups in F-theory, 
  JHEP {\bf 1410}, 16 (2014)
  [arXiv:1405.3656 [hep-th]].


  
\bibitem{Miranda}
R.~Miranda.
\newblock Smooth models for elliptic threefolds.
\newblock In {\em The birational geometry of degenerations ({C}ambridge,
  {M}ass., 1981)}, volume~29 of {\em Progr. Math.}, pages 85--133.
  Birkh{\"a}user Boston, Mass., 1983.
  
  \bibitem{Morrison:1996pp} 
  D.~R.~Morrison and C.~Vafa,
  Compactifications of F theory on Calabi-Yau threefolds. 2.,
  Nucl.\ Phys.\ B {\bf 476}, 437 (1996)
  doi:10.1016/0550-3213(96)00369-0
  [hep-th/9603161].

\bibitem{Morrison:1996na} 
  D.~R.~Morrison and C.~Vafa,
   Compactifications of F theory on Calabi-Yau threefolds. 1, 
  Nucl.\ Phys.\ B {\bf 473}, 74 (1996)
  doi:10.1016/0550-3213(96)00242-8
  [hep-th/9602114].

 
  
  \bibitem{Morrison:2014era} 
  D.~R.~Morrison and W.~Taylor,
  Sections, multisections, and U(1) fields in F-theory,
  Journal of Singularities, 
volume 15 (2016), 126-149.

\bibitem{Nakayama}
N.~Nakayama,  Elliptic fibrations over surfaces. I,  Algebraic geometry and analytic geometry (Tokyo, 1990),  pp. 126--137,  ICM-90 Satell. Conf. Proc.,  Spinger, Tokyo, 1991. 



\bibitem{Neron}A. N\'eron, Mod\`eles Minimaux des Vari\'et\'es Ab\'eliennes sur les Corps Locaux et
Globaux, Publ. Math. I.H.E.S. 21, 1964, pp. 361--482.


\bibitem{RandjbarDaemi:1985wc} 
  S.~Randjbar-Daemi, A.~Salam, E.~Sezgin and J.~A.~Strathdee,
  ``An Anomaly Free Model in Six-Dimensions,''
  Phys.\ Lett.\  {\bf 151B}, 351 (1985).


  \bibitem{Sadov:1996zm} 
  V.~Sadov,
  Generalized Green-Schwarz mechanism in F theory,
  Phys.\ Lett.\ B {\bf 388}, 45 (1996)
  doi:10.1016/0370-2693(96)01134-3
  [hep-th/9606008].
  
  \bibitem{Sagnotti:1992qw} 
  A.~Sagnotti,
  A Note on the Green-Schwarz mechanism in open string theories,
  Phys.\ Lett.\ B {\bf 294}, 196 (1992)
  doi:10.1016/0370-2693(92)90682-T
  [hep-th/9210127].
  
  \bibitem{Schwarz:1995zw} 
  J.~H.~Schwarz,
  Anomaly - free supersymmetric models in six-dimensions,
  Phys.\ Lett.\ B {\bf 371}, 223 (1996)
  
  \bibitem{Sen:1997kw} 
  A.~Sen,
  F theory and the Gimon-Polchinski orientifold,
  Nucl.\ Phys.\ B {\bf 498}, 135 (1997)
  doi:10.1016/S0550-3213(97)00262-9
  
  \bibitem{Szydlo.Thesis}
M.~G. Szydlo.
\newblock {\em Flat regular models of elliptic schemes}.
\newblock ProQuest LLC, Ann Arbor, MI, 1999.
\newblock Thesis (Ph.D.)--Harvard University.
  


\bibitem{Vafa:1996xn} 
  C.~Vafa,
   Evidence for F theory,
  Nucl.\ Phys.\ B {\bf 469}, 403 (1996)


\bibitem{Wazir}
R.~Wazir.
\newblock Arithmetic on elliptic threefolds.
\newblock {\em Compositio Mathematica}, 140(03):567--580, 2004.



\end{thebibliography}
\end{document}